\documentclass{article}

\usepackage{graphicx} 
\usepackage{subfigure} 
\usepackage{placeins} 
\usepackage{comment}
\usepackage{enumerate}
\usepackage{dblfloatfix}
\usepackage[named]{algo}

\usepackage{setspace}

\usepackage[hmargin=1in,vmargin=1in]{geometry}

\usepackage{amsmath}
\usepackage{amssymb}
\usepackage{amsthm}
\usepackage{upgreek}
\usepackage{mathrsfs}
\theoremstyle{plain}
\newtheorem{theorem}{Theorem}
\newtheorem{shortassumption}{A\!\!}

\newtheorem{proposition}{Proposition}

\newtheorem{corollary}{Corollary}
\newtheorem{lemma}{Lemma}

\theoremstyle{definition}
\newtheorem{definition}{Definition}

\theoremstyle{remark}

\newcommand{\B}{\mathbb{B}}
\newcommand{\C}{\mathbb{C}} 
\newcommand{\R}{\mathbb{R}}

\renewcommand{\hat}[1]{\widehat{#1}}

\newcommand{\A}{\mathcal{A}}

\newcommand{\F}{\mathbb{F}}

\renewcommand{\P}{\mathbb{P}}
\newcommand{\E}{\mathbb{E}}
\newcommand{\Z}{\mathbb{Z}}

\newcommand{\myRe}{\text{Re}}

\newcommand{\as}{\text{ a.s.}}
\newcommand{\e}{\epsilon}
\newcommand{\ve}{\varepsilon}

\newcommand{\op}{_{\text{op}}}

\newcommand{\tsum}{\textstyle\sum}
\newcommand{\ts}{\textstyle}

\newcommand{\err}{\text{err}}
\newcommand{\argmin}{\text{argmin}}

\newcommand{\med}{\text{median}}
\newcommand{\opt}{\text{opt}}

\newcommand{\stable}{\text{stable}}
\newcommand{\Log}{\text{Log}_+}
\newcommand{\tpilot}{\hat{t}_{\text{pilot}}}

\newcommand{\topt}{\hat{t}_{\textup{opt}}}

\newcommand{\mnorm}[1]{\left\vert\kern-1.5pt\left\vert\kern-1.5pt\left\vert #1\right\vert\kern-1.5pt\right\vert\kern-1.5pt\right\vert}

\begin{document} 

\begin{center}

{\textbf{\LARGE{Unknown Sparsity in Compressed Sensing:\\ Denoising and Inference}}}

\vspace*{.3in}

{\large{Miles E. Lopes\footnote{In the current journal-length paper, we retain some of the background and introductory material from the conference paper~\cite{lopesICML}, but the results and methods of the current paper are essentially new.}}}\\[0.2cm]

{\large{\texttt{melopes@ucdavis.edu}}}

{\large{
\vspace*{.2in}
Department of Statistics\\[0.2cm]
University of California, Davis\\[0.2cm]

August 10, 2015
}}
\end{center}
~\\
\begin{abstract}

The theory of Compressed Sensing (CS) asserts that an unknown signal $x\in\R^p$ can be accurately recovered from an underdetermined set of $n$ linear measurements with $n\ll p$, provided that $x$ is sufficiently sparse.
However, in applications, the degree of sparsity $\|x\|_0$ is typically unknown, and the problem of directly estimating $\|x\|_0$ has been a longstanding gap between theory and practice. A closely related issue is that $\|x\|_0$ is a highly idealized measure of sparsity, and for real signals with entries not exactly equal to 0, the value $\|x\|_0=p$ is not a useful description of compressibility. In our previous conference paper that examined these problems,~\cite{lopesICML}, we considered an alternative measure of ``soft'' sparsity, $\|x\|_1^2/\|x\|_2^2$, and designed a procedure to estimate $\|x\|_1^2/\|x\|_2^2$ that does not rely on sparsity assumptions.

The present work offers a new deconvolution-based method for estimating unknown sparsity, which has wider applicability and sharper theoretical guarantees. Whereas our earlier work was limited to estimating the quantity $\|x\|_1^2/\|x\|_2^2$, the current paper introduces a family of entropy-based sparsity measures $s_q(x):=\big(\frac{\|x\|_q}{\|x\|_1}\big)^{\frac{q}{1-q}}$ parameterized by $q\in[0,\infty]$. This family interpolates between $\|x\|_0=s_0(x)$ and $\|x\|_1^2/\|x\|_2^2=s_2(x)$ as $q$ ranges over $[0,2]$, and our proposed method allows $s_q(x)$ to be estimated for all $q\in (0,2]$. Two other main advantages of the new approach are that it handles measurement noise with \emph{infinite variance}, and that it yields confidence intervals for $s_q(x)$ with asymptotically \emph{exact} coverage probability (whereas our previous intervals were conservative).

In addition to confidence intervals, we analyze several other aspects of our proposed estimator $\hat{s}_q(x)$. 
An important property of $\hat{s}_q(x)$ is that its relative error converges at the \emph{dimension-free} rate of $1/\sqrt{n}$. This means that using only $n=\mathcal{O}(1)$ measurements, $s_q(x)$ can be estimated to any fixed degree of relative error, even when $p/n\to\infty$. Next, in connection with recovering the full signal $x$, we give new insight into the role of $s_2(x)$ by deriving matching upper and lower bounds on the relative error of the Basis Pursuit Denoising (BPDN) algorithm, at rate $\sqrt{s_2(x)\log(pe/n)/n}$. Finally, since our proposed method is based on randomized measurements, we show that the use of randomization is essential. Specifically, we show that the minimax relative error for estimating $s_q(x)$ with noiseless deterministic measurements is at least of order 1 when $n<p$ and $q\in[0,2]$.\\

\end{abstract}

\section{Introduction}\label{sec:intro}
In this paper, we consider the standard compressed sensing (CS) model, involving $n$ linear measurements $y=(y_1,\dots,y_n)$, generated according to
\begin{equation}\label{setup1}
y=Ax+\sigma\e,
\end{equation}
where $x\in\R^p$ is an unknown signal, $A\in\R^{n\times p}$ is a measurement matrix specified by the user, $\sigma \e\in\R^n$ is a random noise vector, and $n\ll p$. The central problem of CS is to recover the signal $x$ using only the observations $y$ and the matrix $A$. Over the course of the past decade, a large body of research has shown that  this seemingly ill-posed problem can be solved reliably when $x$ is sparse. Specifically, when the sparsity level of $x$ is measured in terms of the $\ell_0$ norm \mbox{$\|x\|_0:=\text{card}\{j : x_j \neq 0\}$}, it is well known that if \mbox{$n\gtrsim \|x\|_0\log(p)$}, then accurate recovery can be achieved with high probability when $A$ is drawn from a suitable ensemble~\cite{donoho2006,candes2006stable,eldarintro,foucartbook}. 
In this way, the parameter $\|x\|_0$ is often treated as being known in much of the theoretical CS literature --- despite the fact that $\|x\|_0$ is usually \emph{unknown} in practice. Due to the fact that the sparsity parameter plays a fundamental role in CS, the issue of unknown sparsity has become recognized as gap between theory and practice \cite{wardCV,eldarStein,willsky,boufounos}. Likewise, our overall focus in this paper is the problem of estimating the unknown sparsity level of $x$ without relying on any sparsity assumptions.

\subsection{Motivations and the role of sparsity}\label{sec:role}
Given that many well-developed methods are available for estimating the full signal $x$, or its support set $S:=\{j\in\{1,\dots,p\}: x_j\neq 0\}$, it might seem surprising that the problem of estimating  $\|x\|_0$ has remained largely unsettled.
Indeed, given an estimate of $x$ or $S$, it might seem natural to estimate $\|x\|_0$ via a ``plug-in rule'', such as $\|\hat{x}\|_0$ or $\text{card}(\hat{S})$. However, it is important to recognize that methods for computing $\hat{x}$ and $\hat{S}$ generally rely on prior knowledge of $\|x\|_0$. Consequently, when using a plug-in rule to estimate $\|x\|_0$, there is a danger of circular reasoning, and as a result, the problem of estimating $\|x\|_0$ does not simply reduce to estimating $x$ or $S$.

To give a more concrete sense for the importance of estimating unknown sparsity, the following list illustrates many aspects of CS where sparsity assumptions play a role, and where it would be valuable to have an ``assumption-free'' estimate of $\|x\|_0$.

\begin{enumerate}

 \item {\textbf{Modeling assumptions and choice of basis.}} In some signal processing applications, sparsity-based methods are not the only viable approach, and it is of basic interest to know whether not a sparse representation is justified by data. For instance, this issue has been actively studied in the areas of face recognition and image classification:~\cite{shi2011face,defenseSparsity,sparsityInSecurity,rigamonti2011sparse}.
In this context, an assumption-free estimate of $\|x\|_0$ would serve as a natural diagnostic tool in model development.

A second issue that is related to model development is the choice of basis used to represent a signal.  Although there are many application-specific bases (e.g. various types of wavelet bases) that often lead to sparse representations, the ability to ``validate'' the choice of basis has clear practical value. In this direction, an estimator of $\|x\|_0$ could be of use in comparing the relative merits of different bases.

\item {\textbf{The number of measurements.}} 
If the choice of $n$ is too small compared to the ``critical'' number $n^*\approx \|x\|_0\log(p)$, then there are known information-theoretic barriers to the accurate reconstruction of $x$~\cite{raskutti}.
At the same time, if $n$ is chosen to be much larger than $n^*$, then the measurement process is wasteful (since there are known algorithms that can reliably recover $x$ with approximately $n^*$ measurements~\cite{eldarintro}). For this reason, it not only important to ensure that $n\geq n^*$, but to choose $n$ close to $n^*$.

To deal with the selection of $n$, a sparsity estimate $\hat{\|x\|}_0$ may be used in two different ways,  depending on whether measurements are collected sequentially, or in a single batch. In the sequential case, an estimate of $\|x\|_0$ can be computed from a small set of ``preliminary'' measurements, and then the estimated value $\hat{\|x\|}_0$ determines how many additional measurements should be collected to recover the full signal. Also, it may not even be necessary to take additional measurements, since the preliminary set may be re-used to compute $\hat{x}$.  Alternatively, if all of the measurements must be taken in one batch, the value $\widehat{\|x\|}_0$ can be used to certify whether or not enough measurements were actually taken.

\item {\textbf{The measurement matrix.}} The performance of recovery procedures depends heavily on the sensing matrix $A$. In particular, the properties of $A$ that lead to good recovery are often directly linked to the sparsity level of $x$. Two specific properties that have been intensively studied are
the  \emph{restricted isometry property of order $k$} (RIP-$k$),~\cite{candes2005}, and the \emph{null-space property of order $k$} (NSP-$k$),\cite{devore,donohonsp}, where $k$ is a presumed upper bound on the sparsity level of the true signal.
Because recovery guarantees are closely tied to RIP-$k$
 and NSP-$k$, a growing body of work has been devoted to certifying whether or not a given matrix satisfies these properties~\cite{daspremont,nemirovski, tangnehorai}.
When $k$ is treated as given, this problem is already computationally difficult. Yet, when the sparsity of $x$ is unknown, we must also remember that such a ``certificate'' is less meaningful if we cannot check that the value of $k$ agrees with the true signal.

 \item {\textbf{Recovery algorithms.}} When recovery algorithms are implemented, the sparsity level of $x$ is often treated as a tuning parameter. For example, if $k$ is a conjectured bound on $\|x\|_0$, then the Orthogonal Matching Pursuit algorithm (OMP) is typically initialized to run for $k$ iterations~\cite{troppOMP}. A second example is the Lasso algorithm, which computes a solution \mbox{$\hat{x}\in\argmin\{ \|y-Av\|_2^2+\lambda\|v\|_1 : v\in\R^p\}$,} for  some choice of $\lambda\geq 0$. The sparsity of $\hat{x}$ is determined by the size of $\lambda$, and in order to select the appropriate value, a family of solutions is examined over a range of $\lambda$ values~\cite{tibshiraniPaths}. 
 In the case of either OMP or Lasso, a sparsity estimate $\hat{\|x\|}_0$ would reduce computation by restricting the possible choices of $\lambda$ or $k$, and it would also ensure that the sparsity level of the solution conforms to the true signal.
 With particular regard to the Lasso, an indirect consequence of our sparsity estimation method (introduced in Section~\ref{sec:proc}) is that it allows for regularization parameter to be adaptively selected when the Lasso problem is written in ``primal form'': $\hat{x}\in \argmin\{ \|y-Av\|_2^2 : v\in\R^p \text{ and } \|v\|_1\leq t\}$. See Section~\ref{sec:app} for further details.

 \end{enumerate}

\subsection{A numerically stable measure of sparsity}
Despite the important theoretical role of the parameter $\|x\|_0$, it has a severe practical drawback of being sensitive to small entries of $x$. In particular, for real signals $x\in\R^p$ whose entries are not exactly equal to 0, the value $\|x\|_0=p$ is not a useful description of compressibility.
In order to estimate sparsity in a way that accounts for the instability of $\|x\|_0$, it is desirable to replace the $\ell_0$ norm with a ``soft'' version. More precisely, we would like to identify a function of $x$ that can be interpreted as counting the ``effective number of coordinates of $x$'', but remains stable under small perturbations.  In the next subsection, we derive such a function by showing that $\|x\|_0$ is a limiting case of a more general sparsity measure based on entropy. 

\subsubsection{A link between $\|x\|_0$ and entropy}\label{sec:entropy}
\noindent Any vector $x\in\R^p\setminus\{0\}$ induces a distribution $\pi(x)\in\R^p$ on the set of indices $\{1,\dots,p\}$, assigning mass $\pi_j(x):=|x_j|/\|x\|_1$ at index $j$.\footnote{It is also possible to normalize $\pi(x)$ in other ways, e.g. $\pi_j(x)=|x_j|^2/\|x\|_2^2$. See the end of Section~\ref{properties} for additional comments.}
Under this correspondence, if $x$ places most of its mass at a small number of coordinates, and $J\sim \pi(x)$ is a random variable in $\{1,\dots,p\}$, then $J$ is likely to occupy a small set of \emph{effective states}. This means that if $x$ is sparse, then $\pi(x)$ has low entropy. From the viewpoint of information theory, it is well known that the entropy of a distribution can be interpreted as the logarithm of the distribution's effective number of states. Likewise, it is natural to count effective coordinates of $x$ by counting effective states of $\pi(x)$ via entropy. To this end, we define the \emph{numerical sparsity}\footnote{ Our terminology derives from the notion of \emph{numerical rank} coined by~\cite{vershyninnumerical}. }
\begin{align}\label{srenyi}
s_q(x) &:= 
 \begin{cases} \exp(H_q(\pi(x)))  & \text{if } \  x\neq 0 \\ 
0 & \mbox{if }\  x=0,
\end{cases}
\end{align}
where $H_q$ is the R\'enyi entropy of order $q\in [0,\infty]$.
 When $q\not\in\{0,1,\infty\}$, the R\'enyi entropy is given explicitly by 
\begin{equation}
H_q(\pi(x)) := \ts\frac{1}{1-q}\log\big(\tsum_{i=1}^p \pi_i(x)^q\big),
\end{equation}
and cases of $q\in \{0,1,\infty\}$ are defined by evaluating limits, with $H_1$ being the ordinary Shannon entropy. 
Combining the last two lines with the definition of $\pi(x)$, we see that for $x\neq 0$ and $q\not\in\{0,1,\infty\}$, the numerical sparsity may be written conveniently in terms of $\ell_q$ norms as
\begin{equation}
s_q(x)=\left(\displaystyle\frac{\|x\|_q}{\|x\|_1}\right)^{\frac{q}{1-q}}.
\end{equation}
As with $H_q$, the the cases of $q\in \{0,1,\infty\}$ are  evaluated as limits:
\begin{align}\label{conv}
 s_0(x)& =\lim_{q\to 0} s_q(x) \ = \ \|x\|_0\\[0.2cm]
 s_1(x)&=\lim_{q\to 1} s_q(x) \ = \ \exp(H_1(\pi(x))) \\[0.2cm]
 s_{\infty}(x)&=\lim_{q\to \infty} s_q(x) = \ \ts\frac{\|x\|_1}{\ \|x\|_{\infty}}.
\end{align}

\subsubsection{Background on the definition of $s_q(x)$}\label{origin} To the best of our knowledge, the definition of numerical sparsity~\eqref{srenyi} in terms of R\'enyi entropy is new in the context of CS. However, numerous special cases and related definitions have been considered elsewhere. For instance, in the early study of wavelet bases, Coifman and Wickerhauser proposed \mbox{$\exp\big(-\sum_{i=1}^p \frac{|x_i|^2}{\|x\|_2^2}\log(\frac{|x_i|^2}{\|x\|_2^2})\big)$} as a measure of effective dimension~\cite{coifmanentropy}. (See also the papers \cite{raoaffine}~\cite{donohoentropy}~\cite{measures}.) The important difference between this quantity and $s_q(x)$ is that the R\'enyi entropy leads instead to a convenient ratio of norms, which will play an essential role in our procedure for estimating $s_q(x)$.

In recent years, there has been growing interest in ratios of norms as measures of sparsity, but such ratios have generally been introduced in an ad-hoc manner, and there has not been a principled way to explain where they ``come from''. To this extent, our definition of $s_q(x)$ offers a way of conceptually unifying these ratios.\footnote{See also our discussion of  analogues of $s_q(x)$ for matrix rank in Section~\ref{rank}.} Examples  of previously studied instances include  $\|x\|_1^2/\|x\|_2^2$ corresponding to $q=2$~\cite{lopesICML},\cite{tangnehorai},~\cite{hoyer}, $\|x\|_1/\|x\|_{\infty}$ corresponding to $q=\infty$~\cite{pilanci}~\cite{demanet}, as well as a ratio of the form $(\|x\|_a/\|x\|_b)^{ab/(b-a)}$ with $a,b>0$, which is implicit in the paper~\cite{barak}.\footnote{This quantity is implicitly suggested in the paper~\cite{barak} by considering a binary vector $x$ with $\|x\|_0=k\geq 1$, and then choosing an exponent $c$ so that $(\|x\|_a/\|x\|_b)^c=k$.}
To see how the latter quantity fits in the scope of our entropy-based definition, one may consider a different normalization of the probability vector $\pi(x)$ discussed earlier. That is, if we put $\pi_j(x)=|x_j|^t/\|x\|_t^t$ for some $t>0$, then it follows that $\exp(H_q(\pi(x))) = (\|x\|_{tq}/\|x\|_t)^{tq/(1-q)}$. Furthermore, if one chooses $t=b$ and $q=a/b$, then the two quantities match.

Outside the context of CS, the use of R\'enyi entropy to count the effective number of states of a distribution has been well-established in the ecology literature for a long time. There, R\'enyi entropy is used to count the effective number of species in a community of organisms. More specifically, if a distribution $\pi$ on $\{1,\dots,p\}$ measures the relative abundance of $p$ species in a community, then the number $\exp(H_q(\pi))$ is a standard measure of the \emph{effective number of species} in the community. In the ecology literature, this number is known as the \emph{Hill index} or \emph{diversity number} of the community. We refer the reader to the papers \cite{hill1973} and~\cite{jost}, as well as the references therein for further details. In essence, the main conceptual ingredient needed to connect these ideas with the notion of sparsity in CS is to interpret the signal $x\in\R^p$ as a distribution on the set of indices $\{1,\dots,p\}$.

\subsubsection{Properties of $s_q(x)$}\label{properties} The following list summarizes some of the most important properties of $s_q(x)$, and clarifies the interpretation of $s_q(x)$ as a measure of sparsity.

 \begin{enumerate}[(i)]

 \item   {\textbf{(continuity).}} Unlike the $\ell_0$ norm, the function $s_q(\cdot)$ is continuous on \mbox{$\R^p\setminus\{0\}$} for all $q>0$, and is hence stable under small perturbations of $x$.
 
 \item {\textbf{(range equal to $[0,p]$).}} For all $x\in\R^p$ and all $q\in[0,\infty]$, the numerical sparsity satisfies
 $$0\leq s_q(x)\leq p.$$
This property follows from the fact that for any $q$, and any distribution $\pi$ on $\{1,\dots,p\}$, the R\'enyi entropy satisfies $0\leq H_q(\pi)\leq \log(p)$.
 
 \item {\textbf{(scale-invariance).}} The property that $\|cx\|_0=\|x\|_0$ for all scalars $c\neq 0$ is familiar for the $\ell_0$ norm, and this generalizes to $s_q(x)$ for all $q\in[0,\infty]$. Scale-invariance encodes the idea that sparsity should be based on relative (rather than absolute) magnitudes of the entries of $x$.

\item {\textbf{(lower bound on $\|x\|_0$ and monotonicity in $q$).}} For any $x\in\R^p$, the function $q\mapsto s_q(x)$ is monotone decreasing on $[0,\infty]$, and interpolates between the extreme values of $s_{\infty}(x)$ and $s_0(x)$. That is, for any $q'\geq q\geq 0$, we have the bounds
\begin{equation}
\ts\frac{\|x\|_1}{\ \|x\|_{\infty}}=s_{\infty}(x)\leq s_{q'}(x)\leq s_q(x)\leq s_0(x)= \|x\|_0.
\end{equation}
In particular, we have the general lower bound
\begin{equation}
s_q(x)\leq \|x\|_0.
\end{equation}
The monotonicity is a direct consequence of the fact that the R\'eny entropy $H_q$ is decreasing in $q$.

 \item {\textbf{(Schur concavity).}}  The notion of \emph{majorization} formalizes the idea that the  coordinates of a vector $x\in\R^p$ are more ``spread out'' than those of another vector $\tilde{x}\in\R^p$. (See the book \cite{olkin} for an in-depth treatment of majorization.)  If $x$ is majorized by $\tilde{x}$, we write $x\prec \tilde{x}$, where larger vectors in this partial order have coordinates that are less spread out.   From this interpretation, one might expect that if $|x|\prec |\tilde{x}|$, then $\tilde{x}$ should be sparser than $x$, where $|x|:=(|x_1|,\dots,|x_p|)$. It turns out that this intuition is respected by $s_q(\cdot)$, in the sense that for any $q\in[0,\infty]$,
 \begin{equation}\label{schur}
 |x|\prec |\tilde{x}| \ \Longrightarrow \ s_q(x)\geq s_q(\tilde{x}).
 \end{equation}
In general, if a function $f$ satisfies $f(x)\geq f(\tilde{x})$ for all $x\prec \tilde{x}$ with $x,\tilde{x}$ lying in a set $S$, then $f$ is said to be \emph{Schur concave} on $S$.  Consequently, line~\eqref{schur} implies that $s_q(\cdot)$ is Schur concave on the orthant $\R_+^p\setminus\{0\}$. This property follows easily from the fact that the R\'enyi entropy is Schur concave on the $p$-dimensional probability simplex.

 \end{enumerate}

\paragraph{The choice of $q$ and normalization.} The parameter $q$ controls how much weight $s_q(x)$ assigns to small coordinates. When $q=0$, an arbitrarily small coordinate is still counted as being ``effective''. By contrast, when $q=\infty$, a coordinate is not counted as being effective unless its magnitude is close to $\|x\|_{\infty}$.  
The choice of $q$ is also relevant to other considerations. For instance, we will show in Section~\ref{sec:recov} that the case of $q=2$ is important  because signal recovery guarantees can be derived in terms of
\begin{equation}
s_2(x)=\frac{\|x\|_1^2}{\|x\|_2^2}.
\end{equation}
In addition, the choice of $q$ can affect the type of measurements used to estimate $s_q(x)$. In this respect, the case of $q=2$ turns out to be attractive because our proposed method for estimating $s_2(x)$ relies on Gaussian measurements --- which can be naturally \emph{re-used} for recovering the full signal $x$. Furthermore, in some applications, the measurements associated with one value of $q$ may be easier to acquire (or process) than another. In our proposed method, smaller values of $q$ lead to measurement vectors sampled from a distribution with heavier tails. Because a vector with heavy tailed i.i.d.~entries will tend to have just a few very large entries, such vectors are approximately sparse. In this way, the choice of $q$ may enter into the design of measurement systems because it is known that sparse measurement vectors can simplify certain recovery procedures~\cite{gilbertIndyk}. 

Apart from the choice of $q$, there is a second degree of freedom associated with $s_q(x)$. In defining the probability vector $\pi_j(x)=|x_j|/\|x\|_1$ earlier, we were not forced to normalize the mass of the coordinates using $\|x\|_1$, and many other normalizations are possible. Also, some normalizations may be  computationally advantageous for the problem of minimizing $s_q(x)$, as discussed below.

\paragraph{Minimization of $s_q(x)$.}

Our work in this paper does not require the minimization of $s_q(x)$, and neither do the applications we propose. Nevertheless, some readers may still naturally be curious about what can be done in this direction.
It turns out that for certain values of $q$, or certain normalizations of the probability vector $\pi(x)$, the minimization of $s_q(x)$ may be algorithmically tractable (under suitable constraints).  The recent paper~\cite{euclidtaxicab} discusses methods for minimizing $s_2(x)$. Another example is the minimization of $s_{\infty}(x)$, which can be reduced to a sequence of linear programming problems~\cite{pilanci}~\cite{demanet}.  Lastly, a third example deals with the normalization $\pi_j(x)=|x_j|^t/\|x\|_t^t$ with $t=1/2$, which leads to $\exp(H_2(\pi(x))=(\|x\|_2/\|x\|_4)^4$. In the paper~\cite{barak}, the problem of minimizing $\|x\|_2/\|x\|_4$ has been shown to have interesting connections with sums-of-squares (SOS) optimization problems --- for which efficient algorithms are available~\cite{lasserre}.

\paragraph{Applications of $s_2(x)$ in imaging systems.} The quantity $s_2(x)$ has been applied successfully in real imaging systems. For instance, in the paper~\cite{kelly2015}, $s_2(x)$ was found to be useful in estimating the parameters of the wavelet power law for natural images. Also, in the papers \cite{hero2013,krishnan2011} apply the quantity $s_2(x)$  in the problem of de-blurring.

\subsubsection{Graphical interpretations} The fact that $s_q(x)$ is a sensible measure of sparsity for non-idealized signals is illustrated in Figure 1 for the case of $q=2$. In essence, if $x$ has $k$ large coordinates and $p-k$ small coordinates, then $s_q(x)\approx k$, whereas $\|x\|_0=p$. In the left panel, the sorted coordinates of three different vectors in $\R^{100}$ are plotted. The value of $s_2(x)$ for each vector is marked with a triangle on the x-axis, which shows that $s_2(x)$ essentially measures the ``elbow'' in the decay plot. This idea can be seen in a more geometric way in the right panel, which plots the  the sub-level sets $\mathcal{S}_c:=\{x \in \R^p : s_2(x)\leq c\}$ with $c=1.1$ and $c=1.9$ where $p=2$. When $c\approx 1$, the vectors in $\mathcal{S}_c$ are closely aligned with the coordinate axes, and hence contain one effective coordinate. As $c\uparrow p$, the set $\mathcal{S}_c$ expands to include less sparse vectors until $\mathcal{S}_p=\R^p$.
\begin{figure*}[h]
\centering
{\includegraphics[angle=0,
  width=.45\linewidth]{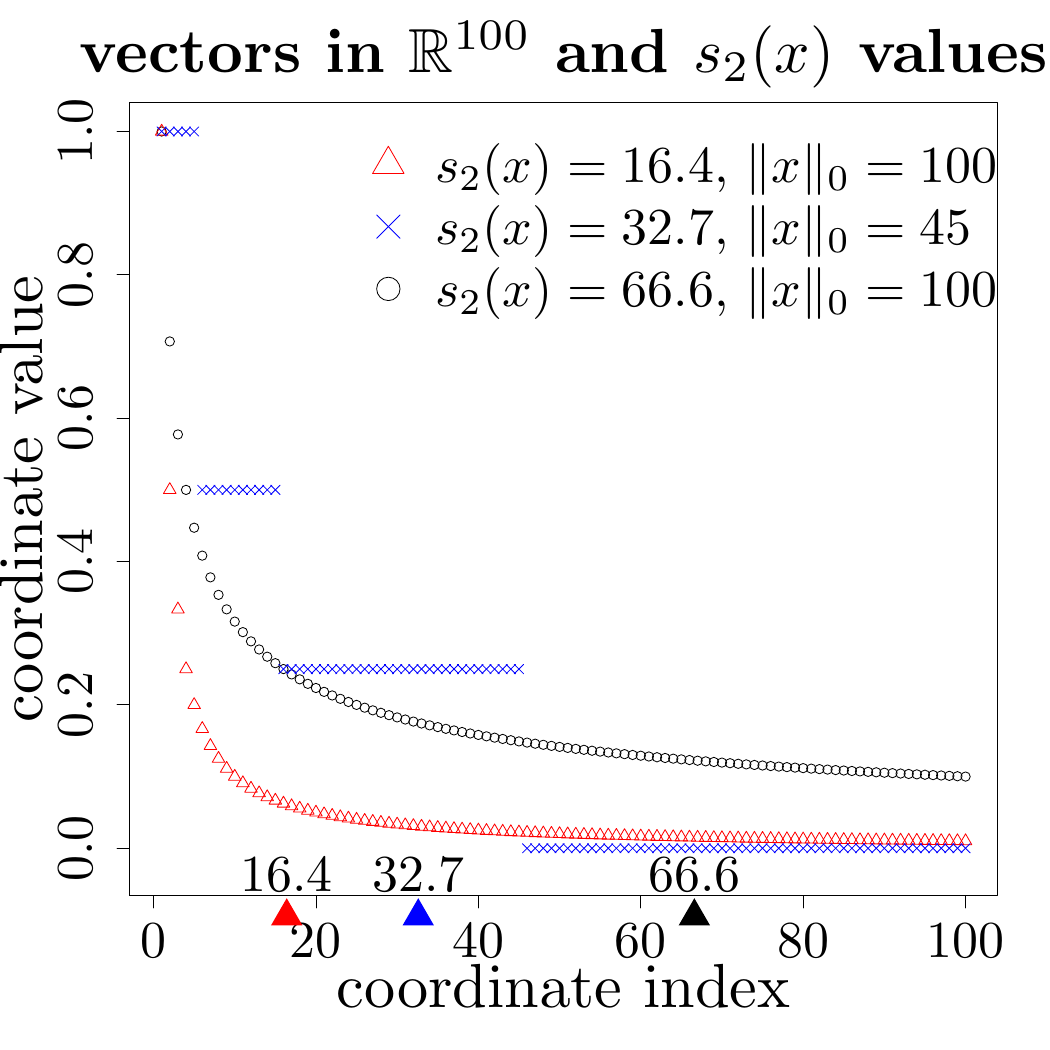}}
{\includegraphics[angle=0,
  width=.45\linewidth]{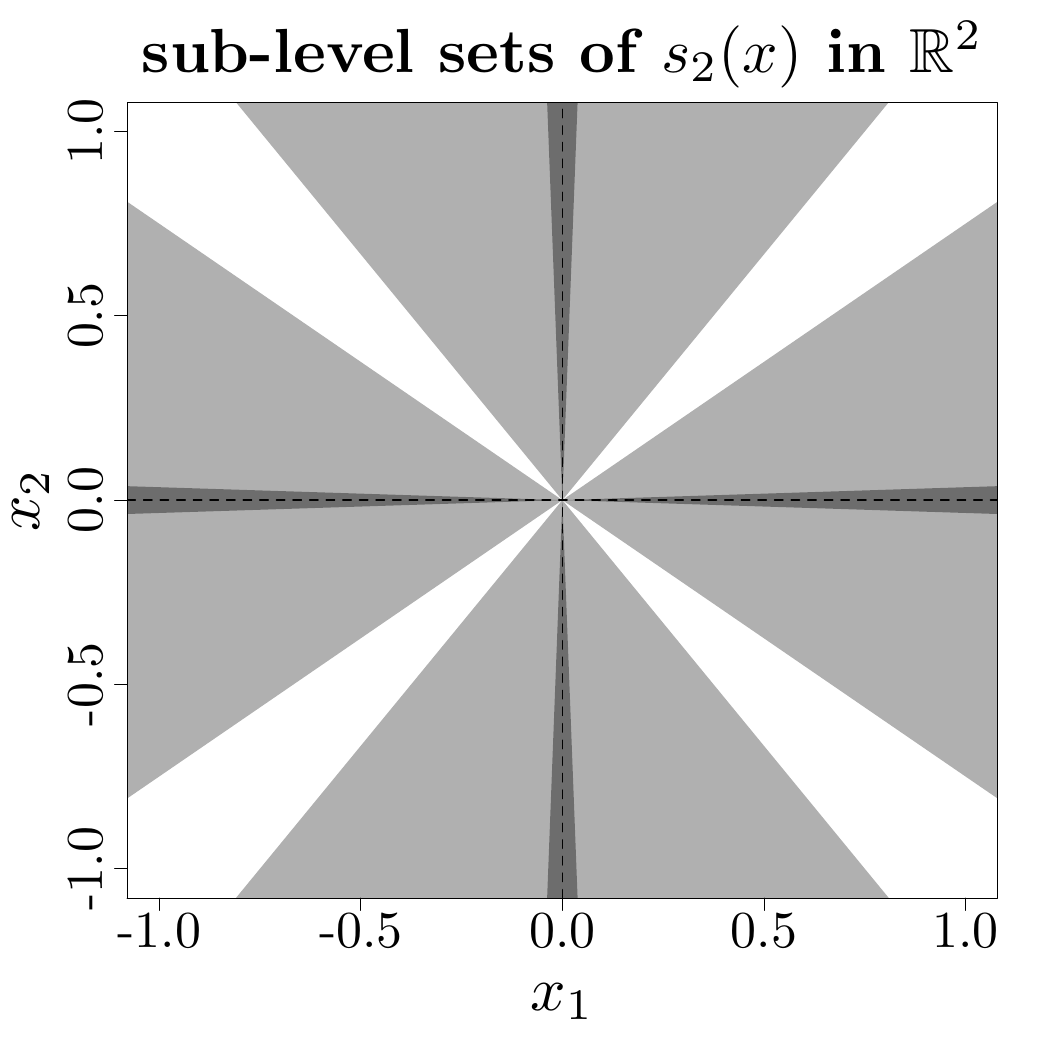}}

\caption{Characteristics of $s_2(x)$. Left panel: Three vectors (red, blue, black) in $\R^{100}$ have been plotted with their coordinates in order of decreasing size (maximum entry normalized to 1). Two of the vectors have power-law decay profiles, and one is a dyadic vector with exactly 45 positive coordinates (red: $x_i \propto i^{-1}$, blue: dyadic, black: $x_i\propto \ts i^{-1/2}$).  Color-coded triangles on the bottom axis indicate that the $s_2(x)$ value represents the ``effective'' number of coordinates. Right panel: The light grey set is given by $\{x \in \R^2 : s_2(x)\leq 1.9\}$, and the dark grey set is given by $\{x \in \R^2 : s_2(x) \leq 1.1 \}$.
}
\label{fig:sparse}
\end{figure*}
\subsubsection{Numerically stable measures of rank and sparsity for matrices}\label{rank}

The framework of CS naturally extends to the problem of recovering 
an unknown matrix \mbox{$X\in \R^{p_1\times p_2}$} on the basis of the 
measurement model
\begin{equation}\label{matrixModel}
\ts y = \A(X) + \sigma \e,
\end{equation}
where $y\in \R^n$, and
$\A$ is a user-specified linear operator from $\R^{p_1\times p_2}$ to $\R^n$.
In recent years, many researchers have explored the recovery of $X$
when it is assumed to have sparse or low rank structure. We refer to the papers~\cite{CandesPlan,venkatinverse}
for descriptions of numerous applications.
In analogy with the previous section, the parameters $\text{rank}(X)$ or $\|X\|_0$ play important theoretical roles, but are very sensitive to perturbations of $X$. Likewise, it is of basic interest to estimate robust measures of rank and sparsity for matrices. Since the  analogue of $s_q(\cdot)$ for measuring matrix sparsity is easily derived by viewing $X$ as a vector in $\R^{p_1p_2}$, we restrict our attention to the more distinct issue of soft measures of rank.

In the context of recovering a low-rank matrix $X$,
the quantity $\text{rank}(X)$ plays the role that the 
norm $\|x\|_0$ does in the recovery of a sparse vector. If we let 
$\varsigma(X)\in\R^p_+$ denote the vector of ordered singular values of 
$X$, the connection can be made explicit by writing
$$\text{rank}(X)=\|\varsigma(X)\|_0.$$
As in our discussion of sparsity, it is of basic interest to consider a numerically stable version of the usual rank function. Motivated by the definition of $s_q(x)$ in the vector case, 
we can also consider 
$$\ts r_q(X):=s_q(\varsigma(X))=\Big(\frac{\|\varsigma(X)\|_q}{\|\varsigma(X)\|_1}\Big)^{\frac{q}{1-q}} =\Big(\frac{\mnorm{X}_q}{\mnorm{X}_1}\Big)^{\frac{q}{1-q}} $$
as a measure of the effective rank of $X$, where $q>0$ and $\mnorm{X}_q:=\|\varsigma(X)\|_q$. (When $q\geq 1$, $\mnorm{X}_q$ is known as the Schatten $q$-norm of $X$.) Essentially all of the properties of $s_q(\cdot)$ described earlier carry over to $r_q(\cdot)$ in a natural way, and so we do not state these in detail. We also note that quantities related to $r_q(X)$, or special instances, have been considered elsewhere as a measure of rank, e.g.   the \emph{numerical rank}\footnote{This can be cast in the framework of $r_q(X)$ by defining the probability vector $\pi(x)$ as $\pi_j(x)=|x_j|^2/\|x\|_2^2$ in the definition of $s_q(x)$ and then choosing $q=\infty$.} $\|X\|_F^2\big/\|X\|\op^2$~\cite{vershyninnumerical}, or the instance $r_2(X)$ ~\cite{lopes,ellStar,negahbanmatrix}.

\subsection{Contributions}
The main contributions of the paper are summarized below in three parts. 

\paragraph{The family of sparsity measures $\{s_q(x)\}_{q\geq 0}$.}

As mentioned in Section~\ref{origin}, our definition of $s_q(x)$ in terms of R\'enyi entropy gives a conceptual foundation for several norm ratios that have appeared elsewhere in the sparsity literature. Furthermore, we clarify the meaning of $s_2(x)$ with regard to signal recovery by showing in Section~\ref{sec:recov} that $s_2(x)$ plays an intuitive role in the performance of the Basis Pursuit Denoising (BPDN) algorithm. Specifically, we show that the \emph{relative} $\ell_2$ error of BPDN can be bounded in a sharp way by the quantity $\sqrt{s_2(x)\log(ep/n)/n}$, which is formally similar to the well-known rate of $\ell_2$ approximation $\sqrt{k\log(p)/n}$ for $k$-sparse signals.

\paragraph{Estimation results, confidence intervals, and applications.}

Our central methodological contribution is a new deconvolution-based approach for estimating $\|x\|_q$ and $s_q(x)$ from linear measurements. The procedure we propose is of particular interest in the way that it blends the tools of \emph{sketching with stable laws} and \emph{deconvolution with characteristic functions}. These tools are typically applied in different contexts, as discussed in Section~\ref{sec:related}. Also, the computational cost of our procedure is small relative to the cost of recovering the full signal by standard methods. 

In terms of consistency, the most important features of our estimator $\hat{s}_q(x)$ are that it does not rely on any sparsity assumptions, and that its relative error converges to 0 at the \emph{dimension-free} rate of $1/\sqrt{n}$ (in probability). Consequently,  only $\mathcal{O}(1)$ measurements are needed to obtain a good estimate of $s_q(x)$, even when $p/n\to\infty$.  As mentioned in Section~\ref{sec:role}, this result naturally suggests a two-stage measurement process: First, a small initial measurement price can be paid to obtain $\hat{s}_q(x)$. Second, the value $\hat{s}_q(x)$ can be used to adaptively select ``just enough'' extra  measurements for recovering the full signal. (Proposition~\ref{propUpper} in Section~\ref{sec:recov} indicates that this number can be chosen proportionally to $\hat{s}_2(x)\log(p)$ when BPDN is used for recovery.)

In addition to proving ratio-consistency, we derive a CLT for $\hat{s}_q(x)$, which allow us to obtain confidence intervals $s_q(x)$ with asymptotically exact coverage probability. A notable feature of this CLT is that it is ``uniform'' with respect to the tuning parameter in our  procedure for computing $\hat{s}_q(x)$. The uniformity is important because it allows us to make an optimal \emph{data-dependent} selection of the tuning parameter and still find the estimator's limiting distribution (see Theorem~\ref{clt} and Corollary~\ref{CIs}).
 In terms of applications, we show in Section~\ref{sec:app} how this CLT can be used in inferential problems related to unknown sparsity, i.e.~testing the null hypothesis that $s_q(x)$ is greater than a given level, and ensuring that the true signal lies in the constraint set of the (primal) Lasso or Elastic net problems with a given degree of statistical significance.

\paragraph{The necessity of randomized measurements.}
At the present time, the problem of constructing deterministic measurement matrices with performance guarantees comparable to those of random matrices is one of the major unresolved theoretical issues in CS~\cite[Section 1.3]{foucartbook}~\cite{calderbank2010construction}. Due to the fact that our proposed method for estimating $s_q(x)$ depends on randomized measurements, one may similarly wonder if randomization is essential to the problem of estimating unknown sparsity. In Section~\ref{sec:neg}, we show that randomization is essential from a worst-case point of view. Our main result in this direction (Theorem~\ref{minimaxnew}) shows that for any deterministic matrix $A\in\R^{n\times p}$, and any deterministic procedure for estimating $s_q(x)$, there is always at least one signal for which the relative estimation error is at least of order 1, even if the measurements are noiseless. This contrasts with performance our randomized method, whose relative error is $\mathcal{O}_P(1/\sqrt{n})$ for any choice of $x$. Furthermore, the result has a negative implication for sparse linear regression. Namely, due to the fact that the design matrix is often viewed as ``fixed and given'' in many regression problems, our result indicates that $s_q(x)$ cannot be consistently estimated in relative error in that context (from a worst-case point of view).

\subsection{Related work}\label{sec:related} Our work here substantially extends our earlier conference paper~\cite{lopesICML} and has connections with a few different lines of research.

\paragraph{Extensions beyond the conference paper~\cite{lopesICML}.} Whereas our earlier work deals exclusively with the sparsity measure $\|x\|_1^2/\|x\|_2^2$, the current paper considers the estimation of the family of parameters $s_q(x)$. The procedure we propose for estimating $s_q(x)$ (as well as our analysis of its performance) include several improvements on the earlier approach. In particular, the new procedure tolerates noise with infinite variance and leads to confidence intervals with asymptotically exact coverage probability (whereas the previous approach led to conservative intervals). Also, the applications of our procedure to tuning recovery algorithms and testing the hypothesis of sparsity are new (Section~\ref{sec:app}). Lastly, our theoretical results in Sections~\ref{sec:recov} and~\ref{sec:neg} are sharpened versions of parallel results in the paper~\cite{lopesICML}.

\paragraph{Sketching with stable laws.} Our approach to estimating $s_q(x)$ is based on the sub-problem of estimating $\|x\|_q$ for various choices of $q$. In order to estimate such norms from linear measurements, we employ the technique of \emph{sketching with stable laws}, which has been developed extensively in the streaming computation literature. (The book~\cite{cormodesurvey} offers an overview, and seminal papers include~\cite{indykstable} and~\cite{alon}). Over the last few years, the exchange of ideas between sketching and CS has just begun to accelerate, as in the papers~\cite{gilbertIndyk}~\cite{lopesICML}~\cite{pinglicounting}~\cite{indyk2013}. Nevertheless, to the best of our knowledge, the present paper and our earlier work~\cite{lopesICML} are the first to apply sketching ideas to the problem of unknown sparsity in CS.

In essence, our use of the sketching technique is based on the fact that if a random vector $a_1\in\R^p$ has i.i.d.~coordinates drawn from a standard symmetric $q$-stable law,\footnote{e.g. Gaussian when $q=2$ and Cauchy when $q=1$.} then the random variable $\langle a_1,x\rangle$ has a  $q$-stable law whose scale parameter is equal to $\|x\|_q$. (A more detailed introduction is given in Section~\ref{sketching}.) Consequently, the problem of estimating $\|x\|_q$ can be thought of as estimating the scale parameter of a stable law convolved with noise.

In the streaming computation literature, the observation model giving rise to $\langle a_1,x\rangle$ is quite different than in CS. Roughly speaking, the vector $x$ is thought of as a massive data stream whose entries can be observed sequentially, but cannot be stored entirely in memory. The core idea is that by computing ``sketches'' $\langle a_1,x\rangle= a_{11}x_11+a_{12}x_2+\cdots$ in a sequential manner,  it is possible to estimate various functions of $x$ from the sketches without having to store the entire stream. Under this framework, a substantial body of work has studied the estimation of $\ell_q$ norms with $q\geq 0$~\cite{cosmacounting}~\cite{cormodezeroin}~\cite{pinglitails}~\cite{pinglijmlr}~\cite{indykstable}~\cite{alon}~\cite{feigenbaum}. However, results in this direction are typically not directly applicable to CS, due to essential differences in the observation model. For instance, measurement noise does not generally play a role in the sketching literature.

\paragraph{Empirical characteristic functions.} As just mentioned, our approach to estimating $\|x\|_q$ and $s_q(x)$ can be thought of as deconvolving the scale parameter of a stable law. Given that stable laws have a simple analytic formula for their characteristic function (and have no general formula for their likelihood function), it is natural to use the empirical characteristic function $\hat{\Psi}_n(t)=\ts\frac{1}{n}\sum_{i=1}^n \exp(\sqrt{-1}t y_i)$ as a foundation for our estimation procedure. With regard to denoising, characteristic functions are also attractive insofar as they factor over convolution, and exist even when the noise distribution is heavy-tailed.

The favorable properties of empirical characteristic functions have been applied by several authors to the deconvolution of scale parameters~\cite{markatoubasic}~\cite{markatouerrors},~\cite{butucea}~\cite{meister}~\cite{matias}. Although the basic approach used in these papers is similar to ours, the results in these works are not directly comparable with ours due to differences in model assumptions. Another significantly distinct aspect of our work deals with the choice of the ``tuning parameter'' $t$ in the function $\hat{\Psi}_n(t)$. This choice is a basic element in most methods based on empirical characteristic functions. In detail, we show how to make an optimal data-adaptive choice $\hat{t}$, and we derive the limiting distribution of the estimator $\hat{s}_q(x)$ that originates from $\hat{\Psi}_n(\hat{t}\,)$. This leads to a significant technical challenge in accounting for the randomness of $\hat{t}$, and in order to do this, we show that the process $\hat{\Psi}_n(\cdot)$ arising from our model assumptions satisfies a uniform CLT in the space $\mathscr{C}(\mathcal{I})$ of continuous complex functions on a compact interval $\mathcal{I}$. (See the paper~\cite{marcus} or the book~\cite{ushakov} for more details concerning weak convergence of empirical characteristic functions.) With regard to the cited line of works concerning deconvolution of scale parameters, it seems that our work is the first to derive the limiting distribution of the scale estimator under a data-dependent choice of tuning parameter.

\paragraph{Model selection and validation in CS.}

Some of the challenges described in Section~\ref{sec:role} can be approached with the general tools of cross-validation (CV) and empirical risk minimization (ERM). This approach has been used to select various parameters in CS, such as the number of measurements $n$~\cite{willsky,wardCV}, the number of OMP iterations $k$~\cite{wardCV}, or the Lasso regularization parameter $\lambda$~\cite{eldarStein}.  At a high level, these methods consider a collection of (say $m$) solutions $\hat{x}^{(1)},\dots,\hat{x}^{(m)}$ obtained from different values $\theta_1,\dots,\theta_m$ of some tuning parameter of interest. For each solution, an empirical error estimate $\hat{\err}(\hat{x}^{(j)})$ is computed, and the value $\theta_{j^*}$ corresponding to the smallest $\hat{\err}(\hat{x}^{(j)})$ is chosen. 

Although methods based on CV/ERM share common motivations with our work here, these methods differ from our approach in several ways.  In particular, the problem of estimating a soft measure of sparsity, such as $s_q(x)$, has not been considered from that angle. Also, the cited methods do not give any theoretical guarantees to ensure that the chosen tuning parameter leads to a solution whose $\ell_0$ sparsity level is close to the true one. (Note that even if CV suggests that an estimate $\hat{x}$ has small error $\|\hat{x}-x\|_2$, it is not necessary for $\|\hat{x}\|_0$ to be close to $\|x\|_0$.) This point is especially relevant in inferential problems, such as identifying a set of important variables or making confidence statements related to an unknown sparsity value. From a computational point view, the CV/ERM approaches can also be costly --- since $\hat{x}^{(j)}$ may need to be computed from a separate optimization problem for  for each choice of the tuning parameter.  By contrast, our method for estimating $s_q(x)$ requires very little computation.

\subsection{Outline}
The remainder of the paper is organized as follows. In Section~\ref{sec:recov}, we formulate a (tight) recovery guarantee for the Basis Pursuit Denoising algorithm directly in terms of $s_2(x)$. Next, in Section~\ref{sec:proc}, we propose estimators for $\|x\|_q$ and $\hat{s}_q(x)$, and in Section~\ref{results} we state consistency results and provide confidence intervals for $\|x\|_q$ and $s_q(x)$. Applications to testing the hypothesis of sparsity and adaptive tuning of the Lasso are presented in Section~\ref{sec:app}. In Section~\ref{sec:neg}, we show that the use of randomized measurements is essential to estimating $s_q(x)$ in a minimax sense. In Section~\ref{sec:simulations}, we present simulations that confirm our theoretical results and demonstrate the effectiveness of the proposed method.
We defer all of the proofs to the appendices.

%

%
%
 %
 %
%
%
%

%
\section{Recovery guarantees in terms of $s_2(x)$}\label{sec:recov}
In this section, we state two simple propositions that illustrate the link between $s_2(x)$ and recovery conditions for the Basis Pursuit Denoising (BPDN) algorithm~\cite{chendonohosaunders}. The main purpose of these results is to highlight the fact that $s_2(x)$ and $\|x\|_0$ play analogous roles with respect to the sample complexity of sparse recovery. 
Specifically, we provide \emph{matching} upper and lower bounds for relative $\ell_2$ reconstruction error of BPDN in terms of $s_2(x)$. These bounds also suggest two applications of the quantity $s_2(x)$. First, the order of the reconstruction error can be estimated whenever $s_2(x)$ can be estimated.  Second, when measurements can be collected sequentially, an estimate of $s_2(x)$ from an initial set of measurements allows for the user to select a number of secondary measurements that adapts to the particular structure of $x$, e.g. $n=\hat{s}_2(x)\log(p)$.

\paragraph{Setup for BPDN.} In order to explain the connection between $s_2(x)$ and recovery, we recall a fundamental result describing the
 $\ell_2$ error rate of the BPDN algorithm
  first obtained in the paper~\cite[Theorem 2]{candes2006}. (Our statement of the result is reformulated to include some refinements from the later papers~\cite{caiRIP}\,[Theorem 3.3] and~\cite{vershyninIntro}\,[Theorem 5.65].) 
 For the first two results of this section, we will use two standard assumptions underlying those earlier works.

 \begin{shortassumption}\label{noise1}
 There is a constant $\e_0$ such that all realizations of the noise vector $\e\in\R^n$ satisfy $\|\e\|_2\leq \e_0$.
 \end{shortassumption}
 %
 \begin{shortassumption}\label{mat1}
 The entries of $A\in\R^{n\times p}$ are an i.i.d. sample from $\frac{1}{\sqrt{n}}G_0$, where $G_0$ is a sub-Gaussian distribution with mean 0 and variance 1.
 \end{shortassumption}
 %
 %
  Since $\e$ and $A$ are both random, probability statements will be made with respect to their joint distribution. When the noise distribution satisfies {\textbf{A\ref{noise1}}}, the output of the BPDN algorithm is a solution to the following convex optimization problem
\begin{equation}\label{BPDN}\tag{BPDN}
\hat{x}\in  \text{argmin} \big\{ \|v \|_1 : \|Av - y\|_2\leq \sigma \e_0, v\in \R^p\big\}.
\end{equation}
As a final piece of notation, for any $T\in \{1,\dots,p\}$, we use $x_{|T}$ to denote the best $T$-term approximation of $x$, which is computed by retaining the largest $T$ entries of $x$ in magnitude, and setting all others to $0$.
\begin{theorem}[\cite{candes2006stable,caiRIP,vershyninIntro}]\label{thmcandes}
Suppose the model~\eqref{setup1} satisfies the conditions~{\textbf{A\ref{noise1}}} and {\textbf{A\ref{mat1}}}. Let $x\in \R^p$ be arbitrary, and fix a number $T\in \{1,\dots,p\}$. Then, there are absolute constants $c_2,c_3>0$, and numbers $c_0,c_1>0$ depending only on the distribution $G_0$, such that the following statement is true. If
\begin{equation}\label{nbound}
 n\geq c_0 T\log(pe/T),
 \end{equation}
then with probability at least $1-2\exp(-c_1n)$, any solution $\hat{x}$ to the problem~\eqref{BPDN} satisfies
\begin{equation}\label{candesTao}
\|\hat{x}-x\|_2 \leq c_{2}\,\sigma \e_0 + c_{3} \,\ts\frac{\|x- x_{|T}\|_1}{\sqrt{T}}.
\end{equation}
\end{theorem}
\paragraph{An upper bound in terms of $s_2(x)$.} Two important aspects of Theorem~\ref{thmcandes} are that it holds for \emph{all} signals $x\in\R^p$, and that it measures sparsity via the $T$-term approximation error $\|x-x_{|T}\|_1$, rather than the idealized $\ell_0$ norm. However, a main limitation is that the detailed relationship between $T$ and the approximation error $\ts\frac{1}{\sqrt{T}}\|x-x_{|T}\|_1$ is typically unknown for the true signal $x$. Consequently, it is not clear how large $n$ should be chosen in line~\eqref{nbound} to ensure that $\|x-\hat{x}\|_2$ is small with high probability. The next proposition resolves this issue by modifying the bound~\eqref{candesTao} so that that the relative $\ell_2$ error is bounded by an explicit function of $n$ and the estimable parameter $s_2(x)$.
\begin{proposition}\label{propUpper}
Assume conditions~{\textbf{A\ref{noise1}}} and~{\textbf{A\ref{mat1}}} hold, and let $x\in\R^p\setminus\{0\}$ be arbitrary.  Then, there is an absolute constant $c_2>0$, and  numbers $c_1,c_3>0$ depending only on the distribution $G_0$, such that the following statement is true.  If $n$ and $p$ satisfy $\log(\frac{pe}{n})\leq n \leq p$, then with probability at least $1-2\exp(-c_1n)$, any solution $\hat{x}$ to the problem~\eqref{BPDN} satisfies  
\begin{equation}\label{bpdnNew}
\ts\frac{\|\hat{x}-x\|_2}{\|x\|_2} \leq c_2 \ts\frac{\sigma \e_0}{\|x\|_2}+ c_3 \sqrt{\ts \frac{s_2(x) 
\log(\frac{pe}{n})}{n}}.
\end{equation}

\end{proposition}
\paragraph{A matching lower bound in terms of $s_2(x)$.}

Our next result shows that the upper bound~\eqref{bpdnNew} is sharp in the case of noiseless measurements. More precisely, for any choice of $A\in \R^{n\times p}$, there is always at least one signal $\tilde{x}\in\R^p\setminus\{0\}$ for which the relative $\ell_2$ error of BPDN is at least $\sqrt{s_2(\tilde{x})\log(pe/n)/n}$, up to an absolute constant.
 In fact, the lower bound is applicable beyond BPDN, and imposes a limit of performance on all algorithms that satisfy the mild condition of being \emph{homogenous} in the noiseless setting. To be specific,  if a recovery algorithm is viewed as a map $\mathcal{R}:\R^n \rightarrow \R^p$ that sends a vector of noiseless measurements $Ax\in\R^n$ to a solution  $\hat{x}= \mathcal{R}(Ax)\in \R^p$, then $\mathcal{R}$ is said to be homogenous if
\begin{equation}\label{homog}
\mathcal{R}(A(cx)) = c \cdot\mathcal{R}(Ax) \text{ \ \ for all } c>0.
\end{equation}
It is simple to verify that the BPDN is homogenous in the case of noiseless measurements, since its solution may be written as\footnote{If this minimization problem does not have a unique optimal solution, we still may still regard BPDN as a well defined function from $\R^n$ to $\R^p$ by considering a numerical implementation that never returns more than one output for a given input.}
\begin{equation}\label{bp}\tag{BP}
\hat{x}\in \argmin\{ \|v\|_1 : Av=y, v\in \R^p\}.
\end{equation}
Apart from the basic condition of homogeneity, our lower bound requires no other assumptions. Note also that the statement of the result does not involve any randomness.
\begin{proposition}\label{propLower}
There is an absolute constant $c_0>0$ for which the following statement is true. For any homogenous recovery algorithm $\mathcal{R}:\R^n \to \R^p$, and any  $A\in \R^{n\times p}$ with $n\leq p$,  there is at least one point $\tilde{x}\in\R^p\setminus\{0\}$ such that
\begin{equation}\label{bpdnLower}
\ts\frac{\|\hat{x}-\tilde{x}\|_2}{\|\tilde{x}\|_2} \geq c_0\sqrt{\ts \frac{s_2(\tilde{x})\log(\frac{pe}{n})}{n}},
\end{equation}
where $\hat{x}=\mathcal{R}(A\tilde{x})$.
\end{proposition}

\section{Estimation procedures for $s_q(x)$ and $\|x\|_q^q$ }\label{sec:proc}

 In this section, we describe a procedure to estimate $s_q(x)$ for an arbitrary non-zero signal $x\in\R^p$, and any $q\in (0,2]\setminus\{1\}$. The procedure uses a small number of measurements, makes no sparsity assumptions, and requires very little computation. The measurements we prescribe may also be re-used to recover the full signal after the parameter $s_q(x)$ has been estimated. 
For readers who are interested only in the high level idea of the procedure, it is enough to read just the subsections~\ref{esteqn} and~\ref{optselect}.
 
\subsection{The deconvolution model}\label{deconvmodel}  Here we describe the model assumptions that our estimation procedure for $s_q(x)$ will be based on. (These are different from the assumptions used in the previous section.)
In scalar notation, we consider linear measurements given by
 \begin{equation}\label{model1}\tag{M}
y_i = \langle a_i,x\rangle +\sigma \e_i, \ \ \ \ \ i=1,\dots,n.
\end{equation} 
\paragraph{Model assumptions.} For the remainder of the paper, we assume $x\neq 0$ unless stated otherwise. Regarding the noise variables $\e_i$, we assume they are generated in an i.i.d.~manner from a distribution denoted by $F_0$. When the  $a_i$ are generated randomly, we assume that $\{a_1,\dots,a_n\}$ is an independent set of random vectors, and also that the sets $\{a_1,\dots,a_n\}$ and $\{\e_1,\dots,\e_n\}$ are independent. The noise variables are assumed to be symmetric about 0 and to satisfy $0<\E|\e_1|<\infty$, but they may have \emph{infinite variance}.
A minor technical condition we place on $F_0$ is that the roots of its characteristic function $\varphi_0$ are isolated (i.e.~no limit points). This condition is satisfied by a broad range of naturally occurring distributions, and in fact, many works on deconvolution assume that $\varphi_0$ has no roots at all.\footnote{It is known that a subset of $\R$ is the zero set of a characteristic function if and only if it symmetric, closed, and excludes 0.~\cite{illinskiZeros,gneitingCuriosities}.}
The noise scale parameter $\sigma>0$ and the distribution $F_0$ are treated as being known, which is a common assumption in deconvolution problems. Also note 
that in certain situations,  it may be possible to directly estimate $F_0$ by using ``blank measurements'' with $a_i=0$. 

\paragraph{Asymptotics.} Following the usual convention of high-dimensional asymptotics, we allow the model parameters to vary as $(n,p)\to\infty$. This means that there is an implicit index $\xi \in \Z_+$, such that $n=n(\xi)$, $p=p(\xi)$ and both diverge as $\xi\to\infty$. It will turn out that our asymptotic results will not depend on the ratio $p/n$, and so we allow $p$ to be arbitrarily large with respect to $n$. We also allow $x=x(\xi)$, $\sigma=\sigma(\xi)$ and $a_i=a_i(\xi)$, but the noise distribution $F_0$ is fixed with respect to $\xi$. When making asymptotic statements about probability, we view the set of pairs $\{(a_1,\e_1),\dots,(a_n,\e_n)\}$ as forming a triangular array with rows indexed by $\xi$, and columns indexed by $n(\xi)$. Going forward, we will generally suppress the index $\xi$.

\subsection{Sketching with stable laws in the presence of noise}\label{sketching}

For any $q\in (0,2]$, the sketching technique offers a way to estimate $\|x\|_q^q$ from a set of randomized linear measurements.
Building on this technique, we estimate $s_q(x)=(\|x\|_q/\|x\|_1)^{q/(1-q)}$ by estimating $\|x\|_q^q$ and $\|x\|_1$ from separate sets of measurements.
The core idea is to generate the measurement vectors $a_i\in\R^p$ using \emph{stable laws}. A standard reference on this class of distributions is the book~\cite{zolotarev}.
\begin{definition}\label{def:stable}
A random variable $V$ has a \emph{symmetric $q$-stable distribution} if its characteristic function is of the form $\E[\exp(\sqrt{-1}tV)] = \exp(-|\gamma t|^{q})$ for some $q\in(0,2]$ and some $\gamma>0$, where $t\in \R$.  We denote the distribution by $V\sim \text{stable}_q(\gamma)$, and $\gamma$ is referred to as the \emph{scale} parameter.
 \end{definition}
 The most well-known examples of symmetric stable laws are the cases
 of $q=2$ and $q=1$. Namely, $\stable_2(\gamma)$ is the Gaussian distribution $N(0,2\gamma^2)$, and $\stable_1(\gamma)$ is the Cauchy distribution $C(0,\gamma)$. To fix some notation, if a vector $a_1=(a_{11},\dots,a_{1p})\in\R^p$ has i.i.d.\!\! entries drawn from $\stable_q(\gamma)$, we write $a_1\sim \stable_{q}(\gamma)^{\otimes p}$. Also, since our work will involve different choices of $q$, we will write $\gamma_q$ instead of $\gamma$ from now on. The connection with $\ell_q$ norms hinges on the following property of stable distributions, which is simple to derive from Definition~\ref{def:stable}.
\begin{lemma} Suppose $x\in\R^p$ is fixed, and $a_1\sim \textup{stable}_q(\gamma_q)^{\otimes p}$ with parameters $q\in (0,2]$ and $\gamma_q>0$.  Then, the random variable $\langle x,a_1\rangle$ is distributed according to $\textup{stable}_q(\gamma_q\|x\|_{q}).$
\end{lemma}
Using this fact, if we generate a set of i.i.d.\!\! measurement vectors $a_1,\dots,a_n$ from the distribution $\text{stable}_q(\gamma_q)^{\otimes p}$ and let $\tilde{y}_i=\langle a_i,x\rangle$, then $\tilde{y}_1,\dots,\tilde{y}_n$ is an i.i.d.\!\! sample from $\text{stable}_q(\gamma_q\|x\|_{q})$. Hence, in the special case of noiseless linear measurements, the task of estimating $\|x\|_q^q$ is equivalent to a well-studied univariate problem: \emph{estimating the scale parameter of a stable law from an i.i.d. sample}. Our work below substantially extends this idea in the noisy case. We also emphasize that the extra deconvolution step is an important aspect of our method that distinguishes it from existing work in the sketching literature (where noise is typically not considered).\\

When generating the measurement vectors from $\stable_q(\gamma_q)$, the parameter $\gamma_q$ governs the ``energy level'' of the $a_i$. For instance, in the case of Gaussian measurements with $a_i \sim \stable_2(\gamma_2)^{\otimes p}$, we have $\E\|a_i\|_2^2 = 2p \gamma_2^2$. In general, for any $q$, as the energy level is increased, the effect of noise is diminished. Likewise, in our analysis, we view $\gamma_q$ as a physical aspect of the measurement system that is known to the user. Asymptotically, we allow $\gamma_q=\gamma_q(\xi)$ to vary as $(n,p)\to\infty$ in order to reveal the trade-off between the measurement energy and the noise level $\sigma$.

 \subsection{The sub-problem of estimating $\|x\|_q^q$}\label{sec:procedure}
 Our procedure for estimating $s_q(x)$ uses two separate sets of measurements of the form~\eqref{model1} to compute estimators $\hat{\|x\|}_1$ and $\hat{\,\|x\|_q^q}$. The respective sizes of each measurement set will be denoted by $n_1$ and $n_q$.
 To unify the discussion, we will describe just one procedure to compute $\hat{\|x\|_q^q}$ for any $q\in(0,2]$, since the $\ell_1$-norm estimator is a special case. The two estimators are then combined to obtain an estimator of $s_q(x)$, defined by 
\begin{equation}
\hat{s}_q(x) := \ts\frac{\left(\widehat{\|x\|_q^q}\right)^{\frac{1}{1-q}}}{\left(\hat{\|x\|}_1\right)^{\frac{q}{1-q}}} ,
\end{equation}
which makes sense for any $q\in (0,2]$ except $q=1$. Of course, the parameters $\|x\|_0$ and $s_1(x)$ can still be estimated in practice by using $\hat{s}_q(x)$ for some value of $q$ that is close to 0 or 1. Indeed, the ability to approximate $\|x\|_0$ and $s_1(x)$ in this way is a basic motivation for studying $s_q(x)$ over a continuous range of $q$. 

\subsubsection{An estimating equation based on characteristic functions}\label{esteqn} Characteristic functions offer a very natural route toward estimating $\|x\|_q^q$. If we draw i.i.d.\! measurement vectors
$$a_i \sim \stable_q(\gamma_q)^{\otimes p}, \ \ \ \ \ i=1,\dots,n_q,$$
then the characteristic function of the measurement $y_i = \langle a_i,x\rangle+\sigma \e_i$ is given by
\begin{equation}\label{chf}
\Psi(t):=\E[\exp(\sqrt{-1}ty_i) ] = \exp(-\gamma_q^q|t|^q \|x\|_q^q) \cdot \varphi_0(\sigma t),
\end{equation}
where $t\in \R$, and we recall that $\varphi_0$ denotes the characteristic function of the noise variables $\e_i$. Note that $\varphi_0$ is real-valued since we assume that the noise-distribution is symmetric about 0. Using the measurements $y_1,\dots,y_{n_q}$, we can approximate $\Psi(t)$ by computing the empirical characteristic function

\begin{equation}\label{psihat}
\hat{\Psi}_{n_q}(t):= \frac{1}{n_q}\sum_{i=1}^{n_q} e^{\sqrt{-1}t y_i}.
\end{equation}
 Next, by solving for $\|x\|_q^q$ in the approximate equation
\begin{equation}\label{psiapprox}
\hat{\Psi}_{n_q}(t) \approx \exp(- \gamma_q^q|t|^q \|x\|_q^q) \cdot \varphi_0(\sigma t),
\end{equation}
we obtain an estimator $\hat{\|x\|_q^q}$. To make the dependence on $t$ explicit, we will mostly use the notation $\hat{\nu}_q(t)=\hat{\|x\|_q^q}$. 

Proceeding with the arithmetic in the previous line leads us to define
\begin{equation}\label{nuhat}
\hat{\nu}_q(t):= \ts\frac{-1}{\gamma_q^q|t|^q} \Log \, \myRe \Big( \frac{\hat{\Psi}_{n_q}(t)}{\varphi_0(\sigma t)}\Big),
\end{equation}
when $t\neq 0$ and $\varphi_0(\sigma t)\neq 0$. Here, the symbol $\myRe(z)$ denotes the real part of a complex number $z$. Also, we define $\Log(r):=\log(|r|)$ for any real number $r\neq 0$, and $\Log(0):=1$ (arbitrarily).
For the particular values of $t$ where $t=0$ or $\varphi_0(\sigma t)=0$, we arbitrarily define $\hat{\nu}_q(t)=1$. The need to use $\Log$ and handle these particular values of $t$ will be irrelevant from an asymptotic point of view. We only mention these details for the technical convenience of having an estimator that is defined for all values of $t\in\R$.

\subsubsection{Optimal selection of the tuning parameter} \label{optselect}A crucial aspect of the estimator~$\hat{\nu}_q(t)$ is the choice of $t\in\R$, which plays the role of a tuning parameter. This choice turns out to be somewhat delicate, especially in situations where $\|x\|_q\to\infty$ as $(n,p)\to\infty$. To see why this matters, consider the equation
\begin{equation}\label{scaledest}
\begin{split}
\ts\frac{\hat{\nu}_q(t)}{\|x\|_q^q}
&= \ts\frac{-1}{(\gamma_q |t| \|x\|_q)^q}\Log \, \myRe \Big( \frac{\hat{\Psi}_{n_q}(t)}{\varphi_0(\sigma t)}\Big). 
\end{split}
\end{equation}
If we happen to be in a situation where $\|x\|_q\to\infty$ as $(n_q,p)\to\infty$ while the parameters $\gamma_q$, $\sigma$, and $t$ remain of order 1, then $\|\gamma_q|t| \|x\|_q\to \infty$ and the empirical charcteristic function will  $\hat{\Psi}_{n_q}$ will tend to 0 (due to line~\eqref{psiapprox}). The scaled estimate $\hat{\nu}_q(t)/\|x\|_q^q$ may then become unstable as it can tend to a limit of the form $\frac{\infty}{\infty}$ in line~\eqref{scaledest}.
Consequently, it is desirable to choose $t$ adaptively so that as $(n,p)\to\infty$,
\begin{equation}\label{constlimit}
\gamma_q t \|x\|_q \to c_0,
\end{equation}
 for some finite constant $c_0>0$, which ensures that $\hat{\nu}_q(t)/\|x\|_q^q$ remains asymptotically stable.  
When this desired scaling can be achieved, the next step is to further refine choice of $t$ so as to minimize the limiting variance of $\hat{\nu}_q(t)$. Our proposed method will solve both of these problems.

  Of course, the ability to choose $t$ adaptively requires some knowledge of $\|x\|_q$, which is precisely the quantity we are trying to estimate!~As soon as we select a data-dependent value, say $\hat{t}$, we introduce a significant technical challenge: Inferences based on the adaptive estimator $\hat{\nu}_q(\hat{t}\,)$ must take the randomness in $\hat{t}$ into account. Our approach is to prove a uniform CLT for the function $\hat{\nu}_q(\cdot)$. As will be shown in the next result, the uniformity will allow us to determine the limiting law of $\hat{\nu}_q(\hat{t}\,)$ as if the optimal choice of $t$ was known in advance of observing any data.
To make the notion of optimality precise, we first describe the limiting law of $\hat{\nu}_q(\hat{t})$ for any data-dependent value $\hat{t}$ that satisfies the scaling condition~\eqref{constlimit} (in probability). A method for constructing such a value $\hat{t}$ will be given the next subsection. Consistency results for the procedure are given in Section~\ref{results}.

To state the uniform CLT, we need to introduce the \emph{noise-to-signal} ratio
\begin{equation}\label{nsr}
\rho_{q}=\rho_q(\xi):= \ts\frac{\sigma}{\gamma_q \|x\|_q},
\end{equation}
which will be related to the width of our confidence interval for $\|x\|_q^q$. Although we allow $\rho_q$ to vary with $(n_q,p)$, we will assume that it stabilizes to a finite limiting value:

\begin{shortassumption}\label{rhoassumption}
For each $q\in(0,2]$, there is a limiting constant $\bar{\rho}_q\geq 0$ such that $\rho_q=\bar{\rho}_q+o(n^{-1/2})$ as $(n_q,p)\to\infty$.
\end{shortassumption}
\noindent This assumption merely encodes the idea that the signal is not overwhelmed by noise asymptotically.
\begin{theorem}[Uniform CLT for $\ell_q$ norm estimator]\label{clt} Let $q\in (0,2]$. Assume that the measurement model~\eqref{model1} and Assumption~\ref{rhoassumption} hold. Let $\hat{t}$ be any function of $y_1,\dots,y_{n_q}$ that satisfies
\begin{equation}\label{condlimit}
\hat{t}\gamma_q\|x\|_q\xrightarrow{ \ \ }_P c_0
\end{equation}
as $(n_q,p)\to\infty$ for some constant $c_0> 0$ with $\varphi_0(\bar{\rho}_q c_0)\neq 0$.  Then, the estimator $\hat{\nu}_q(\hat{t}\,)$ satisfies
\begin{equation}\label{deltalog}
\sqrt{n_q}\Big(\ts\frac{\hat{\nu}_q(\hat{t})}{\|x\|_q^q}-1\Big)\xrightarrow{ \ w \ } N(0,v_q(c_0,\bar{\rho}_q))
\end{equation}
as $(n_q,p)\to\infty$, where the limiting variance is strictly positive and defined according to the formula

\small
	\begin{equation}\label{varformula}
	v_q(c_0,\bar{\rho}_q):= \ts\frac{1}{|c_0|^{2q}}\Big(\ts\frac{1}{2}\frac{1}{\varphi_0(\bar{\rho}_q |c_0|)^2}\exp(2|c_0|^q) +\ts\frac{1}{2}\frac{\varphi_0(2\bar{\rho}_q |c_0|)}{\varphi_0(\bar{\rho}_q |c_0|)^2}\exp((2-2^q) |c_0|^q)- 1\Big).
	\end{equation}
\normalsize
\end{theorem}

\paragraph{Remarks.} This result is proved in Appendix~\ref{app:proc}. Although it might seem more natural to prove a CLT for the difference $\hat{\nu}_q(\hat{t})-\|x\|_q^q$ rather than the ratio $\hat{\nu}_q(\hat{t})/\|x\|_q^q$, the advantage of the ratio is that its appropriate scaling factor is $\sqrt{n_q}$, and hence independent of the size of the (possibly growing) unknown parameter $\|x\|_q^q$. 

Now that the limiting distribution of $\hat{\nu}_q(\hat{t})$ is available, we will focus on constructing an estimate $\hat{t}$ so that the limiting value $c_0$  minimizes the variance function $v_q(\cdot, \bar{\rho}_q)$.  Since the formula for $v_q(c_0 ,\bar{\rho}_q)$ is ill-defined for certain values of $c_0$, the following subsection extends the domain of $v_q(\cdot,\bar{\rho}_q)$ so that minimization can be formulated in a way that is more amenable to analysis. This extension will turn out not to affect the set of minimizers.

\subsubsection{Extending the variance function}\label{sec:varextend}
Based on the previous theorem, our aim is to construct $\hat{t}$ so that as $(n_q,p)\to\infty$,
\begin{equation}\label{optlimit}
\gamma_q \hat{t}\|x\|_q\to_P c^{\star}(\bar{\rho}_q),
\end{equation}
  where $c^{\star}(\bar{\rho}_q)$ denotes a minimizer of $v_q(\cdot,\bar{\rho}_q)$. Since we assume the noise distribution is symmetric about 0, it follows that $\varphi_0$ is a symmetric function, and consequently $v_q(c,\rho)$ is symmetric in $c$. Therefore, for simplicity, we may restrict our attention to choices of $c$ that are non-negative.
  
   An inconvenient aspect of minimizing the function $v_q(\cdot, \rho)$ is that its domain depends on $\rho$ --- since the formula~\eqref{varformula} is ill-defined at values of $c_0$ where $\varphi_0(\rho c_0)=0$. Because we will be interested in minimizing $v_q(\cdot, \hat{\rho}_q)$ for some estimate $\hat{\rho}_q$ of $\bar{\rho}_q$, this leads to analyzing the minimizer of a random function whose domain is also random. To alleviate this complication, we will define an extension of $v_q(\cdot,\cdot)$ whose domain does not depend on the second argument. Specifically, whenever $q\in (0,2)$, Proposition~\ref{varextend} below shows that an extension $\tilde{v}_q$ of $v_q$  can be found with the properties that $\tilde{v}_q(\cdot,\cdot)$ is continuous on  $[0,\infty)\times [0,\infty)$, and $\tilde{v}_q(\cdot,\bar{\rho}_q)$ has the same minimizers as $v_q(\cdot,\bar{\rho}_q)$.
   
When $q=2$, one additional detail must be handled. In this case, it may happen for certain noise distributions that $v_2(c,\bar{\rho}_2)$ approaches a minimum as $c$ tends to 0 from the right\footnote{For instance, it can be checked that this occurs in the presence of noiseless measurements where $\varphi_0\equiv 1$, or when the noise distribution is Gaussian. However, for heavier tailed noise distributions, it can also happen that $v_2(\cdot ,\bar{\rho}_2)$ is minimized at strictly positive values.}, i.e.
\begin{equation}\label{0lim}
\lim_{c\to 0+} v_2(c,\bar{\rho}_2)=\inf_{c>0}v_2(c,\bar{\rho}_2).
\end{equation}
This creates a technical nuisance in using Theorem~\ref{clt} because $v_2(c_0,\bar{\rho}_2)$ is not defined for $c_0=0$. 
There are various ways of handling this ``edge case'', but for simplicity, we take a 
practical approach of constructing $\hat{t}$ so that 
$$\gamma_2 \hat{t} \|x\|_2\to_P \ve_2$$
 for some (arbitrarily) small constant $\ve_2>0$.\footnote{ An alternative solution is to simply avoid $q=2$ and estimate $s_q(x)$ for some $q$ close to $2$.
}

 For this reason, in the particular case of $q=2$, we will restrict the domain of the extended function $\tilde{v}_2(\cdot,\cdot)$ to be $[\ve_2,\infty)\times [0,\infty)$. 
The following lemma summarizes the properties of the extended variance function that will be needed later on.
\begin{lemma}[Extended variance function]\label{varextend}
Suppose that $\varphi_0$ satisfies the assumptions of the model~\eqref{model1}, and let $v_q$ be as in formula~\eqref{varformula}. For each $q\in (0,2)$, put $\ve_q:=0$, and let $\ve_2>0$. For all values $q\in (0,2]$, define the function
\begin{equation}
\tilde{v}_q: [\ve_q,\infty)\times [0,\infty) \to [0,\infty], 
\end{equation}
according to
\begin{equation}
\tilde{v}_q(c,\rho) :=
\begin{cases}
 & v_q(c,\rho) \ \ \ \ \ \ \text{ if } (c,\rho) \text{ \ satisfies \ } \ c\neq 0 \ \textup{ and }  \ \varphi_0(\rho c)\neq 0,\\
  & +\infty \ \ \ \ \ \ \ \ \  \ \ \textup{otherwise}.
 \end{cases}
\end{equation}
 Then, the function $\tilde{v}_q(\cdot,\cdot)$ is continuous on $[\ve_q, \infty)\times [0,\infty)$, and for any $\rho\geq 0$, the function $\tilde{v}_q(\cdot,\rho)$ attains its minimum in the set $[\ve_q,\infty)$.

\end{lemma}
\paragraph{Remarks.} A simple consequence of the definition of $\tilde{v}_q(\cdot,\cdot)$ is that any choice of $c\in [\ve_q,\infty)$ that minimizes $\tilde{v}_q(\cdot,\bar{\rho}_q)$ also minimizes  $v_q(\cdot,\bar{\rho}_q)$. Hence there is nothing lost in working with $\tilde{v}_q$.

\paragraph{Minimizing the extended variance function.}  We are now in position to specify the desired limiting value $c^{\star}(\bar{\rho}_q)$ from line~\eqref{optlimit}. That is, for any $\rho\geq0$, we define
\begin{equation}\label{cstardef}
c^{\star}(\rho) \in \underset{c\geq \ve_q}{\argmin} \ \tilde{v}_q(c,\rho),
\end{equation}
where $q\in (0,2]$ and $\ve_q$ is as defined in Lemma~\ref{varextend}. 
Note that $\tilde{v}_q$ is a known function, and so the value $c^{\star}(\rho)$ can be computed for any given $\rho$. 
However, since the limiting noise-to-signal ratio $\bar{\rho}_q$ is unknown, it will be necessary to work with $c^{\star}(\hat{\rho}_q)$ for some estimate $\hat{\rho}_q$ of $\bar{\rho}_q$, which is discussed below as part of our method for constructing an optimal $\hat{t}$.

\subsubsection{A procedure for optimal selection of $t$}\label{sec:optselect} At a high level, we choose $t$ by first computing a simple ``pilot'' value $\hat{t}_{\text{pilot}}$, and then refining it to obtain an optimal value $\hat{t}_{\text{opt}}$ that will be shown to satisfy
\begin{equation}
\hat{t}_{\text{opt}} \gamma_q \|x\|_q \to_P c^{\star}(\bar{\rho}_q).
\end{equation}
 The pilot value of $t$ will satisfy 
\begin{equation}\label{tpilotlimit}
\hat{t}_{\text{pilot}} \gamma_q \|x\|_q \to_P c_0
\end{equation}
 for some (possibly non-optimal) constant $c_0$. The construction of $\tpilot$ will be given in a moment. The purpose of the pilot value is to derive a ratio-consistent estimator of $\|x\|_q^q$ through the statistic $\hat{\nu}_q(\hat{t}_{\text{pilot}})$. With such an estimate of $\|x\|_q^q$ in hand, we can easily derive a consistent estimator of $\bar{\rho}_q$, namely 
\begin{equation}
\hat{\rho}_q:= \ts\frac{\sigma}{\gamma_q (\hat{\nu}_q(\hat{t}_{\text{pilot}}))^{1/q}}.
\end{equation}
Next, we use $\hat{\rho}_q$ to estimate the optimal constant $c^{\star}(\bar{\rho}_q)$ with  $c^{\star}(\hat{\rho}_q)$, as defined in line~\eqref{cstardef}.
 Finally, we obtain an optimal choice of $t$ using
\begin{equation}
\hat{t}_{\text{opt}}:=\ts\frac{c^{\star}(\hat{\rho}_q)}{ \gamma_q (\hat{\nu}_q(\hat{t}_{\text{pilot}}))^{1/q}}.\end{equation}
The consistency of $\hat{\rho}_q$ and $\hat{t}_{\opt}$ is stated as a proposition in Section~\ref{results}.

\paragraph{Constructing the pilot value.} In choosing a pilot value for $t$, there are two obstacles to consider. First, we must choose $\tpilot$ so that the limit~\eqref{tpilotlimit} holds for some constant $c_0$. Second, we must ensure that the value $c_0$ is not a singularity of the function $v(\cdot,\bar{\rho}_q)$, i.e. $c_0\neq 0$ and  $\varphi_0(\bar{\rho}_q c_0)\neq 0$, for otherwise the variance of $\hat{\nu}_q(\tpilot)$ may diverge as $(n_q,p)\to\infty$. To handle the first item, consider the median absolute deviation statistic
\begin{equation}\label{mad}
\hat{m}_q:=\med(|y_1|,\dots,|y_{n_q}|),
\end{equation}
which is a coarse-grained, yet robust, estimate of $\gamma_q \|x\|_q$.
(The drawback of $\hat{m}_q$ is that it does not deconvolve the effects of noise in estimating $\gamma_q\|x\|_q$.)
If we define
\begin{equation}
\hat{t}_{\text{initial}}:=1/\hat{m}_q,
\end{equation}
then a straightforward argument (see the proof of Proposition~\ref{tpilot} in Section~\ref{results}) shows there is a finite constant $c_1>0$ such that as $(n_q,p)\to\infty$,
\begin{equation}\label{tinitiallimit}
\hat{t}_{\text{initial}}\gamma_q \|x\|_q \to_P c_1.
\end{equation}

Now, only a slight modification of $\hat{t}_{\text{initial}}$ is needed so that the limiting constant $c_1$ avoids the singularities of $v_q(\cdot,\bar{\rho}_q)$. Since every characteristic function is continuous and satisfies $\varphi_0(0)=1$, we may find a number $\eta_0>0$ such that $\varphi_0(\eta)>\ts\frac{1}{2}$ for all $\eta\in [0,\eta_0]$. The value $\ts\frac{1}{2}$ has no special importance. Using $\hat{t}_{\text{initial}}$, we define
$$\tpilot := \hat{t}_{\text{initial}} \wedge \ts\frac{\eta_0}{\sigma},$$
where $a\wedge b = \min\{a,b\}$. Combining the limit~\eqref{tinitiallimit} with assumption {\textbf{A\ref{rhoassumption}}}, it follows that $\tpilot \gamma_q\|x\|_q\to_P c_0$ for some finite constant $c_0>0$, since
\begin{align}\label{adjustedlimit}
\tpilot \gamma_q \|x\|_q &= (\hat{t}_{\text{initial}}\gamma_q \|x\|_q)\wedge \big(\eta_0 \ts\frac{\gamma_q\|x\|_q}{\sigma}\big)\\[0.2cm]
&= c_1\wedge (\ts\frac{\eta_0}{\bar{\rho}_q}) +o_P(1) \\[0.2cm]
&=: c_0+o_P(1).
\end{align}
Furthermore, it is clear that $c_0$ is not a singularity of $v(\cdot,\bar{\rho}_q)$, since $c_0$ is positive, and
$$\varphi_0(\bar{\rho}_q c_0) = \varphi_0\big( (\bar{\rho}_q c_1)\wedge \eta_0\big)>\ts\frac{1}{2},$$
due to the choice of $\eta_0$.
This completes the description of $\tpilot$. 

\subsection{Algorithm for estimating $\|x\|_q^q$ and $s_q(x)$}\label{sec:algorithm}

We now summarize our method by giving a line-by-line algorithm for computing the adaptive estimator $\hat{\nu}_q(\topt)$ of the parameter $\|x\|_q^q$. As described earlier, an estimate for $s_q(x)$ is obtained by combining norm estimates $\hat{\nu}_1$ and $\hat{\nu}_q$. When estimating $s_q(x)$ for $q\in (0,2]$ and $q\neq 1$, we assume that two sets of measurements (of sizes $n_1$ and $n_q$) from the model~\eqref{model1} are available, i.e.
\small
\begin{align}
y_i &= \langle a_i,x\rangle +\sigma\e_i, \text{ \ \ \ \   with \ \ \ \   } a_i \overset{\text{i.i.d.}}{\sim} \stable_q(\gamma_q)^{\otimes p},  \text{ \ \ \ \ for \ \ \ \ } i=1,\dots, n_q,\label{meas1}\\
y_i &= \langle a_i,x\rangle +\sigma \e_i, \text{ \ \ \ \   with \ \ \ \   } a_i \overset{\text{i.i.d.}}{\sim} \stable_1(\gamma_1)^{\otimes p},  \text{ \ \ \ \ for \ \ \ \ } i=n_q+1,\dots, n_q+ n_1.\label{meas2}
\end{align}
\normalsize
Once the estimators $\hat{\nu}_q(\hat{t}_{\opt})$ and $\hat{\nu}_1(\hat{t}_{\opt})$ have been computed with their respective values of $\hat{t}_{\opt}$, the estimator $\hat{s}_q(x)$ is obtained as

\begin{equation}
\hat{s}_q(x) := \ts \frac{\big(\hat{\nu}_q(\hat{t}_{\opt})\big)^{\frac{1}{1-q}}}{\big(\hat{\nu}_1(\hat{t}_{\opt})\big)^{\frac{q}{1-q}}}.
\end{equation}
The algorithm to compute $\hat{\nu}_q(\topt)$ is given below.
\begin{algorithm}{(Estimation procedure for $\|x\|_q^q$, for $q\in (0,2]$).}{
\label{algo:ifforwhile}
\qinput{\vspace{0.1cm}
\hrule
\vspace{0.2cm}
 \noindent \begin{itemize}
\vspace{-.53cm}
\item \ observations  $y_i$ generated with i.i.d. measurement vectors $a_i \sim \stable_q(\gamma_q)^{\otimes p}$, for $i=1,\dots, n_q$
\item \ measurement intensity $\gamma_q$
\item \ noise level $\sigma$
\item \ noise characteristic function $\varphi_0$
\item \ threshold $\ve_q$ defined in Lemma~\ref{varextend}
\end{itemize}
}
\vspace{0.1cm}
\vspace{0.1cm}
\hrule
}
\vspace{0.2cm}
 compute $\hat{t}_{\text{initial}}:=1/\hat{m}_q$ where $\hat{m}_q:=\med(|y_1|,\dots,|y_{n_q}|)$\\
 \vspace{0.3cm}
find $\eta_0>0$ such that $\varphi_0(\eta)>\ts\frac{1}{2}$ for all $\eta\in [0,\eta_0]$\\
\vspace{0.3cm}
compute $\tpilot := \hat{t}_{\text{initial}} \wedge \ts\frac{\eta_0}{\sigma},$\\
\vspace{0.3cm}
compute $\hat{\rho}_q:= \ts\frac{\sigma}{\gamma_q (\hat{\nu}_q(\hat{t}_{\text{pilot}}))^{1/q}}$\\
\vspace{0.3cm}
compute $c^{\star}(\hat{\rho}_q)\in \argmin_{c\geq \ve_q} \ \tilde{v}_q(c,\hat{\rho}_q)$.\\
\vspace{0.3cm}
compute $\hat{t}_{\text{opt}}:=\ts\frac{c^{\star}(\hat{\rho}_q)}{ \gamma_q (\hat{\nu}_q(\hat{t}_{\text{pilot}}))^{1/q}}$\\
\vspace{0.3cm}
\qreturn $\hat{\nu}_q(\topt)$
\vspace{0.3cm}
\hrule
\end{algorithm}

\section{Main results for estimators and confidence intervals}\label{results}
In this section, we first show in Proposition~\ref{tpilot} that the procedure for selecting the $\tpilot$ leads to consistent estimates of the parameters $\|x\|_q^q$,  and $\bar{\rho}_q$. 
Next, we show that the optimal constant $c^{\star}(\bar{\rho}_q)$ and optimal variance $v(c^{\star}(\bar{\rho}_q),\bar{\rho}_q)$ are also consistently estimated. These estimators then lead to adaptive confidences intervals for $\|x\|_q^q$ and $s_q(x)$, described in Theorem~\ref{CI} and Corollary~\ref{CIs}.
\subsection{Consistency results}\label{sec:consist}

\begin{proposition}[Consistency of the pilot estimator]\label{tpilot}
Let $q\in(0,2]$ and assume that the measurement model~\eqref{model1} and {\textbf{A\ref{rhoassumption}}} hold. Then as $(n_q,p)\to\infty$,
\begin{equation}\label{pilotconsist}
\ts\frac{\hat{\nu}_q(\hat{t}_{\text{\emph{pilot}}})}{\|x\|_q^q}\longrightarrow_P 1
\end{equation}
and
\begin{equation}\label{rhoconsist}
\hat{\rho}_q \longrightarrow_P \bar{\rho}_q.
\end{equation}
\end{proposition}

\paragraph{Remarks.} We now turn our attention from the pilot value $\tpilot$ to the optimal value $\topt$. In order to construct $\topt$, our method relies on the $M$-estimator $c^{\star}(\hat{\rho}_q)\in \argmin_{c\geq \ve_q}v_q(c,\hat{\rho}_q)$, where $c^{\star}(\cdot)$ is defined in line~\eqref{cstardef}. Consistency proofs for $M$-estimators typically require that the objective function has a unique optimizer, and our situation is no exception. 
Note that we do not need to assume that a minimizer exists, since this is guaranteed by Lemma~\ref{varextend}.
\begin{shortassumption}\label{uniqueassumption}
The function $v_q(\cdot,\bar{\rho}_q)$ has at most one minimizer in $[\ve_q,\infty)$, where $\ve_q$ is defined in Lemma~\ref{varextend}.
\end{shortassumption}
In an approximate sense, this assumption can be checked empirically by simply plotting the function $v_q(\cdot, \hat{\rho}_q)$. In Section~\ref{misc} of the Appendix, we verify the assumption analytically in the case of $\text{stable}_q$ noise.
Based on our graphical inspection, the assumption appears to hold for a variety of natural parametric noise distributions (e.g. Laplace, uniform$[-1,1]$, and the $t$ distribution). However, outside of special cases, analytic verification seems to be difficult, and even the stable case is somewhat involved.
\begin{proposition}[Consistency of  $c^{\star}(\hat{\rho}_q)$]\label{mest}
Let $q\in(0,2]$ and assume that the measurement model~\eqref{model1} holds, as well as assumptions~{\textbf{A\ref{rhoassumption}}}, and {\textbf{A\ref{uniqueassumption}}}.
Then, as $(n_q,p)\to\infty$,
\begin{equation}\label{optlimitprop}
\ \ \ \ \ \ c^{\star}(\hat{\rho}_q) \ \longrightarrow_P \ c^{\star}(\bar{\rho}_q).
\end{equation}
Furthermore,
\begin{align}
\topt\gamma_q \|x\|_q & \ \longrightarrow_P \ c^{\star}(\bar{\rho}_q),
\end{align}
and
\begin{align}
 \ \ \ \ \  v_q(c^{\star}(\hat{\rho}_q),\hat{\rho}_q)& \ \longrightarrow_P \ v_q(c^{\star}(\bar{\rho}_q),\bar{\rho}_q).
\end{align}
\end{proposition}
\paragraph{Remarks.} This result is proved in Section~\ref{resultsproofs} of the Appendix.

\subsection{Confidence intervals for $\|x\|_q^q$ and $s_q(x)$}\label{sec:CI}
In this subsection, we assemble the work in Proposition~\ref{mest} with Theorem~\ref{clt} to obtain confidence intervals for $\|x\|_q^q$ and $s_q(x)$. In the next two results, we will use $z_{1-\alpha}$ to denote the $1-\alpha$ quantile of the standard normal distribution, i.e. $\Phi(z_{1-\alpha}) = 1-\alpha$. To allow our result to be applied to both one-sided and two-sided intervals, we state our result in terms of two possibly distinct quantiles $z_{1-\alpha}$ and $z_{1-\alpha'}$.
\begin{theorem}[Confidence interval for $\|x\|_q^q$]\label{CI}
Let $q\in(0,2]$ and define the estimated variance
\begin{equation}
\hat{\omega}_q:=v_q(c^{\star}(\hat{\rho}_q),\hat{\rho}_q).
\end{equation}
 Assume that the measurement model~\eqref{model1} holds, as well as assumptions~{\textbf{A\ref{rhoassumption}}}, and~{\textbf{A\ref{uniqueassumption}}}. Then as $(n_q,p)\to\infty$,
\begin{equation}\label{nuclt}
\ts\frac{\sqrt{n_q}}{\sqrt{\hat{\omega}_q}} \Big(\frac{\hat{\nu}_q(\topt)}{\|x\|_q^q} - 1\Big) \xrightarrow{\ w \ } N(0,1),
\end{equation}
and consequently for any fixed $\alpha,\alpha'\in [0,\ts\frac{1}{2}]$,
\begin{equation}\label{nuCIlimit}
\small
\P \Bigg[     \Big( 1-\ts\frac{\sqrt{\hat{\omega}_q} z_{1-\alpha}}{\sqrt{n_q}}\Big)  \cdot     \hat{\nu}_q(\topt)\leq \ \|x\|_q^q \  \leq    \Big( 1+\ts\frac{\sqrt{\hat{\omega}_q} z_{1-\alpha'}}{\sqrt{n_q}}\Big) \cdot        \hat{\nu}_q(\topt)\Bigg]\to 1-\alpha-\alpha'.
\normalsize
\end{equation}
\end{theorem}

\paragraph{Remarks.} This result follows by combining Theorem~\ref{clt} with Proposition~\ref{mest}.  We note that if the limit~\eqref{nuclt} is used directly to obtain a confidence interval for $\|x\|_q^q$, the resulting formulas are somewhat cumbersome. Instead, a simpler confidence interval (given in line~\eqref{nuCIlimit}) is obtained using a CLT for the reciprocal $\|x\|_q^q/\hat{\nu}_q(\topt)$, via the delta method. For a one-sided interval with the right endpoint being $+\infty$, we set $\alpha'=0$, and similarly, we set $\alpha=0$ in the opposite case. 

As a corollary of Theorem~\ref{CI}, we obtain a CLT and a confidence interval for $\hat{s}_q(x)$ by combining the estimators $\hat{\nu}_q(\topt)$ and $\hat{\nu}_1(\topt)$. Since each of the norm estimators rely on measurement sets of sizes $n_q$ and $n_1$, we make the following simple scaling assumption, which enforces the idea that each set should be non-negligible with respect to the other.

\begin{shortassumption}\label{assumepi}
For each $q\in (0,2]\setminus\{1\}$, there is a constant $\bar{\pi}_q\in(0,1)$, such that as $(n_1,n_q,p)\to\infty$, 
\begin{equation}
\ts\frac{n_q}{n_1+n_q} = \bar{\pi}_q+o(n_q^{-1/2}).
\end{equation}
\end{shortassumption}
\begin{corollary}[Confidence interval for $s_q(x)$]\label{CIs} Assume $q\in (0,2]\setminus\{1\}$, and that the conditions of Theorem~\ref{CI} hold, as well as assumption~{\textbf{A\ref{assumepi}}}. Also assume $\hat{s}_q(x)$ is constructed from independent sets of measurements~\eqref{meas1} and~\eqref{meas2}. Letting $\hat{\omega}_q$ be as in Theorem~\ref{CI}, define the quantities
 $$\pi_q:=n_q/(n_1+n_q) \text{  \ \ \ \ and \ \ \ \ } \hat{\vartheta}_q:=\ts\frac{\hat{\omega}_q}{\pi_q}(\ts\frac{1}{1-q})^2+\frac{\hat{\omega}_1}{1-\pi_q}(\ts\frac{q}{1-q})^2.$$
  Then as $(n_1,n_q,p)\to\infty$,
\begin{equation}\label{sclt}
\ts\frac{\sqrt{n_1+n_q}}{\sqrt{\hat{\vartheta}_q}} \Big(\frac{\hat{s}_q(x)}{s_q(x)} - 1\Big) \xrightarrow{\ w \ } N(0,1),
\end{equation}
and consequently for any fixed $\alpha,\alpha'\in [0,\ts\frac{1}{2}]$,
\begin{equation}\label{sCIlimit}
\small
\P \Bigg[     \Big( 1-\ts\frac{\sqrt{\hat{\vartheta}_q} z_{1-\alpha}}{\sqrt{n_1+n_q}}\Big)  \cdot     \hat{s}_q(x)\leq \ s_q(x) \ \leq    \Big( 1+\ts\frac{\sqrt{\hat{\vartheta}_q} z_{1-\alpha'}}{\sqrt{n_1+n_q}}\Big) \cdot        \hat{s}_q(x) \Bigg]\to 1-\alpha-\alpha'.
\normalsize
\end{equation}
\end{corollary}
\paragraph{Remarks.} As in Theorem~\ref{CI}, we chose to present a simpler formula for the confidence interval in line~\eqref{sCIlimit}  by using a CLT for the reciprocal $s_q(x)/\hat{s}_q(x)$.\\

\section{Applications of confidence intervals for $\|x\|_q^q$ and $s_q(x)$}\label{sec:app}

In this section, we give some illustrative applications of our results for $\hat{\|x\|_q^q}$ and $\hat{s}_q(x)$.  First, we describe how the assumption of sparsity may be checked in a hypothesis testing framework. Second, we consider the problem of choosing the regularization parameter for the Lasso  and Elastic Net algorithms (in primal form).
\subsection{Testing the hypothesis of sparsity}
In the context of hypothesis testing, the null hypothesis is typically viewed as a ``straw man'' that the practitioner would like to reject in favor of the ``more desirable'' alternative hypothesis. Hence, for the purpose of verifying the assumption of sparsity, it is natural for the null hypothesis to correspond to a non-sparse signal. More specifically, if $1<\kappa\leq p$ is a given reference value of sparsity, then we consider the testing problem
\begin{equation}
\textbf{H}_0:  s_q(x) \geq \kappa  \ \ \ \ \ \text{ versus } \ \ \ \ \  \textbf{H}_1:  1\leq s_q(x)<\kappa.
\end{equation}
To construct a test statistic, we use the well-known duality between confidence intervals and hypothesis tests~\cite{lehmann}. Consider a one-sided confidence interval for $s_q(x)$ of the form $(-\infty,\hat{u}_{\alpha}]$, with asymptotic coverage probability $\P(s_q(x)\leq \hat{u}_{\alpha})=1-\alpha+o(1)$. Clearly, if $\textbf{H}_0$ holds, then this one-sided interval must also contain $\kappa$ with probability at least $1-\alpha+o(1)$. Said differently, this means that under $\textbf{H}_0$, the chance that $(-\infty,\hat{u}_{\alpha}]$ fails to contain $\kappa$ is at most $\alpha+o(1)$. Likewise, one may consider the test statistic
$$ T:= 1\big\{\hat{u}_{\alpha} < \kappa \big\},$$
and reject $\textbf{H}_0$ iff $T=1$, which gives an asymptotically valid level-$\alpha$ testing procedure. Now, by Corollary~\ref{CIs}, if we choose
$$\hat{u}_{\alpha}:=(1+\ts\frac{\hat{\vartheta}_q z_{1-\alpha}}{\sqrt{n_1+n_q}}\big)\hat{s}_q(x),$$
then the interval $(-\infty,\hat{u}_{\alpha}]$ has asymptotic coverage probability $1-\alpha$.
 The reasoning just given ensures that the false alarm rate is asymptotically bounded by $\alpha$ as $(n_1,n_q,p)\to\infty$. Namely,
$$\P_{\textbf{H}_0}(T=1)\leq \alpha+o(1).$$
It is also possible to derive the asymptotic power function of the test statistic. Let $\hat{\vartheta}_q$ be as defined in Corollary~\ref{CIs}, and note that this variable converges in probability to a positive constant, say $\vartheta_q$ (by Proposition~\ref{mest}). Then, as $(n_1,n_q,p)\to\infty$, the asymptotic power satisfies
\begin{equation}\label{powerfunc}
\P_{\textbf{H}_1}(T=1) = \Phi\Big(\ts\frac{\sqrt{n_1+n_q}}{\vartheta_q}\Big(\ts\frac{\kappa}{s_q(x)}-1\Big)-z_{1-\alpha}\Big)+o(1).
\end{equation}
The details of obtaining this limit are straightforward, and hence omitted. Note that as $s_q(x)$ becomes close to the reference value $\kappa$ (i.e. the detection boundary), the power approaches that of random guessing, $\Phi(-z_{1-\alpha})=\alpha$, as we would expect.
\subsection{Tuning the Lasso and Elastic Net in primal form}\label{tuninglasso} 
In primal form, the Lasso algorithm can be expressed as

\begin{equation}\label{lassoprimal}
\begin{aligned}
& \underset{v\in\R^p}{\text{minimize}}
& & \|y-Av\|_2^2 \\
& \text{subject to}
& & v\in \B_1(r),
\end{aligned}
\end{equation}
where $\B_1(r):=\{v\in\R^p: \|v\|_1\leq r\}$ is the $\ell_1$ ball of radius $r\geq 0$, and $r$ is viewed as the regularization parameter. (Note that the matrix $A$ here may be different from the measurement matrix we use to estimate $s_q(x)$.) If $x$ denotes the true signal, then $\B_1(\|x\|_1)$ is the smallest such set for which the true signal is feasible. Hence, one would expect $r=\|x\|_1$ to be an ideal choice of the tuning parameter. In fact, the recent paper~\cite{chatterjeeConvex} shows that this intuition is correct in a precise sense by quantifying how the mean-squared prediction error of the Lasso deteriorates when the tuning parameter differs from $\|x\|_1$.

When using a data-dependent tuning parameter $\hat{r}$, it is of interest to have some guarantee that the true signal is likely to lie in the (random) set $\B_1(\hat{r}\, )$. Our one-sided confidence interval for $\|x\|_1$ precisely solves this problem. More specifically, under the assumptions of Theorem~\ref{CI}, if we choose $\alpha=0$, and $\alpha'\in[0,\ts\frac{1}{2}]$, then under the choice
\begin{equation}
\hat{r}:= \Big( 1+\ts\frac{\sqrt{\hat{\omega}_1} z_{1-\alpha'}}{\sqrt{n_1}}\Big) \cdot        \hat{\nu}_1(\topt),
\end{equation}
we have as $(n_1,p)\to\infty$
\begin{equation}
\P\big(x\in \B_1(\hat{r})\big) \to 1-\alpha'.
\end{equation}
In fact, this idea can be extended further by adding extra $\ell_q$ norm constraints. A natural example is a primal form of the well known \emph{Elastic Net} algorithm~\cite{elasticnet}, which constrains both the $\ell_1$ and $\ell_2$ norms, leading to the convex program
\begin{equation}\label{elasticprimal}
\begin{aligned}
& \underset{v\in\R^p}{\text{minimize}}
&  \|y-Av\|_2^2 \\
& \text{subject to}
& v\in \B_1(r)\\
&& v\in \B_2(\varrho)
\end{aligned}
\end{equation}
for some parameters $r,\varrho\geq 0$. Here, $\B_2$ is defined in the same way as $\B_1$.  Again, under the assumptions of Theorem~\ref{CI}, if for some $\alpha'\in[0,\ts\frac{1}{2}]$ we put
\begin{equation}
\hat{\varrho} := \Big( 1+\ts\frac{\sqrt{\hat{\omega}_2} z_{1-\alpha}}{\sqrt{n_2}}\Big) \cdot        \hat{\nu}_2(\topt),
\end{equation}
and if independent measurement sets of size $n_1$ and $n_2$ used to construct $\hat{r}$ and $\hat{\varrho}$, then as $(n_1,n_2,p)\to\infty$,
\begin{equation}
\P\Big( x \in \B_1(\hat{r})\cap \B_2(\hat{\varrho})\Big) \to (1-\alpha')^2.
\end{equation}
The same reasoning applies to any other combination of $\ell_q$ norms for $q\in (0,2]$.
%
%
%
%
%
%

%
%
%
%
%
%
%
%
%
%
%
%
%
%
%
%

%
%
%
%
 %
 %
 %

\section{Deterministic measurement matrices}\label{sec:neg}

The problem of constructing deterministic matrices $A$ with good recovery properties (e.g. RIP-$k$ or NSP-$k$) has been a longstanding open direction within CS~\cite[see Sections 1.3 and 6.1]{foucartbook}~\cite{devore2007deterministic}. Since our procedure in Section~\ref{sec:procedure} selects $A$ at random, it is natural to ask if randomization is essential to the estimation of unknown sparsity. In this section, we show that estimating $s_q(x)$ with a deterministic matrix $A$ leads to results that are inherently different from our randomized procedure.

At an informal level, the difference between random and deterministic matrices makes sense if we think of the estimation problem as a game between nature and a statistician. Namely, the statistician first chooses a matrix $A\in\R^{n\times p}$ and an estimation rule $\delta :\R^n\to \R$. (The function $\delta$ takes $y\in\R^n$ as input and returns an estimate of $s_q(x)$.)  In turn, nature chooses a signal $x\in\R^p$, with the goal of maximizing the statistician's error. When the statistician chooses $A$ deterministically, nature has the freedom to adversarially select an $x$ that is ill-suited to the fixed matrix $A$. By contrast, if the statistician draws $A$ at random, then nature does not know what value $A$ will take, and therefore has less knowledge to choose a ``bad'' signal.

In the case of \emph{random} measurements, Corollary~\ref{CIs} implies that our particular estimation rule $\hat{s}_q(x)$ can achieve a relative error \mbox{$|\hat{s}_q(x)/s_q(x) -1|$} on the order of $1/\sqrt{n_1+n_q}$ with high probability for any non-zero $x$.  Our aim is now to show that for any set of noiseless \emph{deterministic} measurements, \emph{all} estimation rules $\delta:\R^n \to \R$ have a worst-case relative error $|\delta(Ax)/s_2(x)-1|$ that is much larger than $1/\sqrt{n_1+n_q}$. Specifically, when $q\in [0,2]$, we give a lower bound that is of order $(1-\frac{n}{p})^2$, which means that in the worst case, $s_q(x)$ cannot be estimated consistently in relative error when $n\ll p$. (The same conclusion holds for $q\in (2,\infty]$ up to a factor of $\sqrt{\log(2p)}$.) More informally, this means that there is always a choice of $x$ that can defeat a deterministic procedure, whereas the randomized estimator $\hat{s}_q(x)$ is likely to succeed under any choice of $x$. 

In stating the following result, we note that it involves no randomness whatsoever --- since we assume here that the observed measurements $y=Ax$ are noiseless and obtained from a deterministic matrix $A$. Furthermore, the bounds are non-asymptotic.
 \begin{theorem}\label{minimaxnew}
Suppose $n<p$, and $q\in [0,2]$. Then, the minimax relative error for estimating $s_q(x)$ from noiseless deterministic measurements $y=Ax$ satisfies
\begin{equation*}
 \inf_{A\in \R^{n\times p}} \: \inf_{\delta:\R^n\to\R} \:\sup_{x\in\R^p\setminus\{0\}}\Big|\ts\frac{\delta(Ax)}{s_q(x)}-1\Big|\geq \ts\frac{1}{2\pi e} \big(1-\ts\frac{n}{p}\big)^2-\ts\frac{1}{2p}.
 \end{equation*}
 Alternatively, if $q\in (2,\infty]$, then
 \begin{equation*}
\ \ \ \ \ \ \ \  \inf_{A\in \R^{n\times p}} \: \inf_{\delta:\R^n\to\R} \:\sup_{x\in\R^p\setminus\{0\}}\Big|\ts\frac{\delta(Ax)}{s_q(x)}-1\Big|\geq  \ts\ts\frac{1}{\sqrt{2\pi e}}\cdot \frac{ 1-(n/p)}{1 + \sqrt{16\log(2p)}}-\ts\frac{1}{2p}.
 \end{equation*}
\end{theorem}
\paragraph{Remarks.} The proof of this result is based on the classical technique of a \emph{two-point prior}. In essence, the idea is that for any choice of $A$, it is possible to find two signals $\tilde{x}$ and $x^{\circ}$ that are indistinguishable with respect to $A$, i.e.
\begin{equation}\label{indist}
A\tilde{x} = Ax^{\circ},
\end{equation}
and yet have very different sparsity levels,

\begin{equation}\label{diffsparse}
s_q(x^{\circ}) \ll s_q(\tilde{x}).
\end{equation}
Due to the relation~\eqref{indist}, the statistician has no way of knowing whether $\tilde{x}$ or $x^{\circ}$ has been selected by nature, and if nature chooses $x^{\circ}$ and $\tilde{x}$ with equal probability, then it is impossible for the statistician to improve upon the trivial estimator $\frac{1}{2}s_q(\tilde{x})+\frac{1}{2}s_q(x^{\circ})$ that does not even make use of the data. Furthermore, since $s_q(x^{\circ})\ll s_q(\tilde{x})$, it follows that the trivial estimator has a large relative error -- implying that the minimax relative error is also large. (A formal version of this argument is given in Section~\ref{app:neg} of the Appendix.)

To implement the approach of a two-point prior, the main challenge is to show that for \emph{any} choice of $A\in\R^{n\times p}$, two vectors satisfying~\eqref{indist} and~\eqref{diffsparse} can actually be found. This is the content of the following lemma.

\begin{lemma}\label{impslem}
Let $A\in \R^{n\times p}$ be an arbitrary matrix with $n< p$, and let $x^{\circ}\in\R^p$ be an arbitrary signal. Then, for each $q\in [0,2]$, there exists a non-zero vector $\tilde{x}\in\R^p$ satisfying $A\tilde{x}=Ax^{\circ}$ and
\begin{equation}\label{newbound}
s_q(\tilde{x}) \geq \ts\frac{1}{\pi e}\cdot(1-\ts\frac{n}{p})^2\cdot p.
\end{equation}
Also, for $q\in (2,\infty]$, there is a non-zero vector $\bar{x}$ satisfying $A\bar{x}=Ax^{\circ}$ and
\begin{equation}\label{sinftybound}
s_{q}(\bar{x}) \geq \frac{\sqrt{\ts\frac{2}{\pi e}} (p-n)}{1 + \sqrt{16\log(2p)}}.
\end{equation}
\end{lemma}

\paragraph{Remarks.} Although it might seem intuitively obvious that every affine subspace contains a non-sparse vector, the technical substance of the result lies in the fact that the bounds hold \emph{uniformly} over all matrices $A\in\R^{n\times p}$. This uniformity is necessary when taking the infimum over all $A\in\R^{n\times p}$ in Theorem~\ref{minimaxnew}. Furthermore, the order of magnitude of the bounds for $q\in [0,2]$ is unimprovable when $n\ll p$, since $s_q(x)\leq p$. Similarly, the bound for $q\in(2,\infty]$ is optimal up to a logarithmic factor. Our proof in Appendix~\ref{app:neg} uses the probabilistic method to show that the desired vectors $\tilde{x}$ and $\bar{x}$ exist. Namely, we put a distribution on the set of vectors $v$ satisfying $Ax=Av$, and then show that the stated bounds hold with positive probability. 
%
%
%
%
%
%
%
%
%
%
\section{Simulations}\label{sec:simulations}
In this section, we describe two sets of simulations that validate our theoretical results, and demonstrate the effectiveness of our estimation method. The first set of simulations, discussed in Section~\ref{sec:sweeps} looks at the expected relative error $\E|\frac{\hat{s}_2(x)}{s_2(x)}-1|$ and how it is affected by the dimension $p$, the sparsity level $s_2(x)$, and the noise-to-signal ratio $\rho_2$. The second set of simulations, discussed in Section~\ref{sec:normal} examines the quality of the normal approximation $\frac{\sqrt{n_1+n_2}}{\sqrt{\hat{\vartheta}_2}}(\frac{\hat{s}_2(x)}{s_2(x)}-1) \xrightarrow{ \ w \ } N(0,1)$ given in Corollary~\ref{CIs}.

\subsection{Error dependence on dimension, sparsity, and noise}\label{sec:sweeps}
\begin{figure*}[!]
\centering
{\includegraphics[angle=0,
  width=.45\linewidth]{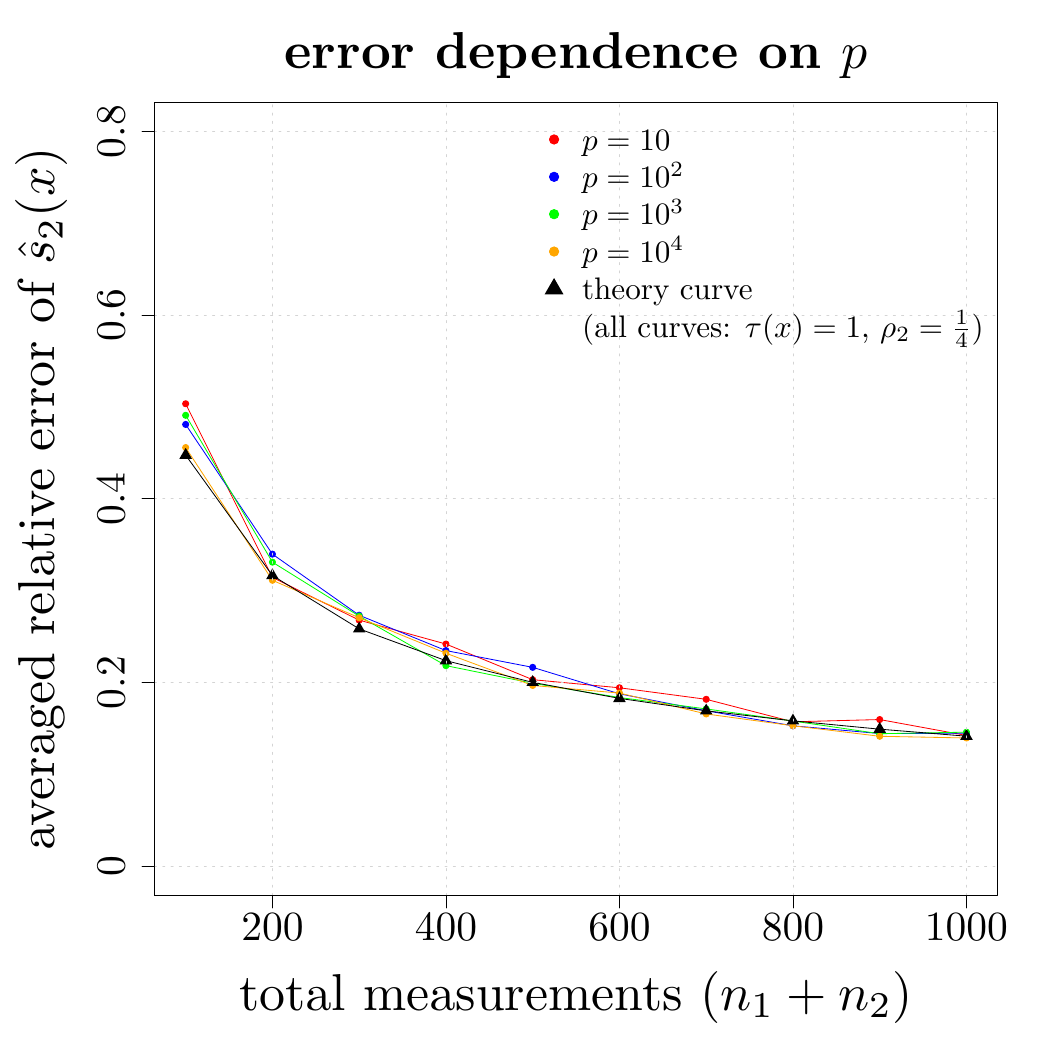}}
{\includegraphics[angle=0,
  width=.45\linewidth]{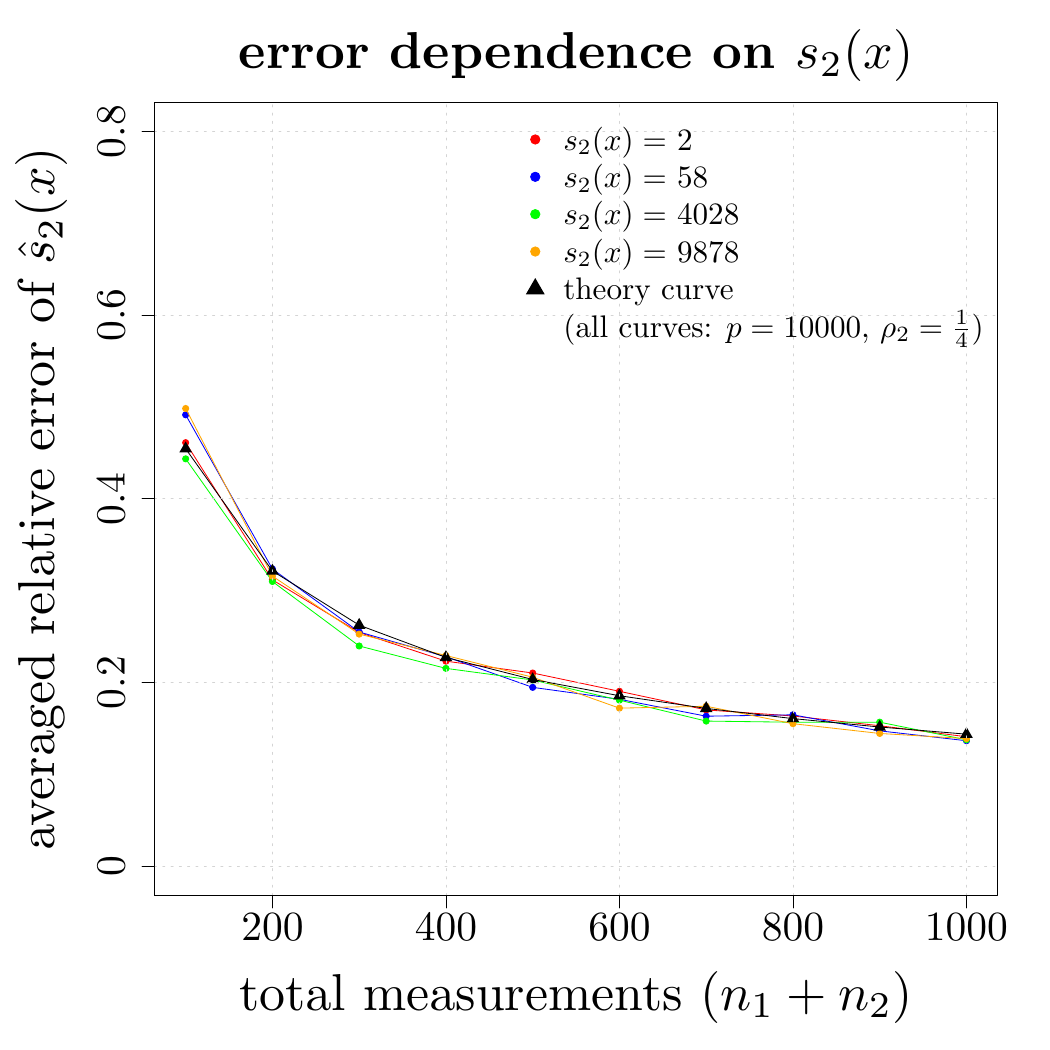}}
  ~\\[0.2cm]
{\includegraphics[angle=0,
  width=.45\linewidth]{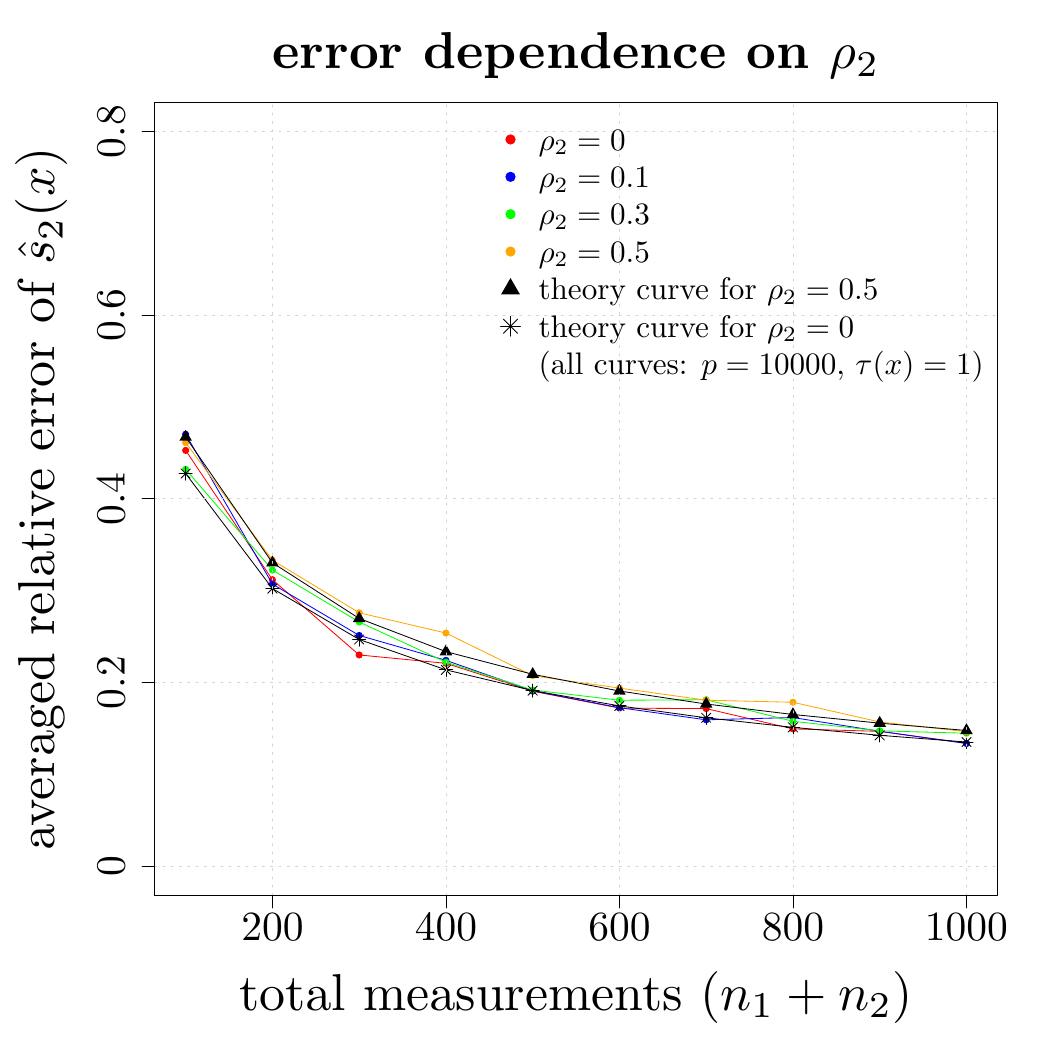}}
\caption{The three plots show that as a function of the number of measurements, the relative error $\E|\frac{\hat{s}_2(x)}{s_2(x)}-1|$ has no observable dependence on $p$ or $s_2(x)$. Also, the dependence on the noise-to-signal ratio $\rho_2$ is mild. The parameter $\tau(x)$ refers to the decay exponent of the signal entries $x_i =c\cdot i^{-\tau(x)}$, with $c$ chosen so that $\|x\|_2=1$. The conditions of the experiments are described in the main text.}
\label{fig:sweeps}
\end{figure*}

\paragraph{Design of simulations.} When estimating $s_2(x)$, our proposed method requires a set of $n_1$ Cauchy measurements, and a set of $n_2$ Gaussian measurements. In each of the three plots in Figure~\ref{fig:sweeps}, we considered a sequence of pairs $(n_1,n_2)= (50,50), (100,100),$ $(150,150),\dots,(500,500)$. For a fixed pair $(n_1,n_2)$, we generated 500 independent vectors of Cauchy measurements $y^{(1)}\in\R^{n_1}$ and Gaussian measurements $y^{(2)}\in\R^{n_2}$ according to
\begin{align}
y^{(1)}&=A^{(1)}x+\sigma \e^{(1)}\label{sim1}\\[0.1cm]
y^{(2)}&=A^{(2)}x+\sigma \e^{(2)}\label{sim2}
\end{align}
where the entries of $A^{(1)}\in\R^{n_1\times p}$ are i.i.d. $\stable_1(1)$ (Cauchy) variables, and the entries of $A^{(2)}\in\R^{n_2\times p}$ are i.i.d. $\stable_2(1)$ (Gaussian) variables. In all three plots, the noise vectors $\e^{(1)}$ and $\e^{(2)}$ were generated with i.i.d. entries from a centered $t$-distribution on 2 degrees of freedom. (We selected this distribution since it has \emph{infinite} variance and therefore reveals the robustness of our method to heavy-tailed noise.)

Applying our estimation method to each pair $(y^{(1)},y^{(2)})$ produced 500 realizations of $\hat{s}_2(x)$ for each $(n_1,n_2)$.  We then averaged the quantity $|\frac{\hat{s}_2(x)}{s_2(x)}-1|$ over the 500 realizations as an approximation of $\E|\frac{\hat{s}_2(x)}{s_2(x)}-1|$. The sample average is plotted in the colors red, blue, green, and orange as a function of $(n_1+n_2)$, depending on the choices of $p$, $s_2(x)$, or $\rho_2$ stated in the legend. Further details for parameter settings are given below.

\paragraph{Qualitative comments.}
There are three high-level conclusions to take away from Figure~\ref{fig:sweeps}. The first is that the black theoretical curves agree well with the colored empirical curves. Second, the average relative error $\E|\frac{\hat{s}_2(x)}{s_2(x)}-1|$ has no observable dependence on $p$ or $s_2(x)$ (when $\rho_2$ is held fixed), as would be expected from Corollary~\ref{CIs}. Third, the dependence on the noise-to-signal ratio is mild. (Note that the theory \emph{does} predict that the relative error for $\rho_2=0.5$ will be somewhat larger than in the noiseless case with $\rho_2=0$.)

In all three plots, the theoretical curves were computed in the following way. From Corollary~\ref{CIs} we have the approximation $|\frac{\hat{s}_2(x)}{s_2(x)}-1| \approx \frac{\sqrt{\vartheta_2}}{\sqrt{n_1+n_2}} |Z|$ where $Z$ is a standard Gaussian random variable, and $\vartheta_2$ is the limit of $\hat{\vartheta}_2$ that results from Corollary~\ref{CIs} and Proposition~\ref{mest}. Since $\E|Z|=\sqrt{2/\pi}$, the theoretical curves are simply $\frac{\sqrt{ \frac{2}{\pi}\vartheta_2}}{\sqrt{n_1+n_2}}$. Note that $\vartheta_2$ depends only on $\rho_2$, and does not depend on $s_2(x)$ or $p$, which explains why there is only one theoretical curve in the top two plots.

\paragraph{Parameter settings.}\label{paramsettings} In all three plots, the signal $x\in\R^p$ was chosen to have entries decaying according to $x_i= c\cdot i^{-\tau(x)}$ with $c$ chosen so that $\|x\|_2=1$. The dimension $p$ was set to 10,000 in all cases, except as indicated in the top left plot, where $p=10,10^2,10^3,10^4$. The decay exponent $\tau(x)$ was set to 1 in all cases, except in the top right plot where $\tau(x)=2,1,1/2, 1/10$ in order to give a variety of sparsity levels, as indicated in the legend. With regard to the generation of measurements, the energy level was set to $\gamma_1=\gamma_2=1$ in all cases, which gives $\rho_2= \sigma/\gamma_2\|x\|_2=\sigma$ in all cases. In turn, we set $\rho_2=\sigma=1/4$ in all cases, except as indicated in the bottom plot where $\rho_2=\sigma=0, 0.1,0.3, 0.5$. Lastly, our algorithm for computing $\hat{s}_2(x)$ involves a choice of two input parameters $\eta_0$ and $\ve_2$ (see Section~\ref{sec:algorithm}). In all cases, we set $\eta_0=0.3$ and $\ve_2=0.3$. The performance of $\hat{s}_2(x)$ did not seem to change much under different choices, except when $\ve_2$ was chosen to be much smaller than the stated value.

\subsection{Normal approximation}\label{sec:normal}

\paragraph{Design of simulations.} Under four different conditions, we generated 3000 instances of the measurement scheme given in lines \eqref{sim1} and~\eqref{sim2}. Hence, each of the four conditions resulted in 3000 samples of the standardized statistic $\frac{\sqrt{n_1+n_2}}{\sqrt{\hat{\vartheta}_q}}\big(\frac{\hat{s}_2(x)}{s_2(x)}-1\big)$. The four different conditions correspond to setting $(n_1,n_2)=(500,500), (1000,1000)$, or choosing the entries of the noise vectors $\e^{(1)}$, $\e^{(2)}$ to be i.i.d. $t$-distributed (centered), with either 2 or 20 degrees of freedom. (Note that noise with 2 degrees of freedom has infinite variance, while noise with 20 degrees of freedom has finite variance.) Further details on parameter settings are given below.

In each of the four conditions, we plotted a histogram from the 3000 samples of the standardized statistic using the Freedman-Diaconis rule for bin widths ({\tt{R}} command {\tt{hist(..., breaks="FD")}}). The default {{\tt{R}} kernel density estimate is plotted with a red dashed curve in each case, and the solid blue curve is the standard normal density.

\paragraph{Qualitative comments.}  Figure~\ref{fig:hists} illustrates that the CLT $\frac{\sqrt{n_1+n_2}}{\sqrt{\hat{\vartheta}_2}}(\frac{\hat{s}_2(x)}{s_2(x)}-1) \xrightarrow{ \ w \ } N(0,1)$ is able to tolerate two challenging issues: the first being that the tuning parameter $\hat{t}_{\text{opt}}$ is selected in a data-dependent way in all four plots, and the second being that the noise distribution has infinite variance in the top two plots.  When comparing the quality of the normal approximation between the four plots, the results of the simulations are intuitive. Namely, increasing the number of measurements $(n_1+n_2)$, and increasing the number of moments of the noise distribution both lead to improvement of the normal approximation (especially in the left tail and center of the standardized statistic's distribution).

\paragraph{Parameter settings.}  The parameter settings used in the simulations for Figure~\ref{fig:hists} are largely the same as those used for Figure~\ref{fig:sweeps}. The signal $x\in\R^p$ was constructed as $x_i=c\cdot i^{-1}$ with $c$ chosen so that $\|x\|_2=1$ in all cases. Also, in all four cases, we set $p=10000$, $\gamma_1=\gamma_2=1$, and $\sigma=0.1$. The parameter settings used for $\eta_0=\ve_2=0.3$ were also the same as in the case of Figure~\ref{fig:sweeps}.

\begin{figure*}[!]
\centering
{\includegraphics[angle=0,
  width=.45\linewidth]{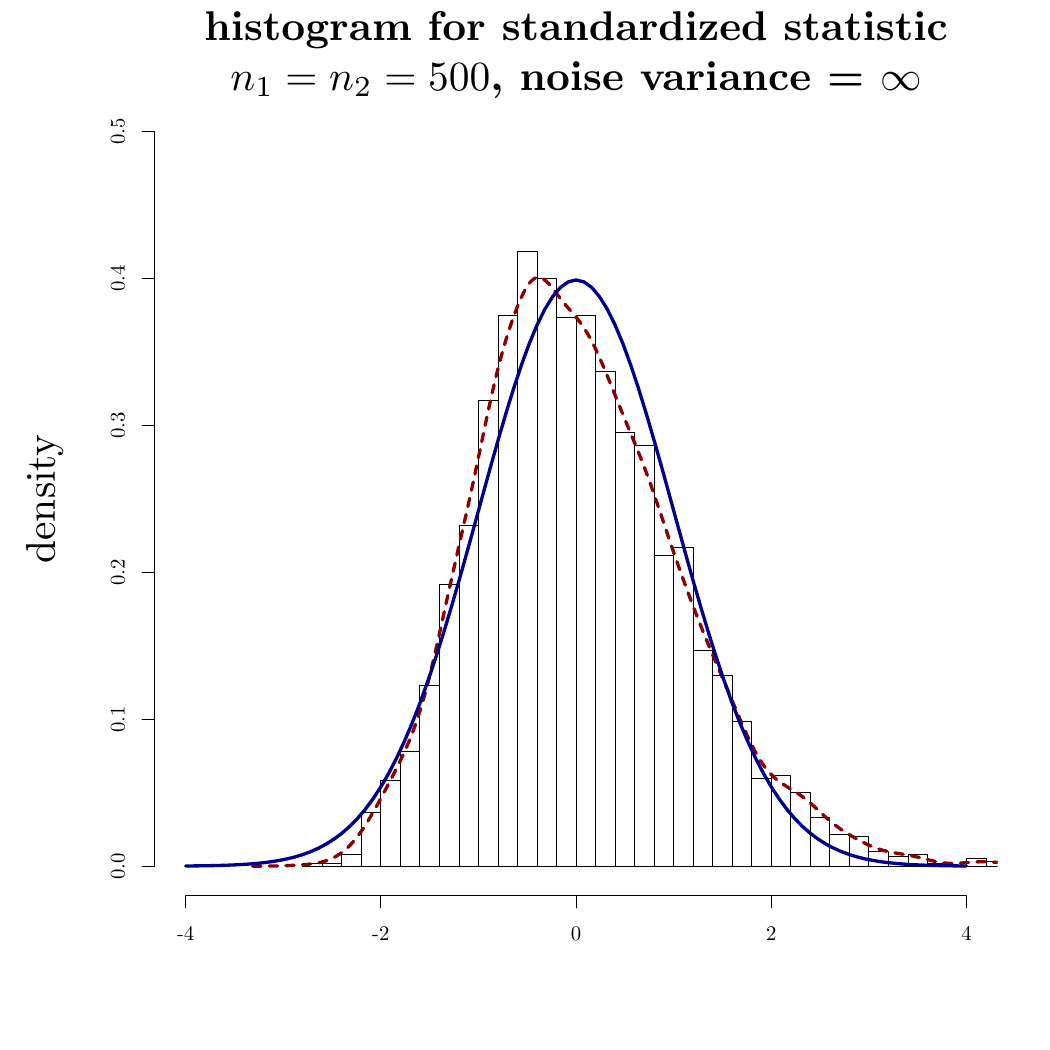}}
{\includegraphics[angle=0,
  width=.45\linewidth]{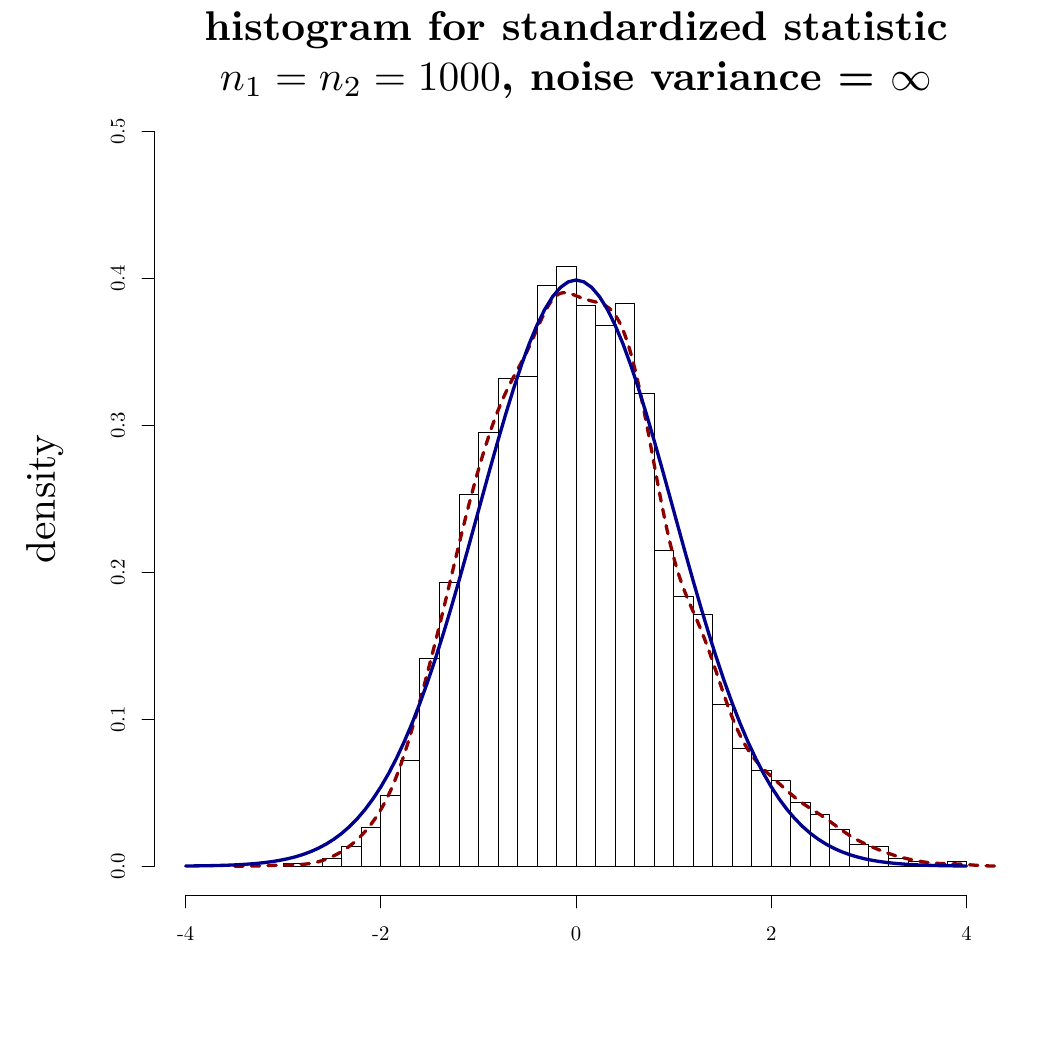}}
  ~\\[0.2cm]
{\includegraphics[angle=0,
  width=.45\linewidth]{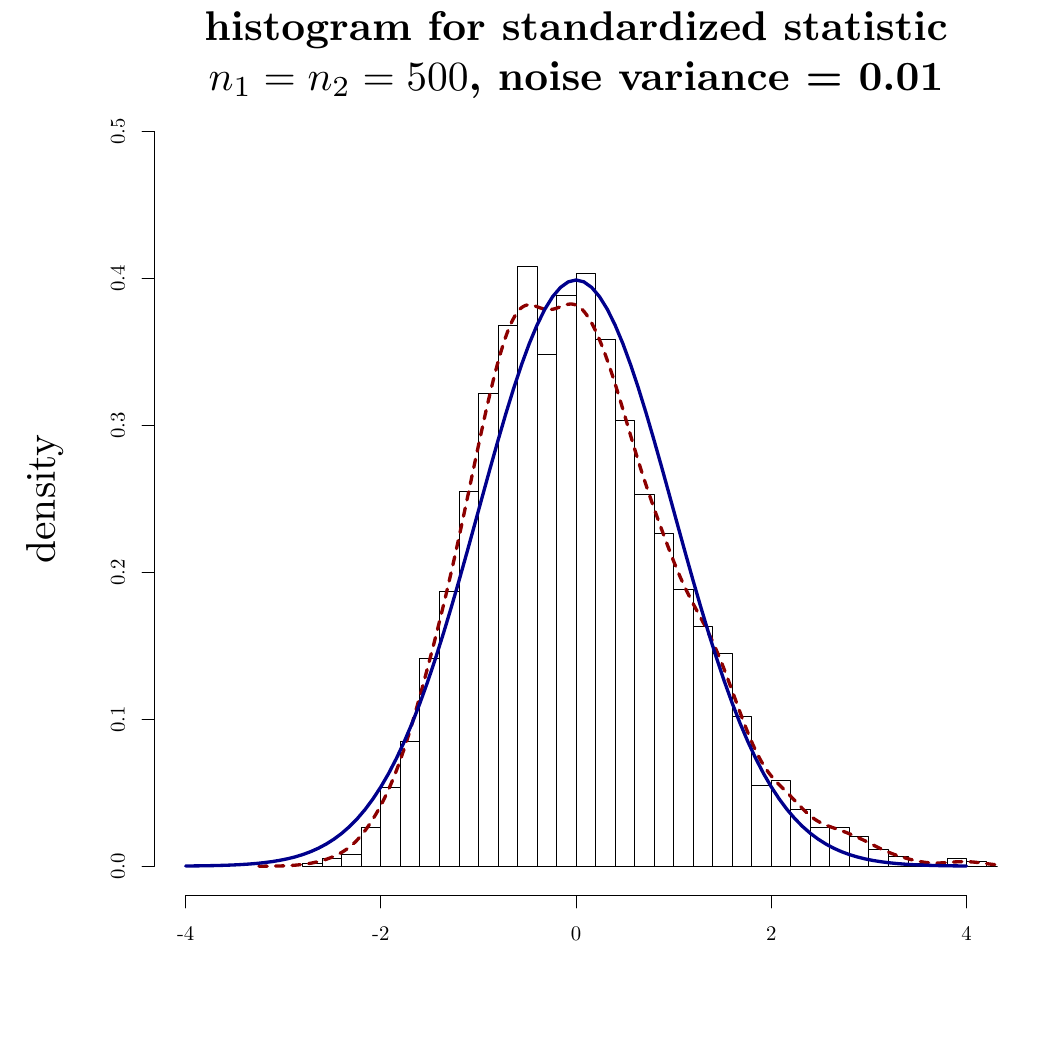}}
  {\includegraphics[angle=0,
  width=.45\linewidth]{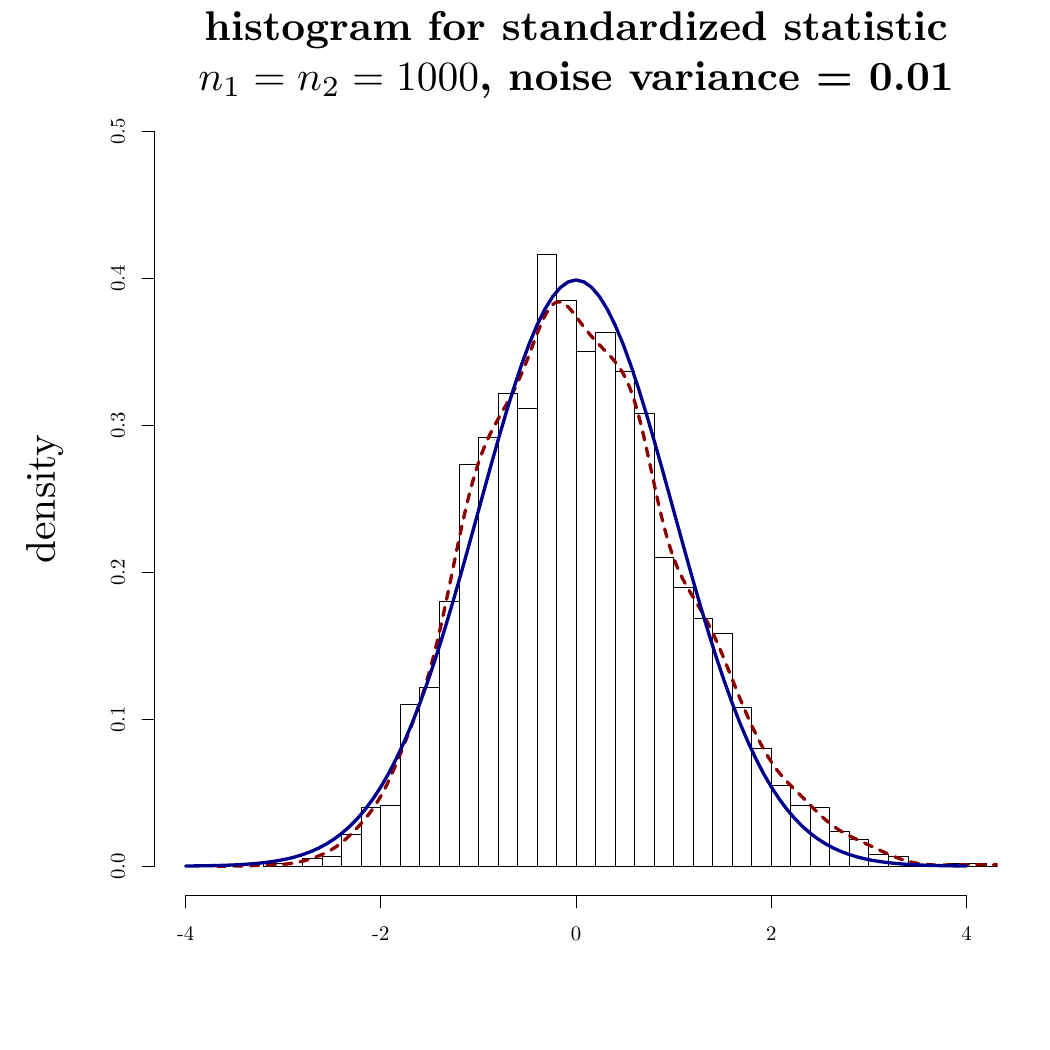}}
\caption{The solid blue curve is the standard normal density in all four plots. The histograms were generated from 3000 realizations of the standardized statistic $\frac{\sqrt{n_1+n_2}}{\sqrt{\hat{\vartheta}_q}}\big(\frac{\hat{s}_2(x)}{s_2(x)}-1\big)$ under four conditions --- depending on the number of measurements $(n_1+n_2)$ and the tails of the noise distribution. The dashed red curve is the default {\tt{R}} kernel density estimate obtained from the same sets of samples. Notably, the normal approximation gives sensible results in the top two plots, even when the noise is $t$-distributed (centered) with 2 degrees of freedom (which has infinite variance). As would be expected, increasing the number of measurements and giving the noise distribution a finite variance lead to some improvement.}
\label{fig:hists}
\end{figure*}

 \section{Appendix}
\appendix
\section{Proofs for Section~\ref{sec:recov}}

\subsection{Proposition~\ref{propUpper}}
\noindent \emph{Proof.} Define the positive number $t:=\frac{n}{\log(\frac{pe}{n})}$, and choose $T=\lceil t \rceil$ in Theorem~\ref{thmcandes}.  We first verify that $T\log(\ts\frac{pe}{T})\leq 3n$ for every $n$. Observe that
\begin{equation}
\begin{split}
T\log(\ts\frac{pe}{T}) &\leq (t+1)\log(\ts\frac{pe}{t})\\[0.3cm]
&=\ts  \frac{n}{\log(\frac{pe}{n})}\log \Big(\ts\frac{pe}{n}\cdot\log(\ts\frac{pe}{n})\Big)+\log \Big(\ts\frac{pe}{n}\cdot\log(\ts\frac{pe}{n})\Big).
\end{split}
\end{equation}
Let $r=p/n$ and recall that we assume $n\leq p$. Simple calculus shows that for all $r\geq 1$, the quantity $\log\big(re\cdot \log(re)\big)$ is at most $1.4 \log(re),$
and so
\begin{equation}
\begin{split}
T\log(\ts\frac{pe}{T})&\leq  \ts\frac{n}{\log(\frac{pe}{n})} \cdot 1.4\log(\frac{pe}{n}) + 1.4\log(\ts\frac{pe}{n})\\
&\leq 1.4 n +1.4n\\
&\leq 3n
\end{split}
\end{equation}
with the second step following from the assumption $\log(\ts\frac{pe}{n})\leq n$. Consequently,  the condition~\eqref{nbound} of Theorem~\ref{thmcandes} is satisfied for every $n$ under this choice of $T$, and we conclude that there is an absolute constant $c_1>0$ such that the bound~\eqref{candesTao} holds with probability at least $1-2\exp(-c_1n)$. To finish the argument, observe that
\begin{equation}
\begin{split}
\ts\frac{1}{\sqrt{T}} \|x-x_{|T}\|_1 &\leq \ts\frac{1}{\sqrt{t}} \|x\|_1 = \ts\frac{1}{\sqrt{n}}\sqrt{\|x\|_1^2 \log(\ts\frac{pe}{n})}.
\end{split}
\end{equation}
Dividing the inequality~\eqref{candesTao} through by $\|x\|_2$ leads to

$$\ts\frac{\|x-\hat{x}\|_2}{\|x\|_2} \leq c_2\ts\frac{\sigma \e_0}{\|x\|_2}+\ts\frac{c_3}{\sqrt{n}}\sqrt{\ts\frac{\|x\|_1^2}{\|x\|_2^2} \log(\ts\frac{pe}{n})},
$$
and the proof is complete.\qed

\subsection{Proposition~\ref{propLower}}
\noindent \emph{Proof.} Let $\B_1(1)\subset\R^p$ be the $\ell_1$ ball of radius 1, and let $\mathcal{R}:\R^n\to\R^p$ be a homogeneous recovery algorithm. Also define the number $\eta := \ts\frac{1}{2}c_1\sqrt{\ts\frac{ \log(pe/n)}{n}}$ where $c_1>0$ is an absolute constant to be defined below. Clearly, for any such $\eta$, we can find a point $\tilde{x}\in \B_1(1)$ satisfying
\begin{equation}\label{sup}
\begin{split}
\|\tilde{x}-\mathcal{R}(A\tilde{x})\|_2\geq \sup_{x\in \B_1(1)} \ts\|x-\mathcal{R}(Ax)\|_2 -\eta.
\end{split}
\end{equation}
Furthermore, we may choose such a point $\tilde{x}$ to satisfy $\|\tilde{x}\|_1=1$. (Note that if $\|\tilde{x}\|_1<1$, then we can use the homogeneity of $\mathcal{R}$ to replace $\tilde{x}$ with the $\ell_1$ unit vector $\tilde{x}/\|\tilde{x}\|_1\in \B_1(1)$ and obtain an even larger value on the left side.) The next step of the argument is to further  lower bound the right side in terms of minimax error, leading to
\begin{equation}
\|\tilde{x}-\mathcal{R}(A\tilde{x})\|_2 \geq \inf_{A\in\R^{n\times p}} \  \inf_{R:\,\R^n\to \R^p} \ \sup_{x\in \B_1(1)} \|x-R(Ax)\|_2 - \eta,
\end{equation}
where the infima are over all sensing matrices $A$ and all recovery algorithms $R$ (possibly non-homogenous). It is known from the theory of Gelfand widths 
that there is an absolute constant $c_1>0$, such that the minimax $\ell_2$ error over $\B_1(1)$ is lower-bounded by

\begin{equation}
 \inf_{A\in\R^{n\times p}} \ \inf_{R:\,\R^n\to \R^p} \ \sup_{x\in \B_1(1)} \|x-R(Ax)\|_2 \geq c_1\sqrt{\ts\frac{\log(pe/n)}{n}}.
 \end{equation}
(See~\cite[Section 3.5]{candesMadrid}, as well as~\cite{kashin}~\cite{garnaev}.) Using our choice of $\eta$, as well as the fact that $\|\tilde{x}\|_1=1$, we obtain
\begin{equation}
\|\tilde{x}-\mathcal{R}(A\tilde{x})\|_2 \geq \ts\frac{1}{2}c_1\sqrt{\ts\frac{\|\tilde{x}\|_1^2\cdot \log(pe/n)}{n}}.
\end{equation}
Dividing both sides by $\|\tilde{x}\|_2$ completes the proof. \qed

\section{Proofs for Section~\ref{sec:proc}}\label{app:proc}

\subsection{Theorem~\ref{clt} -- Uniform CLT for $\ell_q$ norm estimator}

The proof of Theorem~\ref{clt} consists of two parts. First, we prove a uniform CLT for a re-scaled version of $\hat{\Psi}_n(t)$, which is given below in Lemma~\ref{uniformclt}. Second, we extend this limit  to the statistic $\hat{\nu}_q(t)$ by way of the functional delta method, which is described at the end of the section.

\paragraph{Remark on the subscript of $n_q$.} For ease of notation, we will generally drop the subscript from $n_q$ in the remainder of the appendix, as it will not cause confusion.

\paragraph{Weak convergence of the empirical characteristic function.} To introduce a few pieces of notation, let $c\in [-b,b]$ for some fixed $b>0$, and define the re-scaled empirical characteristic function, 
\begin{equation}
\hat{\psi}_{n}(c):= \frac{1}{n}\sum_{i=1}^n e^{\sqrt{-1}\frac{c y_i}{\gamma_q \|x\|_q}},
\end{equation}
which is obtained from the re-scaled observations $\frac{y_i}{\gamma\|x\|_q}\overset{d}{=} S_i+\rho_q\e_i$, where $S_i\sim \stable_q(1)$, and $\e_i\sim F_0$.
The relation between $\hat{\Psi}_n$ and $\hat{\psi}_n$ is given by
\begin{equation}
\hat{\Psi}_n(t) = \hat{\psi}_n(\gamma_q t\|x\|_q).
\end{equation}
The re-scaled population characteristic function is
\begin{equation}
\psi_n(c):=\exp(-|c|^q)\varphi_0(\rho_q c),
\end{equation}
which converges to the function
\begin{equation}
\psi(c):=\exp(-|c|^q)\varphi_0(\bar{\rho}_q c),
\end{equation}
as $\rho_q\to \bar{\rho}_q$.
Lastly, define the normalized process
\begin{equation}
\chi_{n}(c) :=\sqrt{n}\Big(\hat{\psi}_n(c)-\psi(c)\Big),
\end{equation}
and let $\mathscr{C}([-b,b];\C)$ be the space of continuous complex-valued functions on $[-b,b]$ equipped with the sup-norm.
\begin{lemma}\label{uniformclt}
Fix any $b>0$. Under the assumptions of Theorem~\ref{clt}, the random function $\chi_n$ satisfies the limit
\begin{equation}
\chi_n(c) \overset{w}{\longrightarrow} \chi_{\infty}(c) \ \ \  \text{ in } \ \ \  \mathscr{C}([-b,b];\C),
\end{equation}
where $\chi_{\infty}$ is a centered Gaussian process whose marginals satisfy $\text{\emph{Re}} (\chi_{\infty}(c))\sim N(0,\omega(c,\bar{\rho}_q)),$ and
\begin{equation}
\omega(c,\bar{\rho}_q)=\ts\frac{1}{2}+ \ts\frac{1}{2}\exp(-2^q|c|^q)\varphi_0(2\bar{\rho}_q c)- \exp(-2|c|^q)\varphi_0^2(\bar{\rho}_q c).
\end{equation}
\end{lemma}
\noindent \emph{Proof.} It is important to notice that $\hat{\psi}_n(c)$ is not the empirical characteristic function associated with $n$ samples from the distribution of $\psi$ (since $\rho_q\neq \bar{\rho}_q$). The more natural process to work with is
\begin{equation}\label{firstclt}
\tilde{\chi}_{n}(c) :=\sqrt{n}\big(\tilde{\psi}_n(c)-\psi(c)\big),
\end{equation}
where $\tilde{\psi}_n(c)=\frac{1}{n}\sum_{i=1}^n \exp(\sqrt{-1}c y_i^{\circ})$ and $y_i^{\circ}=S_i+\bar{\rho}_q\e_i$. (In other words, $\tilde{\psi}_n$ is the empirical characteristic function associated with $\psi$.) As a first step in the proof, we show that the difference between $\chi_n$ and $\tilde{\chi}_n$ is negligible in a uniform sense, i.e.
\begin{equation}\label{approxerror}
\sup_{c\in [-b,b]} | \chi_n(c) -\tilde{\chi}_n(c)| = o(1) \as 
\end{equation}
To see this, observe that for $c\in [-b,b]$,
\begin{align}
|\chi_n(c) -\tilde{\chi}_n(c)| &= \sqrt{n}|\hat{\psi}_n(c)-\tilde{\psi}_n(c)|\\
&=\ts\frac{1}{\sqrt{n}}\Big|\tsum_{i=1}^n e^{\sqrt{-1}c(S_i+\rho_q\e_i)}- e^{\sqrt{-1}c(S_i+\bar{\rho}_q\e_i)}\Big|\\[0.2cm]
&=\ts\frac{1}{\sqrt{n}}\Big|\tsum_{i=1}^n e^{\sqrt{-1}c(S_i+\rho_q\e_i)}\big(1-e^{\sqrt{-1}c(\bar{\rho}_q-\rho_q)\e_i}\big)\Big|\\[0.2cm]
&\leq \ts\frac{1}{\sqrt{n}}\tsum_{i=1}^n \big|1-e^{\sqrt{-1}c(\bar{\rho}_q-\rho_q)\e_i}\big|\\[0.2cm]
&\leq \ts\frac{1}{\sqrt{n}} \tsum_{i=1}^n |c(\bar{\rho}_n-\rho_q)\e_i|\\[0.2cm]
&\leq  \sqrt{n}|\rho_q-\bar{\rho}_q| \cdot \ts\frac{b}{n}\tsum_{i=1}^n |\e_i|,
\end{align}
where the last bound does not depend on $c$, and tends to 0 almost surely. Here we are using the assumption that $\E|\e_1|<\infty$ and assumption~\textbf{A\ref{rhoassumption}} that $\rho_q=\bar{\rho}_q +o(1/\sqrt{n})$.
Now that line~\eqref{approxerror} has been verified, it remains (by the functional version of Slutsky's Lemma~\cite[p.32]{vaartWellner}) to prove 
\begin{equation}\label{tildelimit}
\tilde{\chi}_n \xrightarrow{ \ w \ } \chi_{\infty} \ \ \  \text{ in } \ \ \  \mathscr{C}([-b,b];\C),
\end{equation}
and that the limiting process $\chi_{\infty}$ has the stated variance formula. We first show that this limit holds, and then derive the variance formula at the end of the proof. (Note that it is clear that the limiting process must be Gaussian due to the finite-dimensional CLT.)

By a result of Marcus~\cite[Theorem 1]{marcus}, it is known that the uniform CLT for empirical characteristic functions~\eqref{tildelimit} holds as long as the limiting process $\chi_{\infty}$ has continuous sample paths (almost surely).\footnote{The paper~\cite{marcus} only states the result when $b=1/2$, but it holds for any $b>0$. See the paper~\cite[Theorem 3.1]{csorgo1981multi}.} To show that the sample paths of $\chi_{\infty}$ are continuous, we employ as sufficient condition derived by Cs\"orgo~\cite{csorgo1981}. Let $F_q$ denote the distribution function of the random variable $y_i^{\circ}=S_i+\bar{\rho}_q\e_i$ described earlier. Also let $\delta>0$ and define the function 
\begin{equation}
g^+_{\delta}(u):=
\begin{cases}
&\log(|u|)\cdot \log(\log(|u|))^{2+\delta} \text{ \ \ \ \ if \ } |u|\geq \exp(1)\\
&0 \text{\ \ \ \ \ \ \ \ \ \ \ \ \ \ \ \ \ \  \ \ \ \ \ \ \ \ \ \ \ \ \ \  \ \ \ if \ }|u|< \exp(1).
\end{cases}
\end{equation}
At line 1.17 of the paper~\cite{csorgo1981}, it is argued that if there is some $\delta>0$ such that
\begin{equation}
\int_{-\infty}^{\infty} g^+_{\delta}(|u|)dF_q(u)<\infty,
\end{equation}
then $\chi_{\infty}$ has continuous sample paths. Next, note that for any $\delta,\delta'>0$, we have
\begin{equation}
g^+_{\delta}(|u|)=\mathcal{O}\Big(|u|^{\delta'}\Big) \ \ \ \text{ as } |u|\to\infty.
\end{equation}
Hence, $\chi_{\infty}$ has continuous sample paths as long as $F_q$ has a fractional moment. To see that this is true, recall the basic fact that if $S_i\sim \text{stable}_q(1)$ then $\E[|S_i|^{\delta'}]<\infty$ for any $\delta'\in(0,q)$. Also, our deconvolution model assumes $\E[|\e_i|]<\infty$, and so it follows that for any $q\in(0,2]$, the distribution $F_q$ has a fractional moment, which proves the limit~\eqref{tildelimit}.\footnote{Here we use the inequality $\E[|U+V|^{a\wedge 1}]\leq \E[|U|^{a\wedge 1}]+ \E[|V|^{a\wedge 1}]$ for generic random variables $U$ and $V$, and any $a>0$.}

Finally, we compute the variance of the marginal distributions $\myRe(\chi_{\infty}(c))$. By the ordinary central limit theorem, we only need to calculate the variance of $\myRe( \exp(\sqrt{-1} y_1^{\circ})$.  The first moment is given by
$$\E[\myRe( \exp(\sqrt{-1} c y_1^{\circ})]=\myRe \,\E[\exp(\sqrt{-1}c y_1^{\circ})] = \exp(-|c|^q)\varphi_0(\bar{\rho}_q c).$$
The second moment is given by
\begin{align}
\E[(\myRe( \exp(\sqrt{-1}c y_1^{\circ}))^2]&=\E[\cos^2(c y_1^{\circ})] \\
&= \E[\ts\frac{1}{2}+\ts\frac{1}{2}\cos(2c y_1^{\circ})]\\
&=\ts\frac{1}{2}+\ts\frac{1}{2}\myRe \, \E[ \exp(\sqrt{-1}\cdot 2 c y_1^{\circ})]\\
&=\ts\frac{1}{2} +\ts\frac{1}{2}\exp(-2^q|c|^q)\varphi_0(2\bar{\rho}_q c).
\end{align}
This completes the proof of Lemma~\ref{uniformclt}.

\qed

\paragraph{Applying the functional delta method.} 
We now complete the proof of Theorem~\ref{clt} by applying the functional delta method to Lemma~\ref{uniformclt} (with a suitable map $\phi$ to be defined below).
In the following, $\mathscr{C}(\mathcal{I})$ denotes the space of continuous real-valued functions on an interval $\mathcal{I}$, equipped with the sup-norm.\\

\noindent \emph{Proof of Theorem~\ref{clt}}. Since $\varphi_0(\bar{\rho}_q c_0)\neq 0$, and the roots of $\varphi_0$ are assumed to be isolated, there is some $\delta_0\in (0,c_0)$ such that over the interval $c\in \mathcal{I}:=[c_0-\delta_0,c_0+\delta_0]$, the value $\varphi_0(\bar{\rho}_q c)$ is bounded away from 0. Define the function $f_0\in \mathscr{C}(\mathcal{I})$ by 
$$f_0(c)=\exp(-|c|^q),$$
 and let $\mathscr{N}(f_0;\ve)\subset \mathscr{C}(\mathcal{I}) $ be a fixed $\ve$-neighborhood of $f_0$ in the sup-norm, such that all functions in the neighborhood are bounded away from 0. Consider the map $\phi:\mathscr{C}(\mathcal{I})\to \mathscr{C}(\mathcal{I})$
 defined according to
\begin{equation}
\phi(f)(c) =
\begin{cases}
& -\ts\frac{1}{|c|^q} \log (f(c)) \ \ \ \ \text{ if } f \in \mathscr{N}(f_0; \ve)\\[0.2cm]
& \ \ \ 1_{\mathcal{I}}(c)  \ \ \ \ \ \ \ \ \ \ \ \ \ \text{ if } f \not\in \mathscr{N}(f_0; \ve),
\end{cases}
\end{equation}
where $1_{\mathcal{I}}$ is the indicator function of the interval $\mathcal{I}$.
The importance of $\phi$ is that it can be related to $\hat{\nu}_q(\hat{t})/\|x\|_q^q$ in the following way. First, let $\hat{c}=\hat{t}\gamma_q\|x\|_q$ and observe that the definition of $\hat{\psi}_n$ gives
\begin{equation}
 \sqrt{n}\big(\ts\frac{\hat{\nu}_q(\hat{t})}{\|x\|_q^q}-1\big)=
 \sqrt{n}\Big(-\ts\frac{1}{|\hat{c}|^q}\Log\big(\myRe\big(\ts\frac{\hat{\psi}_n(\hat{c})}{\varphi_0(\rho_q \hat{c})}\big)\big)+\frac{1}{|\hat{c}|^q}\Log\big(\exp(-|\hat{c}|^q)\big)\Big).
\end{equation}
Next, let $\Pi(\hat{c})$ be the point in the interval $\mathcal{I}$ that is nearest to $\hat{c}$,\footnote{The purpose of introducing $\Pi(\hat{c})$ is that it always lies in the interval $\mathcal{I}$, and hence allows us to work entirely on $\mathcal{I}$.} and define the quantities $\Delta_n$ and $\Delta_n'$ according to
\begin{align}\label{simplify}
\small
\sqrt{n}\big(\ts\frac{\hat{\nu}_q(\hat{t})}{\|x\|_q^q}-1\big)
&= \sqrt{n}\Big(-\ts\frac{1}{|\Pi(\hat{c})|^q}\Log\big(\myRe\big(\ts\frac{\hat{\psi}_n(\Pi(\hat{c}))}{\varphi_0(\rho_q\Pi(\hat{c}))}\big)\big)+\frac{1}{|\Pi(\hat{c})|^q}\Log\big(\exp(-|\Pi(\hat{c})|^q)\big)\Big)+\Delta_n\\[0.3cm]
&= \sqrt{n}\Big(\phi\big(\myRe\big(\ts\frac{\hat{\psi}_n(\cdot)}{\varphi_0(\rho_q\cdot)}\big)\big)(\Pi(\hat{c}))-\phi\big(f_0\big)(\Pi(\hat{c}))\Big) +\Delta_n'+\Delta_n.\label{lastdelta}
\end{align}
\normalsize
In words, $\Delta_n$ is the difference that comes from replacing $\hat{c}$ with $\Pi(\hat{c})$, and $\Delta'_n$ is the difference that comes from replacing terms of the form $-\frac{1}{|\cdot|}\Log(\dots)$ with terms involving $\phi$.

We now argue that both of the terms $\Delta_n$ and $\Delta_n'$ are asymptotically negligible. As a result, we will complete the proof of Theorem~\ref{clt} by showing that the first term in line~\eqref{lastdelta} has the desired Gaussian limit --- which is the purpose of the functional delta method.

To see that $\Delta_n\to_P0$, first recall that $\hat{c}\to_P c_0\in\mathcal{I}$ by assumption. Consequently, along any subsequence, there is a further subsequence on which $\hat{c}$ and $\Pi(\hat{c})$ eventually agree with probability 1. In turn, if $g_n$ is a generic sequence of functions, then eventually $g_n(\hat{c})-g_n(\Pi(\hat{c}))=0$ along subsequences (with probability 1). Said differently, this means $g_n(\hat{c})-g_n(\Pi(\hat{c}))\to_P 0$, and this implies $\Delta_n\to_P0$ since $\Delta_n$ can be expressed in the form  $g_n(\hat{c})-g_n(\Pi(\hat{c}))$.

Next, to see that $\Delta_n'\to_P 0$, notice that as soon as $\hat{\psi}_n(\cdot)/\varphi_0(\rho_q\cdot)$ lies in the neighborhood $\mathscr{N}(f_0;\ve)$, it follows from the definition of $\phi$ that $\Delta_n'=0$. Also, this is guaranteed to happen with probability 1 for large enough $n$, because the function $\hat{\psi}_n(\cdot)/\varphi_0(\rho_q\cdot)$ converges uniformly to $f_0$ on the interval $\mathcal{I}$ with probability 1.\footnote{The fact that $\hat{\psi}_n(\cdot)$ converges uniformly to $\psi(\cdot)$ on $\mathcal{I}$ with probability 1 follows essentially from the Glivenko-Cantelli Theorem~\cite[Equation 1.2]{csorgo1981}. To see that $1/\varphi_0(\rho_q\,\cdot)$ converges uniformly to $1/\varphi_0(\bar{\rho}_q\,\cdot)$ on $\mathcal{I}$ as $\rho_q\to\bar{\rho}_q$, note that since $\E|\e_1|<\infty$, the characteristic function $\varphi_0$ is Lipschitz on any compact interval.} Hence, $\Delta_n'\to 0$ almost surely, but we only need $\Delta_n'=o_P(1)$ in the remainder of the proof.

We have now taken care of most of the preparations needed to apply the functional delta method. Multiplying the limit in Lemma~\ref{uniformclt} through by $1/\varphi_0(\rho_q\cdot)$ and using the functional version of Slutsky's Lemma~\cite[p.32]{vaartWellner}, it follows that
 \begin{equation}\label{tempproc}
\sqrt{n}\big(\myRe\big(\ts\frac{\hat{\psi}_n(\cdot)}{\varphi_0(\rho_q \cdot)}\big)-\exp(-|\cdot|^q)\big) \xrightarrow{ \ w \ } \tilde{z}(\cdot) \ \ \text{ \ in \ \ } \mathscr{C}(\mathcal{I}),
\end{equation}
 where $\tilde{z}(\cdot)$ is a centered Gaussian process with continuous sample paths, and the marginals $\tilde{z}(c)$ have variance equal to
\begin{equation}\label{unnamedvar}
\ts\frac{1}{2}\ts\frac{1}{\varphi_0^2(\bar{\rho}_q c)}+ \ts\frac{1}{2}\exp(-2^q|c|^q)\ts\frac{\varphi_0(2\bar{\rho}_q c)}{\varphi_0^2(\bar{\rho}_q c)}- \exp(-2|c|^q).
\end{equation}
It is straightforward to verify that $\phi$ is Hadamard differentiable at $f_0$ and the Hadamard derivative $\phi'_{f_0}$ is the linear map that multiplies by $-\frac{\exp{|\cdot|^q}}{|\cdot|^q}$. (See Lemmas 3.9.3 and 3.9.25  in~\cite{vaartWellner}.)
Consequently, the functional delta method~\cite[Theorem 3.9.4]{vaartWellner} applied to line~\eqref{tempproc} with the map $\phi$ gives
\begin{align}
  \sqrt{n}\big(\phi\big(\myRe\big(\ts\frac{\hat{\psi}_n(\cdot)}{\varphi_0(\rho_q \cdot)}\big)\big)-\phi\big(\exp(-|\cdot|^q)\big)\big)
&\ \xrightarrow{ \ \  w \ \ } \  \phi'_{f_0}(\tilde{z})(\cdot) \ \ \text{ \ in } \ \ \mathscr{C}(\mathcal{I})\label{linftylim}\\[0.2cm]
& \ \ \ \ \ = \  -\ts\frac{\exp(|\cdot|^q)}{|\cdot|^q} \tilde{z}(\cdot) \\[0.2cm] 
& \ \ \ \ \ =: \ \  z(\cdot). \label{hderiv}
\end{align}
It is clear that $z(\cdot)$, defined in the previous line, is a centered Gaussian process on $\mathcal{I}$, since $\tilde{z}(\cdot)$ is. Combining lines~\eqref{unnamedvar} and~\eqref{hderiv} shows that the marginals $z(c)$ are given by 
$$z(c)\sim N(0,v(c,\bar{\rho}_q)),$$
where
\begin{equation}
v_q(c,\bar{\rho}_q)=\ts\frac{1}{|c|^{2q}}\Big(\ts\frac{1}{2}\frac{1}{\varphi_0(\bar{\rho}_q c)^2}\exp(2|c|^q) +\ts\frac{1}{2}\frac{\varphi_0(2\bar{\rho}_q c)}{\varphi_0(\bar{\rho}_q c)^2}\exp((2-2^q) |c|^q)- 1\Big).
\end{equation}

The final step of the proof essentially involves plugging $\Pi(\hat{c})$ into the limit~\eqref{linftylim} and using Slutsky's Lemma. Since $\Pi(\hat{c})\to_P c_0$, and $z(\cdot)$ takes values in the separable space $\mathscr{C}(\mathcal{I})$ almost surely,
the functional version of Slutsky's Lemma~\cite[p.32]{vaartWellner} gives the following convergence of pairs
\begin{equation}
\Big(  \sqrt{n}\big(\phi\big(\myRe\big(\ts\frac{\hat{\psi}_n(\cdot)}{\varphi_0(\rho_q \cdot)}\big)\big)-\phi\big(f_0\big)\big),\ \Pi(\hat{c}) \ \Big)
\ \xrightarrow{ \ \  w \ \ } \  \big(z(\cdot),c_0\big) \ \ \text{ \ in } \ \ \mathscr{C}(\mathcal{I})\times\mathcal{I}.\label{newlinftylim}
\end{equation}
To finish, note that the evaluation map $\mathscr{C}(\mathcal{I})\times\mathcal{I} \to \R$ defined by $(f,c)\mapsto f(c)$, is continuous. The continuous mapping theorem~\cite[Theorem 1.3.6]{vaartWellner} then gives

\begin{equation}
\sqrt{n}\Big(\phi\big(\myRe\big(\ts\frac{\hat{\psi}_n(\cdot)}{\varphi_0(\rho_q\cdot)}\big)\big)(\Pi(\hat{c}))-\phi\big(f_0\big)(\Pi(\hat{c}))\Big) 
\ \xrightarrow{ \ \  w \ \ } \  z(c_0),
\end{equation}
which is the desired conclusion.
\qed

\subsection{Lemma~\ref{varextend} -- Extending the variance function}\label{app:varextend}

\proof It is simple to verify that $v_q(\cdot,\cdot)$ is continuous at any pair $(c_0,\rho_0)$ for which $c_0\neq 0$ and $\varphi_0(\rho_0 c_0) \neq 0$, and hence $\tilde{v}_q$ inherits continuity at those pairs. To show that $\tilde{v}_q$ is continuous elsewhere, it is necessary to handle two cases.
\begin{itemize}
\item[--] First, we show that for any $q\in (0,2]$, if $(c_0,\rho_0)$ is a pair such that $\varphi_0(\rho_0 c_0)=0$ and $c_0\neq 0$, then $v_q(c_j,\rho_j)\to\infty$ for any sequence satisfying $(c_j,\rho_j)\to (c_0,\rho_0)$ with $\varphi_0(\rho_j c_j)\neq 0$ and $c_j\neq 0$ for all $j$. (Note that  $v_q(\cdot,\cdot)$ is defined in a deleted neighborhood of $(c_0,\rho_0)$ due to the assumption that the roots of $\varphi_0$ are isolated.)
\item[--]  Second, we show that for any $q\in (0,2)$, if $c_0=0$, then $v_q(c_j,\rho_j)\to\infty$ for any sequence $(c_j,\rho_j)\to (0,\rho_0)$, where $\rho_0\geq 0$ is arbitrary, and $c_j>0$ for $j>0$.
\end{itemize}
 To handle the first case where $c_0\neq 0$, we derive a lower bound on $v_q(c,\rho)$ for all $q\in (0,2]$ and all pairs where $v_q$ is defined. Recall the formula
\begin{equation}\label{varformulaagain}
	v_q(c,\rho)= \ts\frac{1}{|c|^{2q}}\Big(\ts\frac{1}{2}\frac{1}{\varphi_0(\rho c)^2}\exp(2|c|^q) +\ts\frac{1}{2}\frac{\varphi_0(2\rho |c|)}{\varphi_0(\rho |c|)^2}\exp((2-2^q) |c|^q)- 1\Big).
	\end{equation}
 The lower bound is obtained by manipulating the factor $\ts\frac{1}{2}\frac{\varphi_0(2\rho c)}{\varphi_0(\rho c)^2}$. Consider the following instance of Jensen's inequality, followed by a trigonometric identity,
\begin{align}
\varphi_0(\rho c)^2 &= (\E[\cos(\rho c \e_1)])^2\\ 
&\leq \E[\cos^2(\rho \e_1)]\\
&=\ts\frac{1}{2} +\ts\frac{1}{2}\E[\cos(2\rho c \e_1)]\\
&=\ts\frac{1}{2}+\ts\frac{1}{2}\varphi_0(2\rho c),
\end{align}
which gives $\frac{1}{2}\frac{\varphi_0(2\rho c)}{\varphi_0(\rho)^2} \geq 1-\ts\frac{1}{2}\ts\frac{1}{\varphi_0(\rho c)^2}$. Letting $\kappa_q:=2-2^q$ we have the lower bound
\begin{equation}\label{varlowerbound}
v_q(c,\rho)\geq \ts\frac{1}{|c|^{2q}}\Big(\ts\frac{1}{2}\ts\frac{1}{\varphi_0(\rho |c|)^2}\Big(\exp(2|c|^q)-\exp(\kappa_q |c|^q)\Big) +\exp(\kappa_q |c|^q)- 1\Big),
\end{equation}
which holds for all $(c,\rho)$ where $v_q(c,\rho)$ is defined. Since the quantity 
$$\big(\exp(2c_0^q)-\exp(\kappa_q c_0^q)\big)$$ 
is positive for all $q\in (0,2]$ and $c_0\neq 0$, it follows that $v_q(c_j,\rho_j)\to\infty$ as $\varphi_0(\rho_jc_j)\to \varphi_0(\rho_0c_0)= 0$. 

Next, we consider the second case where $(c_j,\rho_j)\to (c_0,\rho_0)$ with $c_0=0$ and $\rho_0\geq 0$. Due to the fact that all characteristic functions satisfy $|\varphi_0|\leq 1$, and the fact that $\big(\exp(2|c|^q)-\exp(\kappa_q |c|^q)\big)$ is positive when $c\neq 0$, the previous lower bound gives
\begin{equation}\label{secondlower}
v_q(c,\rho)\geq \ts\frac{1}{|c|^{2q}}\Big(\ts\frac{1}{2}\exp(2|c|^q) +\ts\frac{1}{2}\exp(\kappa_q |c|^q)- 1\Big),
\end{equation}
which again holds for all $(c,\rho)$ where $v_q$ is defined. Note that this bound does not depend on $\rho$. A simple calculation involving L'Hospital's rule shows that whenever $q\in (0,2)$, the lower bound tends to $\infty$ as $c\to 0+$.

To finish the proof, we must show that for any $\rho\geq 0$, the function $v_q(\cdot,\rho)$ attains its minimum on the set $[\ve_q,\infty)$. This is simple because the lower bound~\eqref{secondlower} tends to $\infty$ as $c\to\infty$. \qed

\section{Proofs for Section~\ref{results}}\label{resultsproofs}

\subsection{Proposition~\ref{tpilot} -- consistency of pilot estimator.}

\proof We first show that there is a positive constant $c_1$ such that  $\hat{t}_{\text{initial}}\gamma_q\|x\|_q\to_P c_1$. Note that for each $i=1,\dots,n$, we have $\frac{y_i}{\gamma_q\|x\|_q} \sim S_i+ \frac{\sigma}{\gamma_q\|x\|_q} \e_i$, where the $S_i$ are i.i.d. samples from $\stable_q(1)$. Let $F_n$ denote the distribution function of the random variable  $|S_1+ \frac{\sigma}{\gamma_q\|x\|_q} \e_1|$ and let $\F_n$ be the empirical distribution function obtained from $n$ samples from $F_n$. Then,
\begin{align}
\ts\frac{\hat{m}_q}{\gamma_q\|x\|_q} &= \ts\frac{1}{\gamma_q\|x\|_q}\cdot \med(|y_1|,\dots,|y_n|)\\
&=\ts\med(\frac{|y_1|}{\gamma_q \|x\|_q},\dots,\frac{|y_n|}{\gamma_q \|x\|_q})\\
&= \ts\med(\F_n).
\end{align}
Note also that $F_n\overset{w}{\to}F$, where $F$ is the distribution function of the variable $|S_1+\bar{\rho}_q\e_1|$. Consequently, it follows from a standard that $\med(\F_n)\to_P \med(F)$, and then
\begin{equation}\label{medianlimit}
\ts\frac{\hat{m}_q}{\gamma_q\|x\|_q} \xrightarrow{\ \ }_P \med(F) =: 1/c_1>0.
\end{equation}
(Note that it follows from Anderson's Lemma~\cite{andersonlemma} that $\med(F)\geq \med(|S_1|)$, which is clearly positive.)  
Altogether, we have verified that $\hat{t}_{\text{initial}}=1/\hat{m}_q$ satisfies $\hat{t}_{\text{initial}}\gamma_q\|x\|_q\to_P c_1$. Combining the previous limit with line~\eqref{adjustedlimit} and Theorem~\ref{clt} then proves line~\eqref{pilotconsist}, which in turn implies line~\eqref{rhoconsist}.\qed

\subsection{Proposition~\ref{mest} -- consistency of $c^{\star}(\hat{\rho}_q)$.}
\noindent \emph{Proof.}
As described in Section~\ref{deconvmodel}, we use $\xi$ as a implicit index in our asymptotics, and recall that by assumption~{\textbf{A\ref{rhoassumption}}}, we have $\rho_q=\rho_q(\xi)\to \bar{\rho}_q$ as $\xi\to\infty$. As a preliminary step, we will show $c^{\star}(\rho_q(\xi))$ is a bounded sequence. Namely, we will show there is a fixed compact interval $[\ve_q,c_{\max}]$ such that for all large $\xi$,

\begin{equation}\label{compactreduction}
c^{\star}(\rho_q(\xi))\in [\ve_q,c_{\max}].
\end{equation}
To show this, let $\ell_q(c)$ denote the right hand side of the bound~\eqref{secondlower}, which satisfies
\begin{equation}\label{ellprop1}
\ell_q(c)\leq \tilde{v}_q(c,\rho)
\end{equation}
 for all $q\in(0,2]$, all $\rho$, and all $c>0$. Also let $\bar{v}$ be any number satisfying 
\begin{equation}\label{vbardef}
\tilde{v}_q(c^{\star}(\bar{\rho}_q),\bar{\rho}_q)<\bar{v}.
\end{equation} 
Since $\tilde{v}_q(\cdot,\cdot)$ is continuous, it follows that as $\xi\to\infty$,
\begin{equation}
\tilde{v}_q(c^{\star}(\bar{\rho}_q),\rho_q(\xi))\to \tilde{v}_q(c^{\star}(\bar{\rho}_q),\bar{\rho}_q)
\end{equation}
and so line~\eqref{vbardef} forces us to conclude that
\begin{equation}\label{anotherineq}
\tilde{v}_q(c^{\star}(\bar{\rho}_q),\rho_q(\xi))<\bar{v} \text{ \ for large \ $\xi$}.
\end{equation}
Now, since $\ell_q(c)\to\infty$ as $c\to\infty$, and $\ell_q(\cdot)$ is continuous away from 0, there must be a point $c_{\max}>0$ such that 
\begin{align}
\ell_q(c_{\max})&=\bar{v}, \ \text{and}\\[0.3cm]
 \ell_q(c)&\geq \bar{v} \ \ \text{ for all $c\geq c_{\max}$.} \label{ellprop}
 \end{align}
This shows that $c^{\star}(\rho_q(\xi))$ cannot be greater than $c_{\max}$ when $\xi$ is large, for otherwise~\eqref{ellprop} and~\eqref{ellprop1} imply 
\begin{align}
\bar{v}&\leq \ell(c^{\star}(\rho_q(\xi)))\\[0.2cm]
&\leq v(c^{\star}(\rho_q(\xi)),\rho_q(\xi))\\[0.2cm]
&\leq v(c^{\star}(\bar{\rho}_q(\xi)),\rho_q(\xi)) \text{ \ \  by definition of $c^{\star}(\cdot)$,}
\end{align} contradicting line~\eqref{anotherineq}. Hence, line~\eqref{compactreduction} is true.\\

Since we know that $c^{\star}(\rho_q(\xi))$ is a bounded sequence, we can show that $c^{\star}(\rho_q(\xi))$ converges to $c^{\star}(\bar{\rho}_q))$ if all of its convergent subsequences do. Likewise, suppose there is some $\breve{c}\in[\ve_q,c_{\max}]$, such that along some subsequence $\xi_j\to\infty$,
\begin{equation}\label{brevelimit}
c^{\star}(\rho_q(\xi_j)) \to \breve{c}\in[\ve_q,c_{\max}].
\end{equation}
We now argue that $\breve{c}$ must be equal to $c^{\star}(\bar{\rho}_q).$
Due to the continuity of $\tilde{v}_q(\cdot,\cdot)$ and the limit~\eqref{brevelimit}, we have
\begin{align}
\tilde{v}_q(\breve{c},\bar{\rho}_q) &= \underset{j\to\infty}{\lim} \ \tilde{v}_q(c^{\star}(\rho_q(\xi_j)),\rho_q(\xi_j))\\[0.2cm]
&\leq \underset{j\to\infty}\lim \ \tilde{v}_q(c^{\star}(\bar{\rho}_q),\rho_q(\xi_j))\label{vbound1} \text{ \ \ \ by definition of $c^{\star}(\cdot)$} \\[0.2cm]
&=\tilde{v}_q(c^{\star}(\bar{\rho}_q),\bar{\rho}_q))\\[0.2cm]
&\leq \tilde{v}_q(\breve{c},\bar{\rho}_q),\label{vbound2}
\end{align}
which forces $\tilde{v}_q(\breve{c},\bar{\rho}_q) = \tilde{v}_q(c^{\star}(\bar{\rho}_q),\bar{\rho}_q)$, and the uniqueness assumption~{\textbf{A\ref{uniqueassumption}}} gives $\breve{c}=c^{\star}(\bar{\rho}_q)$, as desired.\qed

\section{Proofs for Section~\ref{sec:neg}}\label{app:neg}
\subsection{Proof of Lemma~\ref{impslem}.}

\noindent \emph{Proof of inequality~\eqref{newbound}.} It is enough to prove the result for $s_2(x)$ since $s_q(x)\geq s_2(x)$ for all $q\in [0,2]$. Let $d$ be the dimension of the null space of $A$, and let $B\in \R^{p\times d}$ be a matrix 
whose columns are an orthonormal basis for the null space of $A$. If $x\neq 0$, then define the scaled matrix $\tilde{B}:=\|x\|_{\infty}B$. (If $x=0$, the steps of the proof can be repeated using $\tilde{B}=B$.)
Letting $z\in \R^{d}$ be a standard Gaussian vector, we will study the random vector
$$\tilde{x} := x+\tilde{B}z,$$
 which satisfies $Ax=A\tilde{x}$ for all realizations 
of $z$. We begin the argument by defining a function $f:\R^p\to \R$ according to
\begin{equation}\label{func}
\textstyle
\begin{split}
f(\tilde{x})&: =\|\tilde{x}\|_1-c(n,p)\|\tilde{x}\|_2,
\end{split}
\end{equation}
where 
\begin{equation}\label{const}
c(n,p):=\ts\frac{1}{\sqrt{\pi e}} \frac{(p-n)}{\sqrt{p}}.
\end{equation}
The essential point to notice is that the event $\{f(\tilde{x})>0\}$ is equivalent to
$$\ts\frac{\|\tilde{x}\|_1^2}{\|\tilde{x}\|_2^2} > c(n,p)^2 
= \ts\frac{1}{\pi e} (1-\ts\frac{n}{p})^2 p,$$
which is the desired bound. (Note that $\tilde{x}$ is non-zero with probability 1.)
Hence, a vector $\tilde{x}$ satisfying the bound~\eqref{newbound} exists if the event $\{f(\tilde{x})>0\}$ occurs with positive probability.
We will prove that the probability $\P(f(\tilde{x})>0)$ is positive by showing $\E[f(\tilde{x})]>0$, which in turn can be reduced to
 upper-bounding
$\E \|\tilde{x}\|_2$, and lower-bounding 
$\E\|\tilde{x}\|_1$. The upper bound on $\E \|\tilde{x}\|_2$ follows from Jensen's inequality and a direct calculation,

\begin{equation}\label{2norm}
\begin{split}
\E\|\tilde{x}\|_2&=\E\|x+Bz\|_2\\
 &< \sqrt{\E \|x+\tilde{B}z\|_2^2}\\
&= \sqrt{\|x\|_2^2+ \|\tilde{B}\|_F^2}\\
&=\sqrt{\|x\|_2^2+\|x\|_{\infty}^2 d}.
\end{split}
\end{equation}
The lower bound on $\E\|\tilde{x}\|_1$ is more involved. If we let $\tilde{b}_i$ denote the $i$th row of $\tilde{B}$,  then the $i$th coordinate of $\tilde{x}$ can be written as $\tilde{x}_i = x_i+\langle \tilde{b}_i,z\rangle$, which is distributed according to $N(x_i, \|\tilde{b}_i\|_2^2)$. Taking the absolute value $|\tilde{x}_i|$ results in a ``folded normal'' distribution, whose expectation can be calculated exactly as
\begin{equation}\label{folded}
\E|\tilde{x}_i| = \|\tilde{b}_i\|_2\sqrt{\ts\frac{2}{\pi}}\exp\Big(\ts\frac{-x_i^2}{2\|\tilde{b}_i\|_2^2}\Big)+|x_i|\Big(1-2\Phi\big(\ts\frac{-|x_i|}{\|\tilde{b}_i\|_2}\big)\Big),
\end{equation}
where $\Phi$ is the standard normal distribution function. Note that it is possible to have $\|\tilde{b}_i\|_2=0$, in which case $\tilde{x}_i=x_i$. This separate case can be easily handled in the rest of the argument.

When $|x_i|/\|\tilde{b}_i\|_2$ is small, the first term on the right side of~\eqref{folded} dominates, and then $\E|\tilde{x}_i|$ is roughly $\|\tilde{b}_i\|_2$. Alternatively, when $|x_i|/\|\tilde{b}_i\|_2$ is large, the second term dominates, and then  $\E|\tilde{x}_i|$ is roughly $|x_i|$. Thus, it is natural to consider the set of indices $\mathcal{I}_1=\{i : \|\tilde{b}_i\|_2\geq |x_i|\}$, and its complement \mbox{$\mathcal{I}_2=\{i: \|\tilde{b}_i\|_2< |x_i|\}$}. This leads us to the following bounds,
\begin{equation}
\begin{split}
\E\|\tilde{x}\|_1&= \sum_{i\in\mathcal{I}_1} \E|\tilde{x}_i|+ \sum_{i\in\mathcal{I}_2} \E|\tilde{x}_i|\\
&\geq \sum_{i\in\mathcal{I}_1} \|\tilde{b}_i\|_2 \sqrt{\ts\frac{2}{\pi}}\exp(-\ts\frac{1}{2})+\displaystyle\sum_{i\in\mathcal{I}_2}|x_i| (1-2\Phi(-1))\\
&\geq \sum_{i\in\mathcal{I}_1} \|\tilde{b}_i\|_2 \sqrt{\ts\frac{2}{\pi e}}+\sum_{i\in\mathcal{I}_2}\|\tilde{b}_i\|_2 (1-2\Phi(-1))\\
&\geq \sqrt{\ts\frac{2}{\pi e}}  \sum_{i=1}^p \|\tilde{b}_i\|_2 \ \ \ \ \text{ using } (1-2\Phi(-1))\geq \sqrt{\ts\frac{2}{\pi e}},\\
&=\sqrt{\ts\frac{2}{\pi e}} \|x\|_{\infty} \sum_{i=1}^p \|b_i\|_2,
\end{split}
\end{equation}
where $b_i$ is the $i$th row of $B\in \R^{p\times d}$. Since the matrix $B\in \R^{p\times d}$ has orthonormal columns, it may be regarded as a submatrix of an orthogonal $p\times p$ matrix, and so the rows $b_i$ satisfy $\|b_i\|_2\leq 1$,  yielding $\|b_i\|_2\geq \|b_i\|_2^2$. Hence,
$$\tsum_{i=1}^p \|b_i\|_2 \geq \tsum_{i=1}^p \|b_i\|_2^2= \|B\|_F^2 = d\geq p-n.$$
Altogether, we obtain the bound
\begin{equation}\label{l1bound}
\E\|\tilde{x}\|_1 \geq \sqrt{\ts\frac{2}{\pi e}} \|x\|_{\infty} (p-n).
\end{equation}
Combining the bounds~\eqref{l1bound} and~\eqref{2norm}, and noting that $d\leq p$, we obtain
\begin{equation}
\begin{split}
\frac{\E\|\tilde{x}\|_1}{\E\|\tilde{x}\|_2}&>  \ts\frac{\sqrt{\ts\frac{2}{\pi e}} \|x\|_{\infty} (p-n)}{\sqrt{\|x\|_2^2+\|x\|_{\infty}^2 p}}
\\[0.3cm]
&= \ts\frac{\sqrt{\ts\frac{2}{\pi e}}(p-n)}{\sqrt{\ts\frac{\|x\|_2^2}{\ \|x\|_{\infty}^2}+ p}}
\\[0.3cm]
&\geq\ts\frac{1}{\sqrt{\pi e}} \frac{(p-n)}{\sqrt{p}}\\[0.2cm]
&=c(n,p),\\
\end{split}
\end{equation}
where we have used the fact that $\ts\frac{\|x\|_2^2}{\ \|x\|_{\infty}^2}\leq p$. This proves $\E[f(\tilde{x})]>0$, giving~\eqref{newbound}.\qed\\

\noindent \emph{Proof of inequality~\eqref{sinftybound}.} It is enough to prove the result for $s_{\infty}(x)$ since $s_q(x)\geq s_{\infty}(x)$ for all $q\in[0,\infty]$. We retain the same notation as in the proof above. Following the same general argument, it is enough to show that
\begin{equation}\label{enoughinfty}
\frac{\E\|\bar{x}\|_1}{\E\|\bar{x}\|_{\infty}} > \bar{c}(n,p),
\end{equation}
where 
\begin{equation}
\bar{c}(n,p):=\frac{\sqrt{\ts\frac{2}{\pi e}} (p-n)}{1 + \sqrt{16\log(2p)}}.
\end{equation}
In particular, we will re-use the bound
\begin{equation}\label{l1boundagain}
\E\|\tilde{x}\|_1 \geq \sqrt{\ts\frac{2}{\pi e}} \|x\|_{\infty} (p-n).
\end{equation}

The new item to handle is an upper bound on $\E\|\tilde{x}\|_{\infty}$.
Clearly, we have $\|\tilde{x}\|_{\infty}\leq \|x\|_{\infty}+\|\tilde{B}z\|_{\infty}$, and so it is enough 
to upper-bound $\E\|\tilde{B}z\|_{\infty}$. We will do this using a version of 
Slepian's inequality. If $\tilde{b}_i$ denotes the $i^{\text{th}}$ row of $\tilde{B}$, 
define the random variable  $g_i=\langle \tilde{b}_i,z\rangle$, and let $w_1,\dots, w_p$ be i.i.d. $N(0,1)$ 
variables. The idea is to compare the Gaussian process $g_i$ with the Gaussian 
process $\sqrt{2}\|x\|_{\infty}w_i$. By Proposition A.2.6 in the book~\cite{vaartWellner}, the inequality
$$\E\|\tilde{B}z\|_{\infty}  = \E\left[ \max_{1\leq i\leq p}|g_i| \right]\leq 2\sqrt{2}\|x\|_{\infty}\,\E \left[\max_{1\leq i\leq p}|w_i|\right],$$
holds as long as the condition $\E(g_i-g_j)^2\leq 2\|x\|_{\infty}^2\, \E(w_i-w_j)^2$ 
is satisfied for all $i,j\in \{1,\dots,p\}$. This can be verified by first noting that $g_i-g_j = \langle \tilde{b}_i-\tilde{b}_j,z\rangle$, which is distributed according to $N(0,\|\tilde{b}_i-\tilde{b}_j\|_2^2)$. Since $\|\tilde{b}_i\|_2\leq \|x\|_{\infty}$ for all $i$, it follows that
\begin{equation}
\begin{split}
\E(g_i-g_j)^2 &= \|\tilde{b}_i-\tilde{b}_j\|_2^2\\
&\leq 4\|x\|_{\infty}^2\\
&=2\|x\|_{\infty}^2\E(w_i-w_j)^2,
\end{split}
\end{equation}
as needed.
To finish the proof, we make use of a standard bound for the expectation of Gaussian maxima 
$$
\E \left[\max_{1\leq i\leq p}|w_i|\right] < \sqrt{2\log(2p)},
$$
which follows from a modification of the proof of Massart's finite class 
lemma~\cite[Lemma 5.2]{massartFiniteClass}.
Combining the last two steps, we obtain
\begin{equation}\label{inftynorm}
\E\|\tilde{x}\|_{\infty}< \|x\|_{\infty} + 2\sqrt{2}\|x\|_{\infty}\sqrt{2\log(2p)}.
\end{equation}
Hence, the bounds~\eqref{l1boundagain} and~\eqref{inftynorm} clearly lead to~\eqref{enoughinfty}. \qed

\subsection{Proof of Theorem~\ref{minimaxnew}}

\proof We begin by making some reductions. First, we claim it is enough to show that 
\begin{equation}\label{red1}
\inf_{A\in \R^{n\times p}} \inf_{\delta:\R^n\to \R} \:\sup_{x\in\R^p\setminus\{0\}}\Big|\delta(Ax)-s_2(x)\Big|\geq \ts\frac{1}{2} \ts\frac{1}{\pi e}\cdot(1-\ts\frac{n}{p})^2\cdot p-\ts\frac{1}{2}.
 \end{equation}
To see this, note that the general inequality $s_2(x)\leq p$ implies
$$\big|\ts\frac{\delta(Ax)}{s_2(x)}-1\big| \geq \ts\frac{1}{p}\big|\delta(Ax)-s_2(x)\big|,$$
and we can optimize both sides with respect $x,\delta,$ and $A$. Next, for any fixed matrix $A\in \R^{n\times p}$, it is enough to show that
\begin{equation}\label{red1}
\inf_{\delta:\R^n\to \R} \:\sup_{x\in\R^p\setminus\{0\}}\Big|\delta(Ax)-s_2(x)\Big|\geq \ts\frac{1}{2}\ts\frac{1}{\pi e}\cdot(1-\ts\frac{n}{p})^2\cdot p-\ts\frac{1}{2},
 \end{equation} 
as we may take the infimum over all matrices $A$ without affecting the right hand side. To make a third reduction, it is enough to prove the same bound when $\R^p\setminus\{0\}$ is replaced with any subset, as this can only make the supremum smaller. In particular, we replace $\R^p\setminus\{0\}$ with the two-point subset   $\{e_1,\tilde{x}\}$, where $e_1=(1,0,\dots,0)\in\R^p$, and by Lemma 1, there exists $\tilde{x}$ to satisfying $Ae_1=A\tilde{x}$, with
$$s_2(e_1)=1, \text{  \  and \ \ } s_2(\tilde{x})\geq \ts\frac{1}{\pi e}\cdot(1-\ts\frac{n}{p})^2\cdot p.$$

We now complete the proof by showing that the lower bound \eqref{red1} holds for the two-point problem, i.e.
\begin{equation}\label{red2}
\inf_{\delta:\R^n\to \R} \:\sup_{x\in\{e_1,\tilde{x}\}}\Big|\delta(Ax)-s_2(x)\Big|\geq \ts\frac{1}{2}\ts\frac{1}{\pi e}\cdot(1-\ts\frac{n}{p})^2\cdot p - \ts\frac{1}{2},
 \end{equation} 
 and we will accomplish this using the classical technique of constructing a Bayes procedure with constant risk. For any decision rule $\delta:\R^n\to \R$, any $A\in\R^{n\times p}$, and any point \mbox{$x\in \{e_1,\tilde{x}\}$,} define the (deterministic) risk function 
 $$R(x, \delta):=\big| \delta(Ax)-s_2(x)\big|.$$
 Also, for any prior $\pi$ on the two-point set $\{e_1,\tilde{x}\}$, define
 $$r(\pi,\delta):= \int R(x,\delta) d\pi(x).$$ 
  By Propositions 3.3.1 and 3.3.2 of~\cite{bickelDoksum}, the inequality~\eqref{red2} holds if there exists a prior distribution $\pi^*$ on $\{e_1,\tilde{x}\}$ and a decision rule $\delta^*:\R^n\to\R$ with the following three properties:
\begin{enumerate}
\item The rule $\delta^*$ is Bayes for $\pi^*$, i.e. $r(\pi^*,\delta^*)= \inf_{\delta} r(\pi^*,\delta)$.
\item The rule $\delta^*$ has constant risk over $\{e_1,\tilde{x}\}$, i.e. $R(e_1,\delta^*)= R(\tilde{x},\delta^*)$.
\item The constant value of the risk of $\delta^*$ is at least $ \ts\frac{1}{2}\ts\frac{1}{\pi e}\cdot(1-\ts\frac{n}{p})^2\cdot p-\ts\frac{1}{2}$.
\end{enumerate} 
To exhibit $\pi^*$ and $\delta^*$ with these properties, we define $\pi^*$ to be the two-point prior that puts equal mass at $e_1$ and $\tilde{x}$, and we define $\delta^*$ to be the trivial decision rule that always returns the average of the two possibilities, namely  $\delta^*(Ax)\equiv\frac{1}{2}(s_2(\tilde{x})+s_2(e_1))$ for all $x\in\{e_1,\tilde{x}\}$.  It is simple to check the second and third properties.
To check that $\delta^*$ is Bayes for $\pi^*$,  the triangle inequality gives
\begin{equation}
\begin{split}
r(\pi^*,\delta)&=\ts\frac{1}{2}\Big|\delta(A\tilde{x})-s_2(\tilde{x})\Big|+\frac{1}{2}\Big|\delta(Ae_1)-s_2(e_1)\Big|,\\[0.2cm]
&\geq\ts \frac{1}{2}\big|s_2(\tilde{x})-s_2(e_1)\big|\\[0.2cm]
&=\ts\frac{1}{2}\big| \delta^*(A\tilde{x})-s_2(\tilde{x})\big|+\ts\frac{1}{2}\big|\delta^*(Ae_1)-s_2(e_1)\big|\\[0.2cm]
&=r(\pi^*,\delta^*),
\end{split}
\end{equation}
which holds for every $\delta$, implying that $\delta^*$ is Bayes for $\pi^*$.
\qed

\section{A unique minimizer for the variance function with stable noise}\label{misc}
In this section, we aim to show that when the noise distribution is $\text{stable}_q(1)$, the variance function $\tilde{v}_q(\cdot,\bar{\rho}_q)$ has a unique minimizer in $[\ve_q,\infty)$.\footnote{Recall that $\ve_q=0$ for $q\in(0,2)$ and $\ve_2>0$.} Note that since  $\varphi_0$ has no roots in this case, the extended variance function $\tilde{v}_q(c,\bar{\rho}_q)$ agrees with $v_q(c,\bar{\rho}_q)$ for all $c\neq 0$. Furthermore, when $q\in(0,2)$ the minimizer cannot occur at $c=0$ due to Lemma~\ref{varextend}. Hence, for all $q\in (0,2]$ it is enough to check that $v_q(\cdot,\bar{\rho}_q)$ has a unique minimizer in $(0,\infty)$.

Recall that the characteristic function for $\text{stable}_q(1)$ is
\begin{equation}
\varphi_0( t) = \exp(-|t|^q),
\end{equation}
and it follows that for any $q\in (0,2]$, the variance function is given by
	\begin{equation}\label{varformulalast}
	\begin{split}
	v_q(c,\bar{\rho}_q)&= \ts\frac{1}{|c|^{2q}}\Big(\ts\frac{1}{2}\frac{1}{\varphi_0(\bar{\rho}_q |c|)^2}\exp(2|c|^q) +\ts\frac{1}{2}\frac{\varphi_0(2\bar{\rho}_q |c|)}{\varphi_0(\bar{\rho}_q |c|)^2}\exp((2-2^q) |c_0|^q)- 1\Big)\\
	&=	%
	 \ts\frac{1}{|c|^{2q}}\Big(\ts\frac{1}{2}\exp((2(\bar{\rho}_q^q+1)|c|^q)+\ts\frac{1}{2}\exp((2-2^q)(\bar{\rho}_q^q+1)|c|^q)- 1\Big).\\[0.2cm]
	\end{split}
	\end{equation}
Now consider the monotone change of variable $u:=|c|^q$ for positive $c$, and notice that $v_q(c,\bar{\rho}_q)=f(u)/u^2$ where
\begin{equation}
f(u):=\ts\frac{1}{2}\exp(2(\bar{\rho}_q^q+1)u)+\ts\frac{1}{2}\exp((2-2^q)(\bar{\rho}_q^q+1)u)-1.
\end{equation}
The following lemma demonstrates the desired claim by showing that $u\mapsto f(u)/u^2$ is strictly convex on $(0,\infty)$. (We omit the simple derivative calculations involved in checking that this $f(u)$ satisfies the conditions of the lemma.) 
\begin{lemma}
Let $f:[0,\infty)\to \R$  be a $4$-times differentiable function such that $f(0)\geq 0$, $f'(0)\geq 0$, and $f^{(4)}(u)>0$ for all $u> 0$. 
Then, the function $u\mapsto \frac{f(u)}{u^2}$ is strictly convex on $(0,\infty)$.
\end{lemma}
\noindent \emph{Proof.}
Let $h(u)=\frac{f(u)}{u^2}$, and let $\psi(u)=u^4 h^{\prime\prime}(u)$. To show that $h''(u)$ is strictly positive on $(0,\infty)$, it suffices to show that $\psi(u)> 0$ for all $u>0$. By direct calculation,
$$\psi(u)=u^2f^{\prime\prime}(u)-4uf'(u)+6f(u),$$
and so the assumption $f(0)\geq 0$ implies $\psi(0)\geq 0$. Consequently, it is enough to show that $\psi$ is strictly increasing on $(0,\infty)$. Since,
$$\psi'(u)= u^2 f^{(3)}(u)-2uf^{\prime\prime}(u)+2f'(u),$$
the assumption $f'(0)\geq 0$ implies $\psi'(0)\geq 0$, and so it is enough to show that $\psi^{\prime}$ is strictly increasing on $(0,\infty)$. Differentiating $\psi'$ leads to a notable cancellation, giving
$$\psi^{\prime\prime}(u)=u^2 f^{(4)}(u),$$
and so the assumption on $f^{(4)}(u)>0$ for all $u>0$ completes the proof.\qed
{\bibliography{thesis_references_cauchy_submission}}

\bibliographystyle{alpha}

\end{document}